\documentclass[11pt,a4paper]{article}  
\usepackage{jheppub} 
\usepackage{amsmath} 
\usepackage{subcaption}
\usepackage{amssymb}
\usepackage{epsfig}
\usepackage{color} 
\usepackage{xcolor}
\usepackage{bm}
\usepackage{afterpage}

\title{Charmonium-like resonances with $J^{PC}=0^{++},2^{++}$ in\\ coupled  $D\bar D$, $D_s\bar D_s$ scattering on the lattice}

\subheader{MITP/20-065}


\author[a,b,c]{Sasa Prelovsek}
\author[a]{Sara~Collins}
\author[d,e,f]{Daniel~Mohler}
\author[d,e]{M.~Padmanath}
\author[a]{Stefano Piemonte}
\affiliation[a]{Institute for Theoretical Physics, University of Regensburg, 93040 Regensburg, Germany}
\affiliation[b]{Faculty of Mathematics and Physics, University of Ljubljana, Slovenia}
\affiliation[c]{Jozef Stefan Institute, Ljubljana, Slovenia}
\affiliation[d]{Helmholtz-Institut Mainz, Johannes Gutenberg-Universit\"at, D-55099 Mainz, Germany}
\affiliation[e]{GSI Helmholtzzentrum f\"ur Schwerionenforschung, 64291 Darmstadt, Germany}
\affiliation[f]{Johannes Gutenberg-Universit\"at Mainz, D-55099 Mainz, Germany} %

\emailAdd{sasa.prelovsek@ijs.si}
\emailAdd{pmadanag@uni-mainz.de}

\abstract{
We present the first lattice investigation of coupled-channel $D\bar D$ and $D_s\bar D_s$ scattering in the $J^{PC}=0^{++}$ and $2^{++}$ channels. The
scattering matrix for partial waves $l=0,2$ and isospin zero is determined using multiple
volumes and inertial frames via  L\"uscher's formalism. Lattice QCD
ensembles from the CLS consortium with $m_{\pi}\simeq280$ MeV,  $a \simeq 0.09 $ fm and   $L/a=24,~32$ are
utilized. The resulting scattering matrix suggests the existence of three charmonium-like
states with $J^{PC}=0^{++}$ in the energy region ranging from  slightly below $2m_D$ up to
$4.13~$GeV. We find a so far unobserved $D\bar D$ bound state just below
threshold and a $D\bar D$ resonance likely related to $\chi_{c0}(3860)$,
which is believed to be $\chi_{c0}(2P)$. In addition, there is an indication for a
narrow $0^{++}$ resonance just below the $D_s\bar D_s$ threshold with a large
coupling to $D_s\bar D_s$ and a very small coupling to $D\bar D$. This resonance is
possibly related to the narrow $X(3915)$/$\chi_{c0}(3930)$ observed in
experiment also just below $D_s\bar D_s$. The partial wave $l=2$ features a
resonance likely related to $\chi_{c2}(3930)$.  We work with several assumptions, such as the omission of $J/\psi\omega$, $\eta_c\eta$ and three-particle channels. Only statistical uncertainties  are quantified, while the extrapolations to the physical quark-masses and the continuum limit are challenges for the future. 
 }

\begin{document}
\maketitle

\section{Introduction}\label{sec:introduction}

Since the discovery of the $J/\psi$ meson in 1970s a multitude of
charmonium bound states and resonances have been found  with energies
ranging up to almost 5~GeV. A simple $c\bar{c}$ quark model
provides a reasonable description of the levels below the strong decay
thresholds and also some of the states above, however, there are
clearly too many states to fit into this picture. Some mesons,
such as the charged $Z_c$ states certainly have additional quark
content, while for other states the interpretation is not so clear.
On the theory side the nature of these states is being explored in
tetraquark, molecular, and hybrid meson models, among others, while
on the experimental side insight is provided by establishing their
quantum numbers, decay modes and widths. Lattice QCD studies of the
charmonium spectrum have a significant role to play in terms of guiding
experimental searches, determining the quantum numbers of the states
not well established as well as investigating their internal structure.

In this work we focus on the isoscalar channel $I(J^{PC})=0(0^{++})$
 in the
region up to 4.13~GeV for which there are a number of open questions.  The
ground state, $\chi_{c0}(1P)$, found well below the $D\bar{D}$ threshold is interpreted as the $^{3}1P_0$
$c\bar{c}$ level of the quark model and is the only well
established state. In the energy region around 3.9~GeV, above the
threshold, one expects a corresponding excited state. So far, three
hadrons have been observed with the possible assignment of
$J^{PC}=0^{++}$: the $X(3860)$, a broad resonance detected by
Belle~\cite{Chilikin:2017evr,pdg}, and two narrow resonances just below the
$D_s\bar{D}_s$ threshold --- the $\chi_{c0}(3930)$ discovered in the
$D\bar{D}$ channel by LHCb~\cite{chic03930,Aaij:2020hon} and the $X(3915)$ observed
through it's decay into $J/\psi\omega$~\cite{Aubert:2007vj,Uehara:2009tx,delAmoSanchez:2010jr,Lees:2012xs}~(with
the assignment of $J^{PC}=0^{++}$ or $2^{++}$). While the latter two
resonances could be the same state, their narrowness may indicate
exotic content, where    $X(3915)$  has been interpreted as  $c\bar{c}s\bar{s}$ meson in Ref. 
\cite{Lebed:2016yvr}. Predictions
have also been made for an additional, as yet unobserved, bound state just
below the $D\bar{D}$ threshold~\cite{Gamermann:2006nm,Gamermann:2007mu}.

The determination of the low lying charmonium spectrum on the lattice is
relatively straightforward, with the energy levels being directly
accessed from correlation functions measured on the configurations
generated in the Monte-Carlo simulation. Systematics arising from
finite lattice spacing and simulating with unphysical light (sea)
quark masses must be addressed by carrying out a continuum and quark-mass
extrapolation.  Near and above threshold, the analysis is considerably
more challenging with information on the masses and (for resonances)
also the widths being inferred from scattering amplitudes which can be 
obtained from the finite volume spectra via the L\"uscher
method~\cite{Luscher:1986pf,Luscher:1990ux,Luscher:1991cf}. Two-particle interpolators must be included in the basis of operators for
the construction of the correlation functions in order to reliably
determine these spectra. Simulating charmonia in flight provides
additional levels with which to probe the scattering matrix, however,
the identification of the continuum spin and parity quantum numbers of
the levels is complicated due to the reduced symmetry on the
lattice. In addition, for the energy range of interest both the
$D\bar{D}$ and $D_s\bar{D}_s$ thresholds must be considered
leading to a coupled-channel scattering analysis. 

So far, the coupled-channel scattering matrix has been extracted  for several  light-meson systems, for example,      $\pi K,\eta
K$~\cite{Dudek:2014qha,Wilson:2014cna}, $\pi \eta$,
$K\bar{K}$~\cite{Dudek:2016cru} and $\pi\pi, K\bar{K},
\eta\eta$ \cite{Briceno:2017qmb} by the Hadron Spectrum Collaboration.  In the heavy sector,
there has been one investigation of $D\pi,D\eta,D_{s}\bar{K}$
scattering in isospin-1/2~\cite{Moir:2016srx} and a recent analysis of
the $Z_c(3900)$ via $D^*\bar{D}, J/\psi\pi$
scattering~\cite{Chen:2019iux}.
The HALQCD Collaboration has also investigated the $Z_c(3900)$ using a different approach which involves solving the Schr\"odinger equation with potentials determined on the lattice \cite{Ikeda:2016zwx,Ikeda:2017mee}.
Pioneering works such as these were
limited to a single lattice spacing and unphysical light-quark masses.

The
charmonium  scalar channel has previously been studied by
some of the authors considering only $D\bar{D}$ scattering with total momentum zero~\cite{Lang:2015sba}.  
 Here we present a lattice study of  scattering  in the coupled-channels
  $D \bar{D}$ and   $D_s\bar{D}_s$  with
  quantum numbers $I=0$ and $J^{PC}=0^{++},2^{++}$. This represents the first
  determination of the coupled-channel scattering matrix  from lattice QCD
  in the charmonium system with isospin zero. 
 Two lattice volumes are employed for the charmonium system  at rest   and in flight.    
 This analysis uses the same lattice setup as our previous
article on the identification of the spin and parity of the single
hadron spectrum~\cite{Padmanath:2018tuc} and the investigation of
single channel $D\bar{D}$ scattering for $J^{PC}=1^{--}$ and
$3^{--}$~\cite{Piemonte:2019cbi}.
While the present study represents a significant improvement on
previous work, some simplifications remain and a comparison of the
results for the masses and widths with experiment is qualitative.
Within the energy range of interest, additional scattering channels,
such as the $J/\psi\omega$, $\eta_c\eta$ and those involving three
particles, could in principle also be  relevant.  The effects of 
these channels will be investigated in the future, along with
systematics associated with finite lattice spacing and unphysical
light quark masses.

\vspace{0.2cm}

  The remainder of the paper is organized as follows. We begin by
  reviewing the essential general aspects of one-channel and two-channel
  scattering in Section~\ref{sec:generalities}.  The details of the lattice setup and methodology are
  given in Section~\ref{sec:lattice-details} and the single- and
  two-meson interpolators used in the correlation functions are
  discussed in Section~\ref{sec:operators}. 
  Simplifying assumptions made in this study are  summarized in  Section~\ref{sec:caveats}. 
  The first step in extracting
  the scattering amplitudes is to compute the finite-volume spectra
  from the correlation functions. Our analysis and the final spectra
  are presented in Section~\ref{sec:En}.  An overview of determining
  the scattering amplitudes from the lattice eigen-energies is
  provided in Section \ref{sec:SfromE}.  Our results for the
  $J^{PC}=0^{++}$ and $2^{++}$ channels are detailed in
  Section~\ref{sec:channels} and the relation to states observed in
  experiment is discussed in Section~\ref{sec:summary}. Finally,
  Section \ref{sec:conclusions} presents our conclusions. More details are given in several Appendices.

\section{Generalities on scattering matrices, poles, hadron masses and widths }\label{sec:generalities}
 
 The masses and widths of strongly-decaying resonances should be inferred from
 the study of scattering processes where these resonances appear. In this section, we briefly review relevant concepts regarding scattering matrices, complex energy planes,
    pole singularities, hadron masses, and their decay widths. The first part 
    lists definitions and notations for the scattering amplitudes, the
    phase space factors, {\it etc.}. The second 
    part discusses naming conventions for various Riemann sheets, pole
    singularities in the complex energy plane
    and their relation to the hadron properties.
   
    \subsection{Scattering matrices for real energies}
    
    The unitary scattering  amplitude $S$  for {\it one-channel scattering} ($D\bar D$ or $D_s\bar D_s$)  of spin-less particles in partial wave $l$  is generally parametrized in terms of the energy-dependent phase shift $\delta(E_{cm})$,
 \begin{align}
 \label{S-elastic}
 S&=1+2~ i~ ~\rho  t=e^{2i\delta}~,  \\
 t^{-1}&= \cot\delta ~\rho-i~\rho =\frac{2}{E_{cm} ~p^{2l}} ~\tilde K^{-1}   - i ~\rho~ \quad \mathrm{with}\quad  \tilde K^{-1}=p^{2l+1} \cot\delta~,\nonumber  
  \end{align}
 where $\rho\equiv 2p/E_{cm}$, $p$ denotes the momentum of the scattering
 particles in the center-of-momentum frame and $t$ is the scattering amplitude.  The factors $p^{-2l}$ in front of $\tilde K^{-1}$ lead to smooth behavior close to the threshold. In the case of $t$ exhibiting simple Breit-Wigner type behavior, $\tilde K^{-1}/E_{cm}$ falls linearly  as a function of $E_{cm}^2$,
 \begin{equation}
 \label{bw}
 t(E_{cm})=\frac{-E_{cm}~\Gamma(E_{cm})}{E_{cm}^2-m_R^2+i~E_{cm}~\Gamma(E_{cm})}~,\quad \Gamma(E_{cm})=g^2 ~\frac{p^{2l+1}}{E_{cm}^2}~,\quad \frac{\tilde K^{-1}}{E_{cm}}=\frac{m_R^2-E_{cm}^2}{g^2}~.
 \end{equation}
 The phase shift equals $\pi/2$ at $E_{cm}=m_R$, while the width
 $\Gamma(E_{cm})$ is parametrized in terms of the coupling $g$ and the phase
 space.  $S$, $t$, $\tilde K$ and $\delta$ depend on $E_{cm}$ and partial wave
 $l$ (the dependence on $l$ is not written explicitly). 
    
For {\it coupled-channel scattering} of  $D\bar D$ and $D_s\bar D_s$
  in partial wave $l$, the  scattering matrices $S$ are energy-dependent $2\times 2$ unitary matrices,
   \begin{align}
 \label{S-coupled}
 S_{ij}&=\delta_{ij}+2~ i~ \sqrt{\rho_i\rho_j}  ~t_{ij}=\left\{
                \begin{array}{ll}  \eta ~e^{2i\delta^i}\qquad\quad\;\;\qquad i=j\nonumber\\
 \sqrt{1-\eta^2}  ~e^{i(\delta^i+\delta^j)}\quad i\not = j
 \end{array}
              \right.~,
 \nonumber \\
 (t^{-1})_{ij}&= \frac{2}{E_{cm} ~p_i^l p_j^l} ~(\tilde K^{-1})_{ij}   - i ~\rho_i~ \delta_{ij}~, \\
 \rho_i&\equiv 2p_i/E_{cm}=\sqrt{1-(2m_i)^2/E_{cm}^2}~,\quad i,j~=\ 1 ~ (D\bar
         D),\ 2~  (D_s\bar D_s)~.\nonumber
 \end{align}
 The momenta of $D$ and
 $D_s$ in the center-of-momentum frame are denoted by $p_{1}$ and $p_2$,
 respectively. $t$ is the scattering matrix and $\tilde K(E_{cm})$ is a real symmetric matrix.  
 We follow the definition of $t$ by the Hadron Spectrum Collaboration  (e.g. \cite{Dudek:2014qha}) and the definition of $\tilde K$ from Ref. \cite{Brett:2018jqw}\footnote{Note that unlike in Ref. \cite{Brett:2018jqw} we do not divide our energy levels or physical quantities by the mass of the scattering particles.}.

 \subsection{Continuation to complex $E_{cm}$, Riemann sheets and poles }
  
 In experiment and lattice QCD simulations the scattering matrices
 $S(E_{cm})$ are determined for real energies. The theoretical interpretation in terms of (virtual) bound states
 and resonances is conventionally made via the poles in the $t$-matrix, 
 analytically continued to the complex $s$-plane. 
 The feature that leads to interesting physics is the square root branch cut related to
 $\rho=2p/E_{cm}=\sqrt{1-(2m)^2/E_{cm}^2}$ starting from the threshold connecting the physical 
 Riemann sheet (or sheet I), conventionally chosen to have $\mathrm{Im}(\rho)>0$, 
 to the unphysical Riemann sheet (or sheet II), which has $\mathrm{Im}(\rho)<0$. For a two channel system, there will be four 
 Riemann sheets, such that
  \begin{align}
 \label{sheets}
 &\mathrm{sheet}~ I~~~: \mathrm{Im}(\rho_D)\!>\!0,\mathrm{Im}(\rho_{D_s})\!>\!0\ , \ \mathrm{sheet}~II: \mathrm{Im}(\rho_D)\!<\!0,\mathrm{Im}(\rho_{D_s})\!>\!0~, \quad (\rho_i=2p_i/E_{cm})~, \nonumber \\
 &\mathrm{sheet}~ III: \mathrm{Im}(\rho_D)\!<\!0,\mathrm{Im}(\rho_{D_s})\!<\!0\ ,\ \mathrm{sheet}~ IV: \mathrm{Im}(\rho_D)\!>\!0,\mathrm{Im}(\rho_{D_s})\!<\!0~.
  \end{align}

\begin{figure}[t]
	\begin{center}
	  \centerline{\includegraphics[width=0.23\textheight]{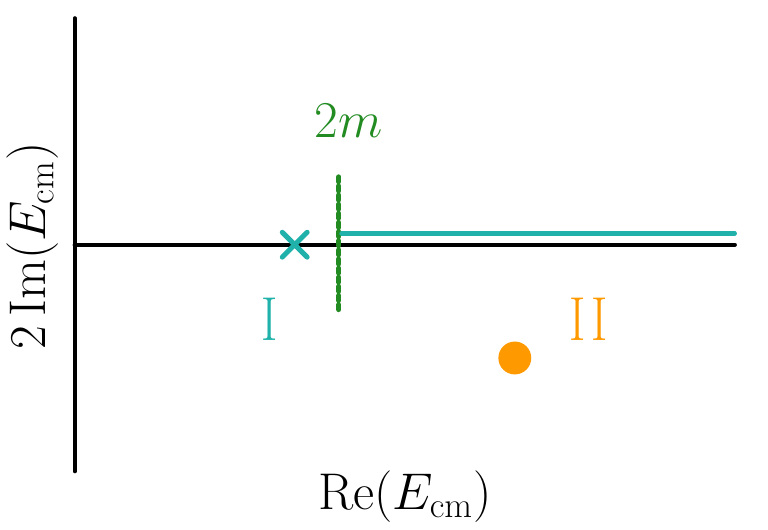}
            \hspace{1cm}
            \includegraphics[width=0.23\textheight]{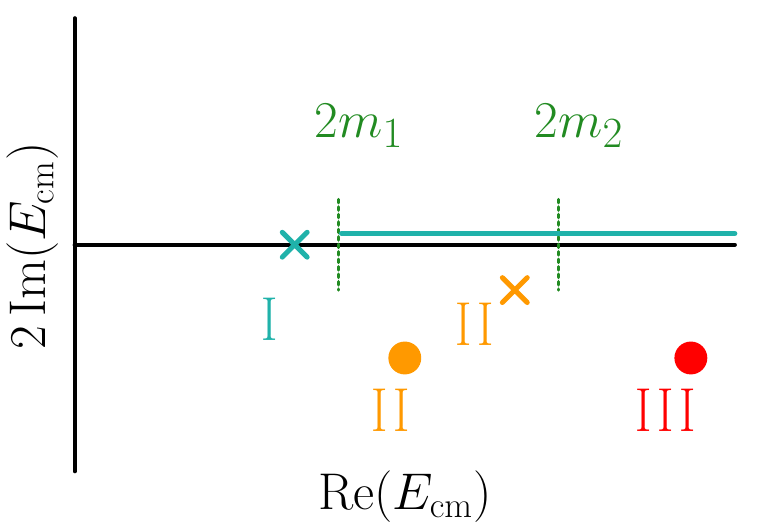}
            }
		\caption{Sketch of the pole locations in the scattering matrix
                  $t$  that typically affect the experimental  rates on the
                  physical axes (denoted by the cyan line) for one-channel
                  scattering (left) and two coupled channels (right). The Roman numbers indicate the Riemann sheets where the poles are located according to Eq.~(\ref{sheets}). Poles immediately below a threshold, indicated by crosses, can also have observable effects on the physical axes above the respective threshold.}
	\label{fig:sketch}
	\end{center}
  \end{figure}

Bound states, virtual bound states and resonances are related to pole singularities of $t$ in the complex $s$-plane. 
 These poles affect the physical axes, indicated by the cyan line in Fig.~\ref{fig:sketch}, along 
which the experimental measurements are made. Fig.~\ref{fig:sketch} presents a schematic picture of various pole 
locations in our study, that can affect scattering amplitudes/matrices along the physical axes for 
one-channel and two-channel scattering. The  location of the poles are related to the masses and widths via  $E^{p}_{cm}=m-\tfrac{i}{2}\Gamma$ for resonances and  $E^{p}_{cm}=m$ for the (virtual) bound states. 
 
In the close vicinity of the pole, the scattering matrix has the energy dependence 
\begin{equation}
\label{residues}
t_{ij}\sim \frac{c_i~c_j}{(E_{cm}^p)^2-E_{cm}^2}\quad \mathrm{for}\quad  E_{cm}\simeq E_{cm}^p~, \qquad i=\ 1 ~ (D\bar D),\ 2~  (D_s\bar D_s)~,
\end{equation}
 and the residue ($c_ic_j$) can typically be factorized into the couplings $c_i$ (\footnote{The couplings $c_i$ should not be confused with the coupling $g$ (\ref{bw}) that will be used to parametrize the full width of a resonance. }), whose relative size is related to the branching ratios of a resonance (associated with the pole) to both channels $i=1,2$.  

\section{Lattice  details}\label{sec:lattice-details}

We employ two ensembles generated with  $N_f=2+1$ non-perturbatively
$\mathcal{O}(a)$ improved Wilson dynamical fermions 
provided by the Coordinated Lattice Simulations (CLS) consortium
\cite{Bruno:2014jqa,Bali:2016umi}. The quark masses $m_{u/d,s}$ are chosen along a
trajectory that approaches the physical point holding the average quark mass, $2m_{u/d}+m_s $,
constant. The ensembles, denoted H105 and U101, have an
inverse gauge coupling $\beta=6/g^2=3.4$, corresponding to a lattice spacing 
$a=0.08636(98)(40)$~fm and volumes $N_T\times N_L^3=128\times 24^3$ and 
$96\times 32^3$, respectively \cite{Bruno:2016plf}. Open
boundary conditions in time  are imposed \cite{Luscher:2012av} and the sources
of the correlation functions are placed in
the bulk away from the boundary. We remark that these correlation functions do
not show any effects related to the finite time extent in the time regions analyzed. 
For H105 we use replica r001 and r002 for which
the issue of negative strange-quark determinants described in Ref.~\cite{Mohler:2020txx} is not of practical
relevance. For our analysis we use 255~(492) configurations   on
two replicas  for ensemble U101~(H105).    

\begin{table}[tb]
  \begin{center}
    \begin{tabular}{cccccc}
      \hline
  $m_\pi$~[MeV]& $m_K$~[MeV] & $m_D$~[MeV] & $m_{D^*}$~[MeV] & $m_{D_s}$~[MeV] & $aM_{av}$~[MeV]\\
      \hline
      \hline
      280(3) & 467(2) & 1927(2) & 2050(2) & 1981(1) & 3103(3)\\
      \hline
  \end{tabular}
  \end{center}
  \caption{Hadron masses in physical units for the gauge
    configurations used in this project, where $M_{av}=(m_{\eta_c}+3m_{J/\psi})/4$. The hadrons containing charm quarks
    are from  $\kappa_c=0.12315$.}
  \label{tab:masses}
\end{table}

The masses of the pion, kaon, $D$ and $D_s$ mesons  determined on the larger
ensemble are shown in Table~\ref{tab:masses}. Note that the chosen quark-mass
trajectory leads to a larger than physical $m_{u/d}$ and a smaller than
physical $m_s$. This means that the splitting between the $D\bar{D}$ and $D_s\bar D_s$ thresholds  is smaller than in experiment, emphasizing the need for a
coupled-channel analysis.  We employ the charm-quark hopping parameter $\kappa_c=0.12315$
corresponding to a charm-quark mass $m_c$ and
spin-averaged $1S$-charmonium mass $M_{av}$ that
are slightly larger than their physical values. For estimates of the statistical uncertainty we use the bootstrap method with (asymmetric) error bars resulting from
the central 68\% of the samples. Further details are
collected in Appendix~\ref{app:errors}. The correlation matrices are averaged
over several source-time slices and momentum polarizations to increase the
statistical precision. Note that all quoted uncertainties are statistical only, and that results quoted in MeV have been obtained using the central value of the lattice scale without propagating its statistical or systematic
uncertainties into the results.

For hadrons with charm quarks, non-negligible discretization effects
are observed when computing the dispersion relation on lattices with
$a\approx0.086$~fm. A comparison of the finite momentum lattice
  energies and the continuum dispersion relation for the $D$ meson on
  the two ensembles utilised in this work is given in Table II of
  Ref.~\cite{Piemonte:2019cbi}. The deviations found are small but
  statistically significant. A similar picture is observed for the
  $D_s$ meson.  Note that, these deviations may spoil the
  finite-volume analysis outlined in Section~\ref{sec:SfromE}, which
  assumes the continuum dispersion relation.  In particular, it is
  important to ensure that if the energy shifts observed with respect
  to nearby non-interacting two hadron levels are zero then the
  resulting phase shift arising from the finite-volume analysis is
  also zero.  In order to achieve this and mitigate the affect of the
  discretisation effects we adopt the analysis strategy described in
  Sect.~IV.B. of Ref.~\cite{Piemonte:2019cbi}.  Below we reiterate the
  most important details of the method.

First the energy shift of each interacting
eigenstate with respect to a nearby non-interacting two-hadron level $H_1(\vec p_1)H_2(\vec p_2)$ is computed
\begin{equation}\label{delE}
(\Delta E^{\textrm{lat}})_s=(E^{\textrm{lat}})_s-(E^{\textrm{lat}}_{H_1(\vec p_1)})_s-(E^{\textrm{lat}}_{H_2(\vec p_2)})_s~,
\end{equation}
where $\vec p_{1,2}=\vec n_{1,2} \tfrac{2\pi}{L}$,  $\vec p_1+\vec p_2=\vec
P$ and $s$ denotes the bootstrap sample. Here, $(E^{\textrm{lat}})_s$ is the energy of the interacting two-hadron
system, while $(E^{\textrm{lat}}_{H_i(\vec p_i)})_s$ is the  energy of a single
hadron (either $D$ or $D_s$ meson in this paper) with momentum $\vec p_i$
measured on the lattice. We then use
\begin{equation}\label{Ecalc}
(E^{\textrm{calc}})_s=(\Delta E^{\textrm{lat}})_s+\left(E^{\textrm{cont}}_{H_1(\vec p_1)}\right)_s+\left(E^{\textrm{cont}}_{H_2(\vec p_2)}\right)_s
\end{equation}
as input to the quantization condition (see Eq.~(\ref{qc})) for each
bootstrap sample $s$. The energies $(E^{\textrm{cont}}_{H_i(\vec
  p_i)})_s$ are computed from the continuum dispersion relation using
the lattice momenta $\vec p_{1,2}$ and the single-hadron ($D$ and
$D_s$) masses at rest. The resulting energies $E^{\textrm{calc}}$ are
equal to $E^{\textrm{lat}}$ in the naive continuum limit
$a\rightarrow0$ by construction.  The non-interacting levels are
chosen via an analysis of the overlap factors   by identifying those levels
that are dominated by the corresponding two hadron interpolators.\footnote{The overlap factor
refers to the overlap of an operator $O$ with an eigenstate $|n\rangle$, $\langle 0 |O|n\rangle$.} In the case
where more than one suitable nearby level was identified, we found the
results obtained for $E^{\textrm{calc}}$ were consistent. A comparison of $E^{calc}$ with $E^{lat}$ is presented in Appendix \ref{app:discretization}. 

For further details of the lattice methodology, in particular of
the setup for computing the quark propagators with the (stochastic) distillation method
\cite{Peardon:2009gh,Morningstar:2011ka} we refer the reader to our previous papers  \cite{Padmanath:2018tuc,Piemonte:2019cbi}.

 \section{Interpolators  }\label{sec:operators}

The main aim of this work is to investigate the  coupled-channel $D\bar D$-$D_s\bar D_s$ scattering amplitudes and
cross-sections in the channel $I(J^{PC})=0(0^{++})$ in the energy range encompassing the $D\bar D$ threshold up to $4.13$
GeV.   Following   L\"uscher's
approach \cite{Luscher:1986pf,Luscher:1990ux,Luscher:1991cf,Briceno:2014oea}, this requires a reliable extraction
of the finite-volume charmonium spectrum below 4.13~GeV on several
different volumes and/or in different momentum frames. In this study, we consider the charmonium spectrum in four different lattice irreducible representations~(irreps) $\Lambda$:
\begin{align}
\mbox{i)  }  & \Lambda^{P}=A_1^{+}~,  & |\vec P|^2=0~,\qquad & J^P[\lambda]=0^+[0]~, \nonumber\\
\mbox{ii)  } & \Lambda=A_1~,  & |\vec P|^2=1~,\qquad & J^P[\lambda]=0^+[0],~2^+[0]~,   \nonumber\\
\mbox{iii)  }& \Lambda=A_1~,  & |\vec P|^2=2~,\qquad & J^P[\lambda]=0^+[0],~2^+[0],~2^{\pm}[2]~, \nonumber \\
\mbox{iv)  }& \Lambda=B_1~, & |\vec P|^2=1~,\qquad & J^P[\lambda]=2^{\pm}[2]~. 
\label{irreps}
\end{align}

The squared momenta $|\vec P|^2$ in the lab frame are given in units of $(2\pi/L)^2$. Charge conjugation $C\!=\!+$
is a good quantum number in all frames and hence is suppressed for brevity. On
the right of Eq.~(\ref{irreps}), we list all relevant states with quantum numbers $J^P[\lambda]$ contributing to the respective irreps. Here $\lambda$
refers to the helicity of the state. The first three irreps are relevant for an investigation of the $J^P=0^+$ channel. The irreps in the moving lab frames also receive contributions from states with
$J^P[\lambda]=2^+[0]$ and $2^\pm[2]$   
within the energy range of interest\footnote{States with other $J^P$, such as $3^+$, contribute at higher energies.
We assume these higher lying states to have negligible influence in the energy range considered.}.  The analysis
of the spectrum in the $B_1$ irrep constrains the parameters for $D\bar D$
scattering with $l=2$. This partial wave inevitably contributes to the finite-volume spectrum of irreps $A_1$ with $P>0$. We utilize a large basis of single-meson
as well as two-meson interpolators in the above irreps to reliably determine the relevant low energy spectrum.

\begin{table}[tb]
\centering
\begin{tabular}{c|c|c|c|c||c|c|c|c|c}
  \hline
          & $|\vec P|^2$ & $\Lambda^{(P)C}$ & $O$            & $N_{ops}$ &         & $|\vec P|^2$ & $\Lambda^{(P)C}$  & $O$           & $N_{ops}$ \\\hline
  $O_h$   & $0$          & $A_1^{++}$ & $\bar cc$            &    7      & $Dic_4$ & $1$          & $B_1^{+}$ & $\bar cc$            &    11     \\
          &              &            & $ D(0)\bar D(0)$      &    2      &         &              &           & $D(2)\bar D(1)$      &    1      \\\cline{6-10}
          &              &            & $D(1)\bar D(1)$      &    1      & $Dic_2$ & $2$          & $A_1^{+}$ & $\bar cc$            &    28     \\
          &              &            & $D_s(0)\bar D_s(0)$  &    2      &         &              &           & $D(2)\bar D(0)$      &    2      \\
          &              &            & $D^*(0)\bar D^*(0)$  &    2      &         &              &           & $D(1)\bar D(1)$      &    2      \\
          &              &            & $J/\psi(0)\omega(0)$ &    2      &         &              &           & $D(2)\bar D(2)$      &    1      \\\cline{1-5}
  $Dic_4$ & $1$          & $A_1^{+}$  & $\bar cc$            &    17     &         &              &           & $D(3)\bar D(1)$      &    1      \\
          &              &            & $D(1)\bar D(0)$      &    2      &         &              &           & $D_s(2)\bar D_s(0)$  &    1      \\
          &              &            & $D(2)\bar D(1)$      &    1      &         &              &           & $D_s(1)\bar D_s(1)$  &    1      \\
          &              &            & $D_s(1)\bar D_s(0)$  &    2      &         &              &           & $J/\psi(2)\omega(0)$ &    3      \\
          &              &            & $J/\psi(1)\omega(0)$ &    2      &         &              &           &                      &           \\\hline
\end{tabular}
\caption{The single- and two-meson interpolators utilized in each lattice irrep $\Lambda^{(P)C}$ considered in
this study. We use the simplified notation $M_1(\vec p_1^{\,2})M_2(\vec p_2^{\,2})$ for the two-meson interpolators  with the momentum $\vec{p}_i$ of each meson~($i=1,2$) given in units of $2\pi/L$. The full expressions are
omitted for brevity. $N_{ops}$ indicates the number of operators of each type employed.}\label{tab:pspsops}
\end{table}

As in our previous publications \cite{Padmanath:2018tuc, Piemonte:2019cbi}, we construct the single-meson interpolators following the
procedure in Refs.~\cite{Dudek:2010wm,Thomas:2011rh}, using up to two gauge covariant derivatives. Table \ref{tab:pspsops}
lists the number of single-meson operators employed in each of the finite-volume
irreps considered.  The procedure discussed in Ref.~\cite{Padmanath:2018tuc} guides us in assigning the
quantum numbers $J^P[\lambda]$ to the extracted energy levels  and aids us in selecting the levels  relevant  for the  amplitude analysis.

The $D \bar D$  as well as $D_s \bar{ D}_s$ interpolators are constructed following the same procedure as in Ref. \cite{Piemonte:2019cbi}. 
The  momentum combinations  implemented in this study are given in
Table~\ref{tab:pspsops}. The two operators for $D_{(s)}(0)\bar D_{(s)}(0)$ differ in terms of the gamma matrices employed:  $\gamma_5$ or $\gamma_t \gamma_5$ for each meson. Similarly, for $D^*(0)\bar D^*(0)$ and $J/\psi(0)\omega(0)$,  two operators are constructed by employing $\gamma_i$ or $\gamma_t \gamma_i$ for the spin structure. Only one eigenstate related to $J/\psi(0)\omega(0)$ or $D^*(0)\bar D^*(0)$ is expected in the non-interacting limit. 
 We also include two-meson operators involving spin 1 mesons, such as $J/\psi\omega$ and $D^* \bar D^*$ (see Table ~\ref{tab:pspsops}).  For non-zero momenta, the construction of such operators needs additional care and we follow the induced representation method described in Appendix A2 of Ref. \cite{Dudek:2012gj}. In the $|\vec P|^2=2$ frame, for example, we implement three linearly independent $J/\psi(2) \omega(0)$ operators and observe three almost-degenerate eigenstates.
These operators are not included when extracting the finite volume spectrum for the amplitude analysis,  as  discussed in Section \ref{sec:caveats}.

\section{Assumptions and simplifications in the present study}\label{sec:caveats}

This study is performed using lattice gauge ensembles with two different
physical volumes at a single lattice spacing and at unphysical quark masses~(the resulting masses of key hadrons are given in Table~\ref{tab:masses}).   
As a consequence, only a qualitative comparison of the results can be made with experiment.

Unlike for light hadrons \cite{Briceno:2017max}, scattering studies in
the charmonium sector are still at an early stage. For the physical
states we are interested in, a three-particle channel and
multiple two-particle channels are open and all could, in
principle, be relevant. One possible approach is to simulate at very heavy pion (and kaon) mass,
such that the number of relevant  decay modes is reduced to
a few two-hadron modes, which can then be fully
  explored. This approach has the disadvantage that the quark
masses are far removed from their physical values, making a comparison
to experiment a challenge. We opt for a strategy where we simulate at a moderate pion mass of $280$~MeV and take into account the
scattering channels expected to be most relevant for the physics close
to the open-charm threshold(s). Some additional 
   channels are neglected (as discussed below), however, our assumptions about which thresholds are relevant  can be relaxed successively in future calculations. 

Neglecting certain scattering channels in our study is relevant in two
  different ways, which could be seen as two different assumptions. Some channels
  are already neglected in constructing the correlator matrices. This
  implicitly assumes that the neglected multi-hadron correlators would simply
  yield additional energy levels rather than significantly modifying the
  extracted spectrum.  Additionally, we assume that the resulting energy levels can
be analyzed with the (coupled-channel) formalism for the channels we  deem to
be dominant, which might fail if there is significant coupling to neglected channels. Beyond the scattering channels investigated explicitly, our current study includes
$J/\psi \omega$ and (some) $D^*\bar D^*$ operators in the interpolator basis. Due to
the poor signal obtained for light isoscalar mesons, the energy levels close to
the non-interacting $J/\psi \omega$ levels are not very precisely determined
and would not provide strong constraints on the scattering matrix. In particular,
almost all energy levels dominated by $J/\psi \omega$ interpolators fall
within one standard deviation of the non-interacting $J/\psi \omega$
energies, and --- apart from the additional energy levels which appear --- the other finite-volume
energies do not shift significantly when including these
interpolators.  
Section~\ref{sec:En} will
present the finite volume spectrum up to 4.13~GeV based on all the operator types
in Table \ref{tab:pspsops}, apart from the $J/\psi\omega$ operators~(see Fig.~\ref{fig:Ecm}).
Note that for our lattices the $J/\psi \omega$ threshold is located at
approximately $3.95$~GeV.
 
 We also neglect the $\eta_c\eta$ channel which has a threshold of around
  $3.54$~GeV. We remark that this decay channel has not been
observed for any of the experimental candidates mentioned in the
  introduction (and discussed in more detail in Sect.~\ref{sec:summary}). 
     Operators with more than two hadrons
  are also not implemented. The lowest three-hadron threshold is for the decay
  into $\chi_{c0}\pi\pi$ at $4.02~$GeV. This threshold is within the energy region we consider, close to the upper end. 
   
 The analysis of $D\bar D$ scattering with  $l\!=\!2$     assumes that the coupling   to the  channel $D_s\bar D_s$ with $l=2$ is negligible in the analyzed energy region and hence is omitted.  We also neglect  the coupling to  $D \bar D^*$ with $l\!=\!1$, which  contributes to irrep $B_1$. The $D\bar D^*$ threshold opens at $4.0~$GeV, while    the lowest non-interacting level $D^*(2)\bar D(1)$ would appear  at  $E_{cm}\simeq 4.2~$GeV and $4.1~$GeV on the $N_L=24$ and $32$ ensembles, respectively, which is at the upper limit of the analyzed region  (see Fig.~\ref{fig:Ecm}).  We also assume negligible effect of the $D^*\bar D^*$ channel with threshold at $4.1~$GeV. 
  
As in all studies of charmonium-like resonances to date, charm annihilation Wick contractions are omitted. 
All the remaining contraction diagrams arising from the single- 
and two-meson operators in our basis~(shown in Fig.~1 of Ref.~\cite{Lang:2015sba}) are computed following 
the procedures described in our previous publications \cite{Padmanath:2018tuc,Piemonte:2019cbi}. 
 
We stress that we determine the finite-volume spectra at a
single lattice-spacing and are therefore
unable to quantify the uncertainty associated with the lattice discretization. In particular,
the uncertainty arising from the heavy
  quark discretization may be non-negligible. As discussed in
  the previous section, the dispersion relation deviates from the
  continuum relation in our study and spin-splittings
  are also likely to be affected~\cite{ElKhadra:1996mp,Oktay:2008ex}. In general, lattice spacing effects in heavy-light mesons and charmonium are different
   with the net result that even at physical light-quark masses the open-charm thresholds
  can be shifted with respect to the measured charmonium states at finite lattice spacing.

  \section{Determination of the finite-volume spectrum}\label{sec:En}

\begin{figure}[tbh]
	\begin{center}
	\includegraphics[width=0.24\textwidth]{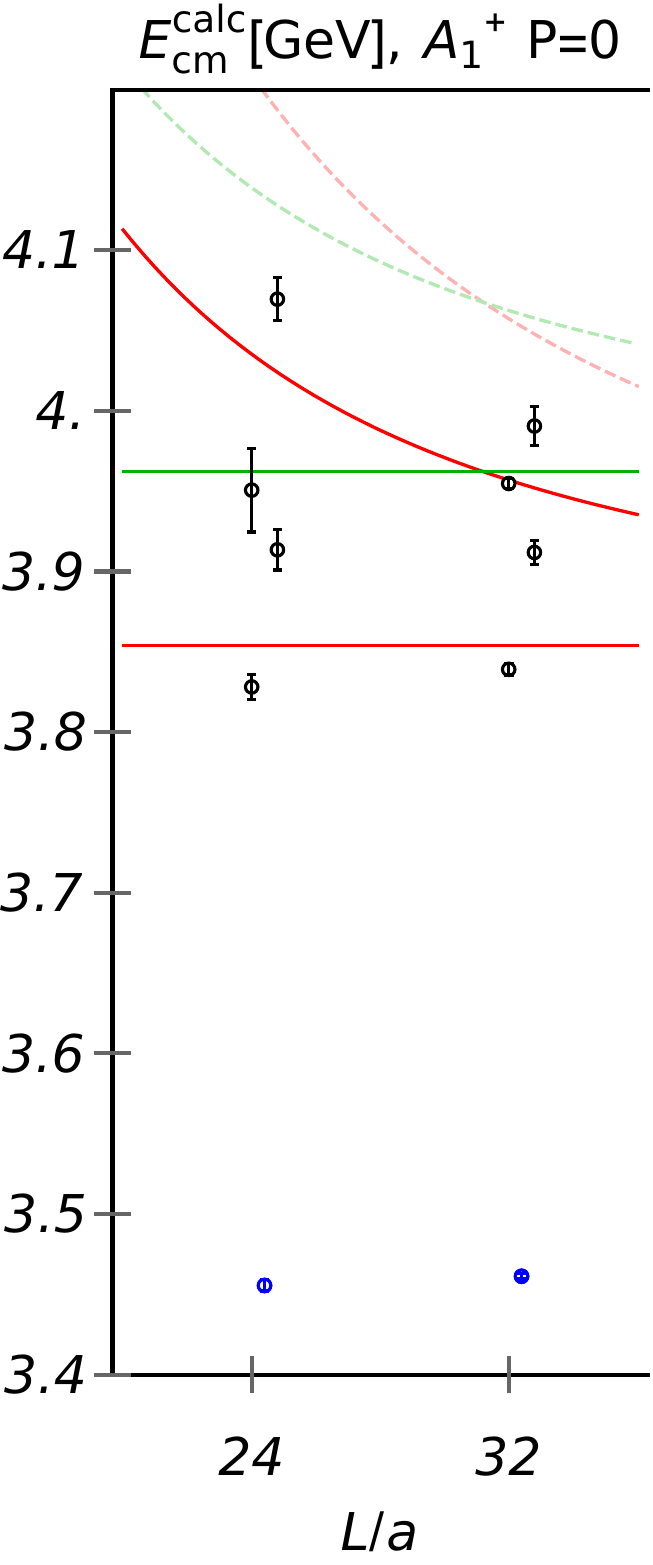}   
        \includegraphics[width=0.24\textwidth]{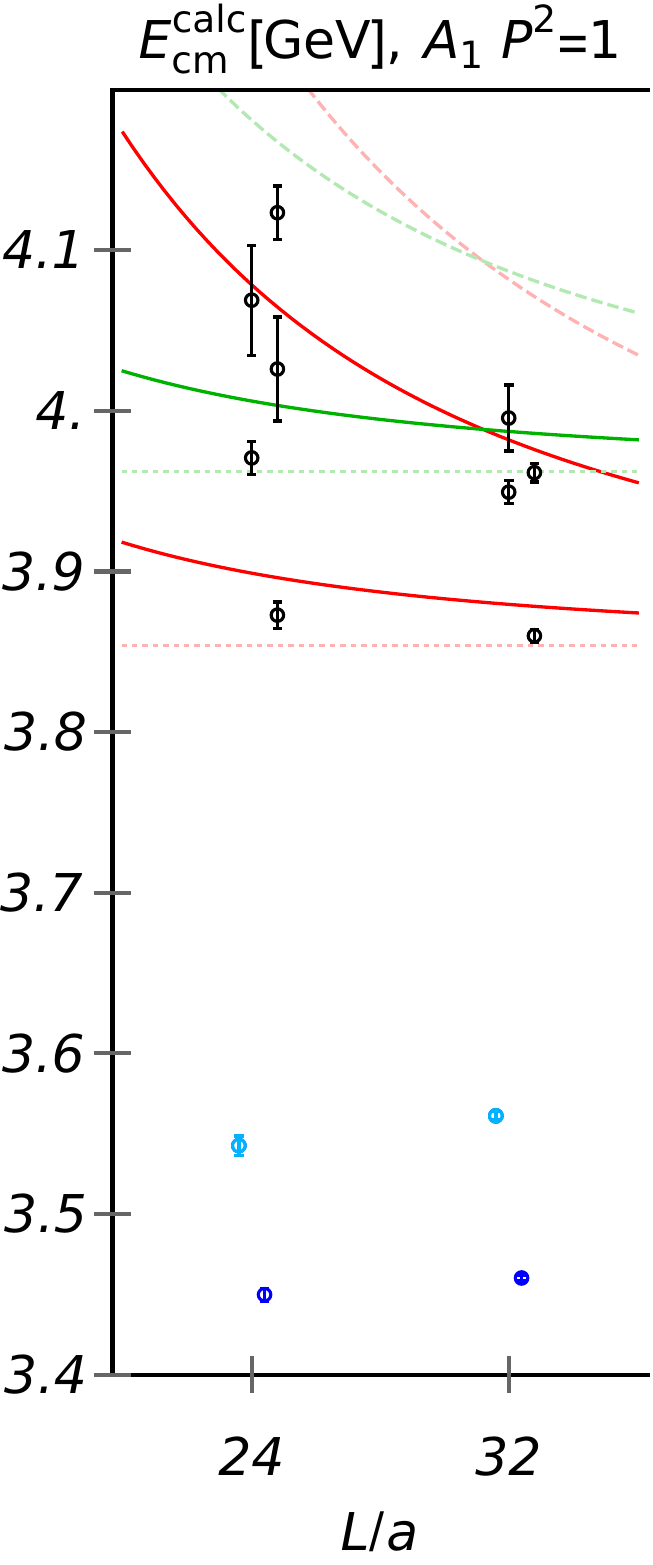} 
        \includegraphics[width=0.24\textwidth]{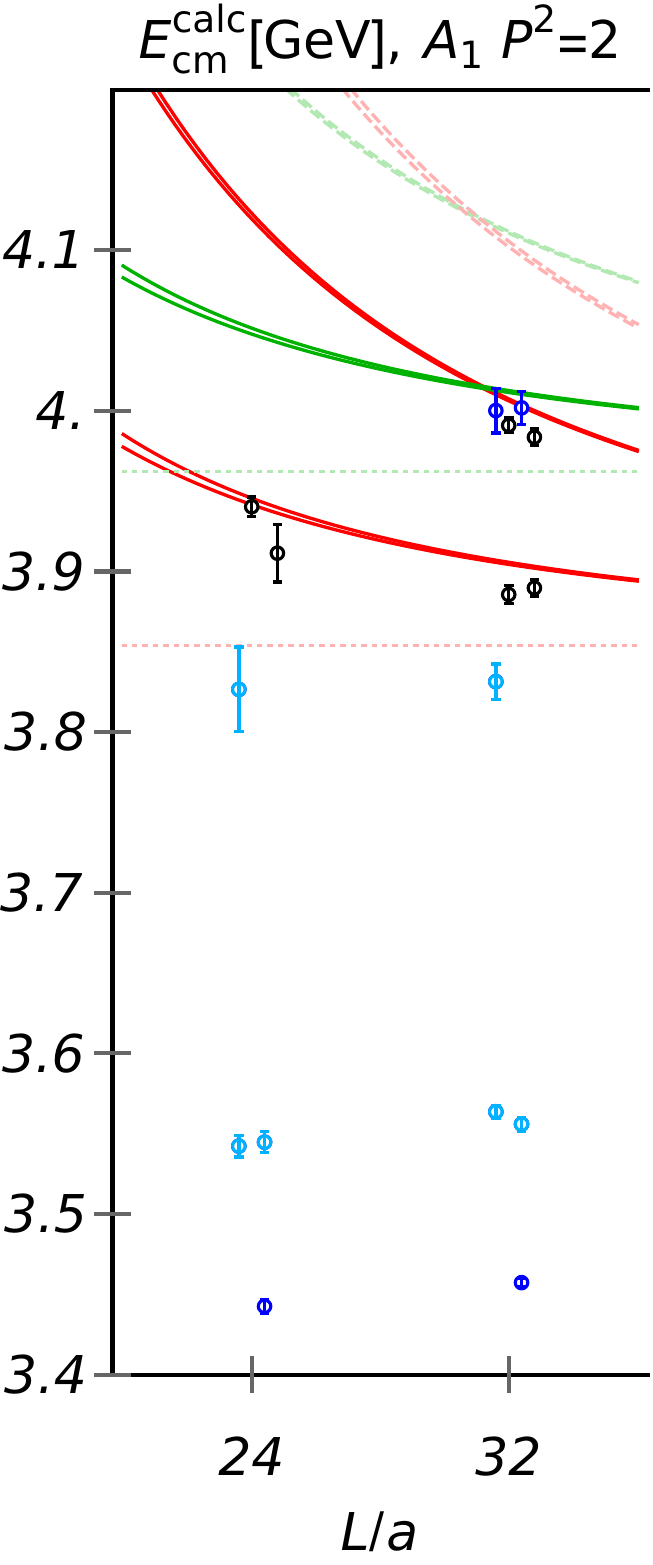} 
        \includegraphics[width=0.24\textwidth]{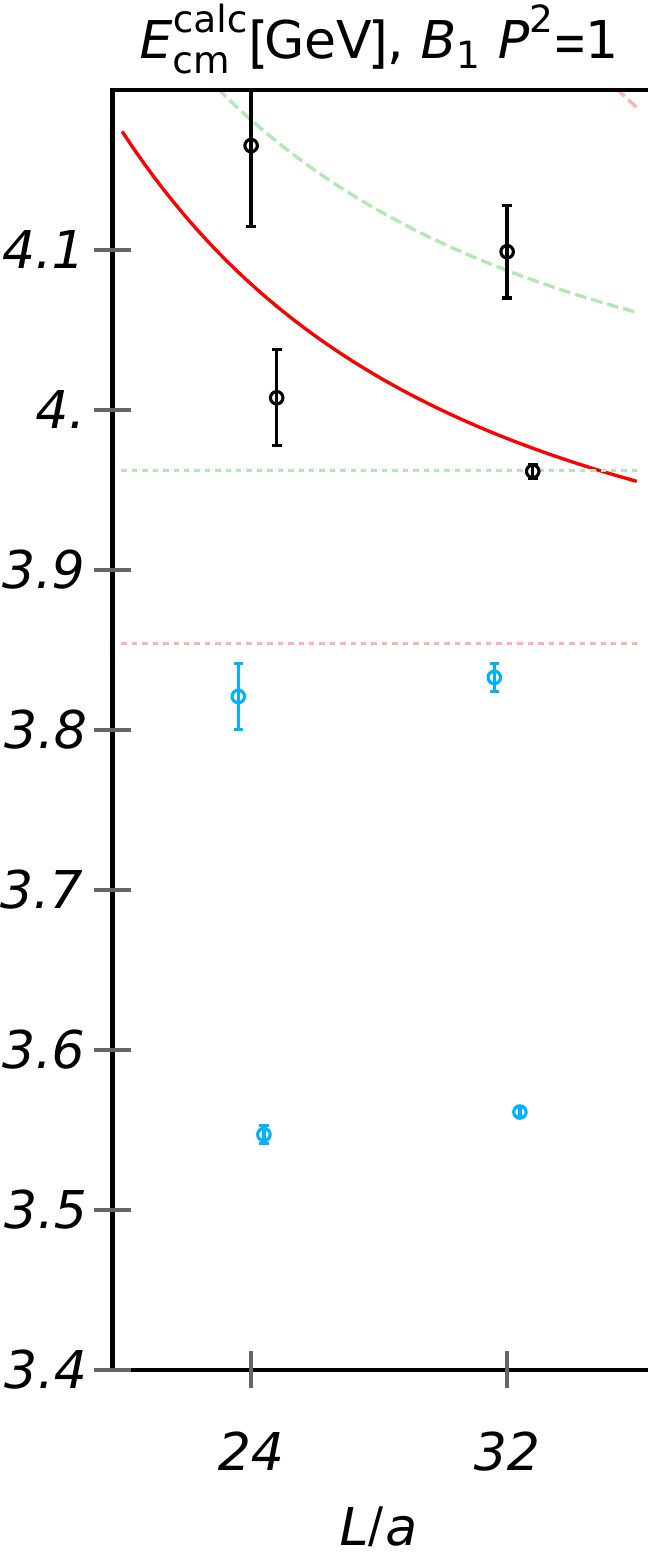}  
		\caption{  The eigen-energies in the center-of-momentum frame ($E_{cm}$) for  the charmonium-like system with $I\!=\!0$ and $C\!=\!+1$. Results are presented for irreducible representations $\Lambda^P=A_1^{(+)},~B_1$ and  total
   momenta $|\vec P|^2=0,1,2$, which  give information on the  channels with $J^{PC}=0^{++},2^{\pm+}$. 
The data points correspond to the eigen-energies obtained from the lattice simulation: the black circles are used to extract the coupled-channel scattering matrices for $D\bar D - D_s\bar D_s$, while  the blue circles are omitted from the scattering analysis. 
		The solid and dashed red (green) lines correspond to discrete $D\bar D$ ($D_s\bar D_s$)   eigen-energies in the non-interacting limit: solid lines correspond to the operators that are implemented, while dashed lines correspond to the lowest-lying energies from  operators that are not implemented.  Dotted lines represent thresholds. The data points indicated by the light blue circles correspond to ground-state charmonia with $J^{PC}=2^{++}$ and  $2^{-+}$, which appear at $m\simeq 3.56~$GeV and $3.83~$GeV, respectively.   Some  points are  shifted horizontally slightly for clarity.     }
	\label{fig:Ecm}
	\end{center}
\end{figure}

  This section presents the eigen-energies $E_n$ that will be used to
  determine the scattering matrices. The energies are obtained from the
  correlation matrices $C_{ij}(t)=\langle O_i(t)O_j^\dagger (0)\rangle $ via
  the widely-used variational  method. This involves solving the
    generalized eigenvalue problem 
    $C(t)u^{(n)}(t)\!=\!\lambda^{(n)}(t)C(t_0)u^{(n)}(t)$ for the
    eigenvalues $\lambda^{(n)}(t)$ and the eigenvectors $u^{(n)}(t)$~\cite{Michael:1985ne,Luscher:1990ck,Blossier:2009kd}. We use the reference time $t_0/a=3$ or $4$. The eigen-energies are
  extracted from 1-exponential fits to the eigenvalues $\lambda ^{(n)}(t)=A_n
  e^{-E_n t}$ with the fit range, in most cases, starting between timeslices $10$ and $12$.  
  
    The finite-volume spectrum of the charmonium system  with isospin $I\!=\!0$ and
   $C\!=\!+1$  is shown in Fig.~\ref{fig:Ecm}.  We present the spectrum
   in the center-of-momentum (cm) frame    $E_{cm,n}=[E_n^2-\vec
   P^2]^{1/2}$ (\footnote{ Here, the energy in the lattice frame $E_n$ stands for
     $E_n^{calc}$ obtained according to Eq. (\ref{Ecalc}) or Eq. (17) of
    Ref. \cite{Piemonte:2019cbi}.
   })  for  irreducible representations $\Lambda^P=A_1^{(+)},~B_1$ and  total
   momenta $|\vec P|^2=0,1,2$. These irreps give information on the charmonium(like)
   states and $D_{(s)}\bar D_{(s)}$ scattering in the channels with
   $J^{PC}=0^{++},2^{\pm+} $ (see Eq. \ref{irreps}). The energies
   indicated by the black-circles are used to  extract information on
   $D_{(s)}\bar D_{(s)}$ scattering. These energies are near or above the $D\bar D$ threshold and are precise enough to  reliably resolve the energy-shifts with respect to the non-interacting energies of $D_{(s)}\bar D_{(s)}$~(indicated by the solid lines). The light-blue circles are the energy levels related to ground-state charmonia with $J^P= 2^{\pm}$.

\section{Determining scattering matrices from  lattice finite-volume  energies}\label{sec:SfromE}
 
 The bound states and resonances are  inferred from the scattering matrices as briefly reviewed in Section \ref{sec:generalities}.   The  infinite-volume scattering matrix $S(E_{cm})$  is related to 
the   finite-volume two-hadron spectrum   for real   energies $E_{cm}$ above the
threshold and somewhat below  it through the well-known L\"uscher relation
\cite{Luscher:1986pf,Luscher:1990ux,Luscher:1991cf}. The eigen-energies of the
coupled channel $D\bar D-D_s\bar D_s$ system given in the previous section
provide information on the $2\times 2$ scattering matrix $S(E_{cm})$ for these
coupled channels via the generalization of this formalism
\cite{Doring:2011vk,Hansen:2012tf,Briceno:2014oea}, considered (for other channels), for example, in
Refs.~\cite{Wilson:2014cna,Dudek:2016cru,Moir:2016srx}. The $S$
matrix can be expressed in terms of a real  function $\tilde K(E_{cm})$ for
one-channel scattering (\ref{S-elastic}) and a real symmetric $2\times 2$
matrix $\tilde K(E_{cm})$ for two coupled channels (\ref{S-coupled}), as
detailed in the next section. $\tilde K$  uniquely determines $S$, while
both depend also on the partial wave $l$. We use the spectrum from the
previous section to  determine $\tilde K(E_{cm})$ using the publicly available
package {\it TwoHadronsInBox} \cite{Morningstar:2017spu}.

 The relation between discrete lattice eigen-energies  $E_{cm}$  and $\tilde K$-matrix for coupled-channel scattering is  referred to as the {\it quantization condition }\cite{Morningstar:2017spu} 
  \begin{equation}
  \label{qc}
 \det[\tilde K^{-1}_{l;ij}(E_{cm}) ~\delta_{ll^\prime}-B^{\vec P,\Lambda}_{ll^\prime;i}(E_{cm})~\delta_{ij}]=0~.
\end{equation}
 Both terms in the determinant are matrices in the space of partial waves
 $l,l^\prime$ and channels $i,j$ ($D\bar D$, $D_s\bar D_s$ or both), and the determinant is evaluated over both indices.  $\tilde K_{l;ij}~\delta_{ll^\prime}$ is  an unknown matrix in channel space that depends on the  partial wave $l$; it is diagonal in $l$ since the good quantum numbers  in continuum scattering of spin-less particles
 (such as $D$ and $D_s$)   are $J$, $S\!=\!0$ and $l\!=\!J\!-\!S\!=\!J$.
 The $B^{\vec P,\Lambda}_{ll^\prime;i}(E_{cm})$ are known box-functions
 \cite{Morningstar:2017spu} that are in general non-diagonal in the partial
 wave index.  
 
 In one-channel scattering and when only  partial wave $l$  contributes, 
 relation (\ref{qc}) simplifies to $\tilde
 K^{-1}(E_{cm})=B^{\vec P,\Lambda}(E_{cm})$, since the argument of the
 determinant is a $1\times 1$ matrix. The values of $K^{-1}(E_{cm})$ will be shown as points in figures for one-channel scattering.
  For two coupled channels, for the case when only  partial wave $l$  contributes, the
  determinant  equation (\ref{qc}) provides one relation between
  $\tilde K_{11}(E_{cm})$,  $\tilde K_{22}(E_{cm})$ and $\tilde
  K_{12}(E_{cm})$ for each energy level, complicating the determination of
  those functions. Therefore, we  follow the strategy proposed in
Ref. \cite{Doring:2011vk}, where  the  $\tilde K_{ij}(E_{cm})$ are parametrized as functions of
the  energy. In this strategy, the $\tilde K$-matrix elements are determined by requiring 
that relation (\ref{qc}) is simultaneously  satisfied for all relevant lattice
energies $E_{cm}$.

 We will focus on certain interesting and rather narrow energy regions, where
 a  linear dependence on $s$ is expected to be a good approximation
 \begin{equation}
 \frac{ \tilde K_{ij}^{-1}(s)}{\sqrt{s}}=a_{ij} + b_{ij} s~,\quad s=E_{cm}^2~.
  \label{linear}
  \end{equation}
  Such a parametrization is equivalent to a Breit-Wigner parametrization in the resonance region and is also similar to the well-known effective range expansion $K_{ij}^{-1}(s)=c_{ij} + d_{ij} p^2$ near threshold, where  $p$ is the momentum of the scattering particles in the center-of-momentum frame. We determine the parameters $a_{ij}$ and $b_{ij}$ following the strategy discussed above, using the determinant residual method proposed in \cite{Morningstar:2017spu}, which is briefly described in Appendix \ref{app:omega}.  A~posteriori, we always verify that the resulting parametrization predicts via Eq. (\ref{qc}) the same number of eigen-energies observed in the actual simulation in the relevant energy range; this is shown in the $\Omega$ plots for some fits in Appendix \ref{app:channels}.  This procedure will be followed for the extraction of the coupled-channel scattering matrix as well as for one-channel scattering. 
 
This study is based on the parametrization in
  Eqn.~(\ref{linear}). Alternatively, one could parametrize $\tilde
  K_{ij}(s)$ itself with common pole terms in both channels, such as
  those tabulated in Table IV of Ref. \cite{Wilson:2015dqa}. We have
  performed fits with different parametrizations of
  $\tilde K_{ij}(s)$ (single pole, double pole, triple pole, poles
  with polynomial terms, etc.). We find that fits (for coupled $D\bar
  D-D_s\bar D_s$ scattering) with a single-pole in the higher energy
  region are not consistent with our data. Including two or more
  poles/resonances leads to fits with six or more parameters. We
  observed that the data used in this work is insufficient to
  accommodate such a large number of parameters and hence such an analysis is beyond scope of this work. An investigation of the model-independence of the findings presented here requires
  extending the lattice calculation to include a larger set of
  ensembles with high statistics.

  The  box-function $B^{\vec P,\Lambda}_{ll^\prime;i}(E_{cm})$ can  have
  off-diagonal elements for $l^\prime \not = l$ due to the lack of rotational
  symmetry in a  finite box.   This will result in
  contributions from multiple partial waves in the quantization condition
  Eq.~(\ref{qc}) for a given lattice irrep $\Lambda$.  We consider partial
  waves $l=0$ and $l=2$ and ignore contributions from
  $l\geq 3$, which is  a reasonable assumption in the energy region considered
  for the respective irreps. In this case the only non-diagonal elements
  $B_{ll^\prime}$ among the $A_1(P^2\not =0)$ irreps that are nonzero are $B^{P=001,A_1}_{02}$ and $B^{P=110,A_1}_{02}$.  These will be taken into account in the analysis of Section~\ref{sec:DD-DsDs-with2}.

\section{Results for various channels and energy regions}\label{sec:channels}

In this section we present our results for the scattering matrices, pole positions, masses, and widths 
of $J^{PC}=0^{++}$ and $2^{++}$ charmonium(like) states in various energy regions and with varying assumptions. The energy range from slightly below $2m_D$ up to 4.13 GeV is divided into smaller intervals, where the elements of the coupled $D\bar D- D_s\bar D_s$ scattering matrix are separately parametrized according to Eq.~(\ref{linear}) or as a constant. The details of the parametrizations and the results are presented in separate subsections below, 
while information on the energy levels considered in each case is 
given in Appendix \ref{app:channels}.

A single description of the whole energy region requires a
finite-volume analysis involving many more parameters, which results
in more challenging and unstable fits. Such an analysis is beyond the
scope of the current investigation. Our inferences and conclusions are
based on the finite-volume analysis of separate energy regions. 
 Similar parametrizations to those employed for the separate energy regions, are employed 
 collectively to a wider energy range in Appendix~\ref{sec:wholeE} as an additional  consistency check.

\begin{figure}[h!]
	\begin{center}
	\includegraphics[width=0.41\textwidth]{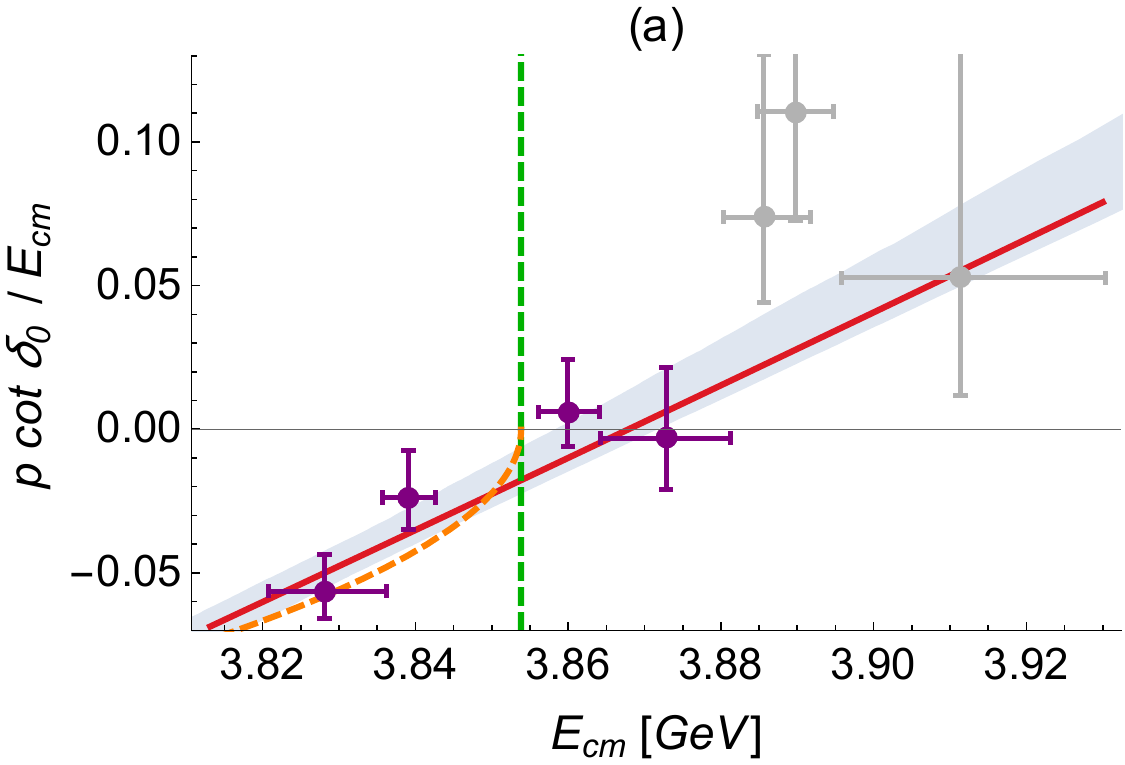} $\qquad$
	 \includegraphics[width=0.41\textwidth]{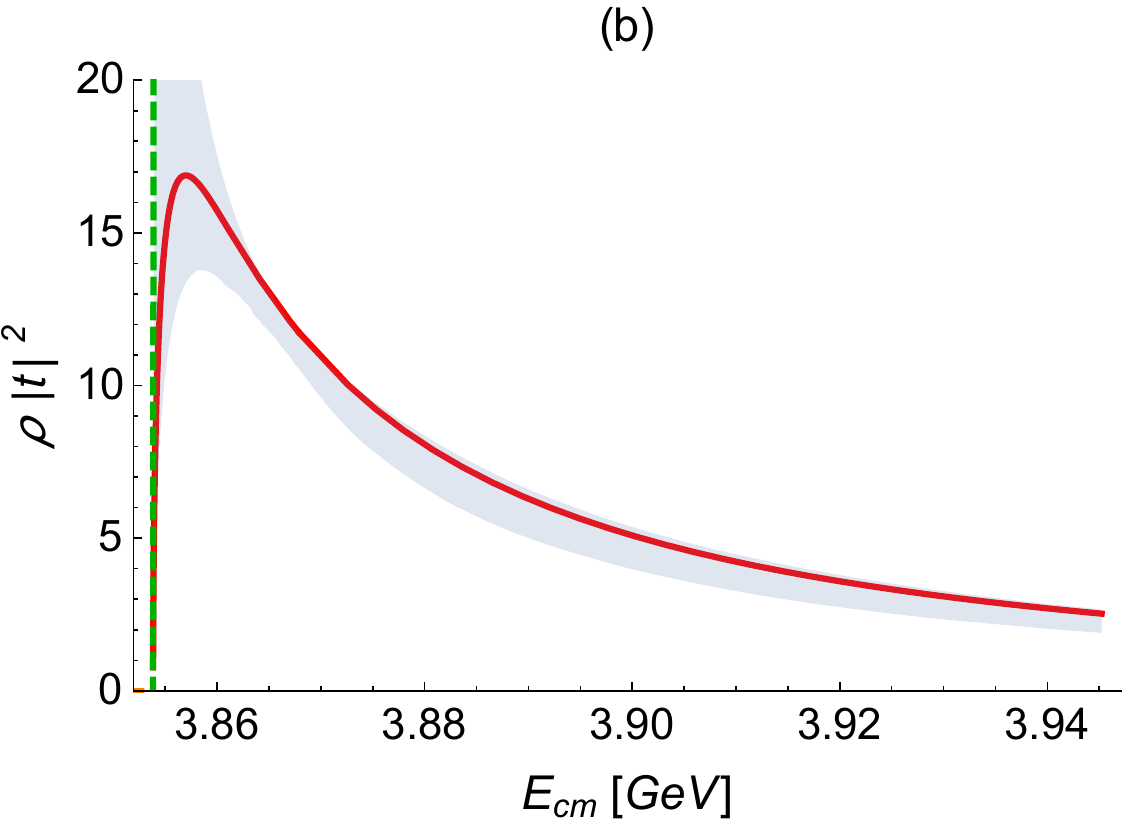}   $\quad $
	 \includegraphics[width=0.41\textwidth]{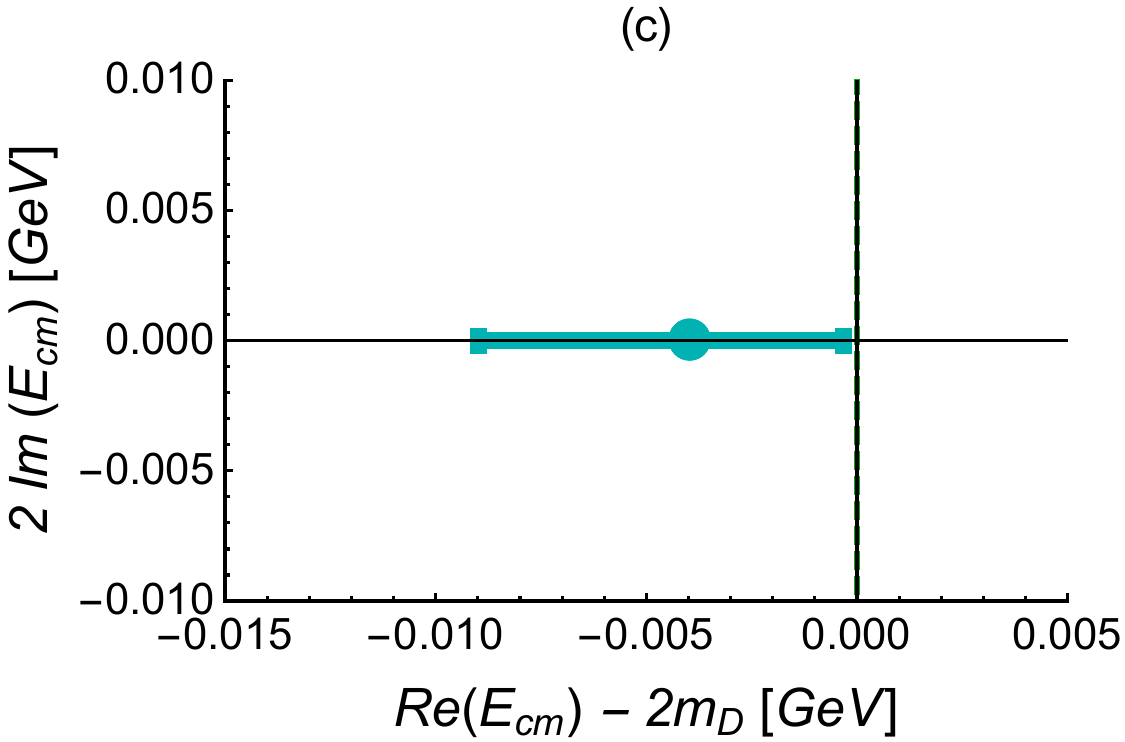}
		\caption{  $D\bar D$ scattering in partial wave $l\!=\!0$
                  near threshold. The green dashed line denotes the $D\bar D$
                  threshold in the simulation where $m_D\!\simeq\! 1927$ MeV.
                  (a) The violet crosses show the quantity
                  $p\cot \delta/E_{cm}$, related to the scattering phase shift
                  $\delta$, as a function of center-of-mass energy
                  $E_{cm}$. The red line indicates the parametrization
                    of Eqs~(\ref{eq:ddparam}) and (\ref{DD-params}). The orange line represents $ip/E_{cm}$. A bound
                  state is located at the energy where the red and orange curves intersect. (b) The quantity $\rho |t|^2$  that is proportional to the number of $D\bar D$ events in experiment $N_{D\bar D} \propto p\sigma\propto \rho |t|^2$ ($\rho=2p/E_{cm}$).  The gray bands represent the uncertainty as defined in Appendix~\ref{app:errors}. (c) Position of the  pole of the scattering matrix  on sheet I: the real component  corresponds to the binding energy, presented in Eq.~(\ref{chic0DD-lat}). }
	\label{fig:DD}
	\end{center}
\end{figure}

 \subsection{$D\bar D$   scattering with $l=0$ near threshold } \label{sec:DD}

  The narrow energy region near the $D\bar D$ threshold is significantly below
  the $D_s\bar D_s$ threshold and can be treated in a one-channel approach. We employ 
  the parametrization in Eq.~(\ref{linear}) 
 \begin{equation}
  \frac{p\cot \delta_{D\bar D}^{l=0}}{E_{cm}}=\frac{(\tilde K^{-1})_{11}^{l=0}}{E_{cm}}=a_{11}^{\prime}+ b_{11}^{\prime} E_{cm}^2~,\label{eq:ddparam}
  \end{equation}
  which is equivalent to the effective range expansion $p\cot\delta_0=1/a_0+r_0 p^2/2$ near threshold. Four lattice energy levels with $E_{cm}$  closest to $2m_D$ (listed in Appendix \ref{app:DD}) are utilized to determine the parameters via the quantization condition (\ref{qc}). We find
 \begin{equation}
 \label{DD-params} 
    \begin{array}{cc}
 a_{11}^\prime=-2.4413 \pm 0.2986\ \  \ \ \\
  b_{11}^\prime=(0.8519 \pm 0.1067)\  a^2\\
\end{array}~,\quad \mathrm{cor}=
   \left[
 \begin{array}{cc}
 1. & -0.98 \\
   & 1. \\
\end{array}
\right]~,
\quad \tfrac{\chi^2}{d.o.f.}=1.5~,
\end{equation}
where cor is the correlation matrix defined in Appendix  \ref{app:errors}.  The fit is shown in Fig.~\ref{fig:DD}a.  This scattering matrix leads to a bound state at the energy $E_{cm}\!=\!m$  when the scattering matrix $t$ (\ref{S-elastic}) has a pole on the real axis  below threshold on sheet I  
\begin{align}
\label{bound-state}
  t&= \frac{1}{\rho(\cot\delta -i)}=\infty\quad \to\quad  \cot\delta=i~,\\
  p\cot \delta/E_{cm}&=ip/E_{cm}=-|p|/E_{cm}~,\quad (p=i|p|)~.
\end{align}
The lhs of the second
  equation is shown as the red line in the figure, while the rhs is indicated by the orange
  line. The bound state occurs at the value
  $E_{cm}\!=\!m$, where the two curves intersect. The slope of $p\cot
\delta$ at the intersection, is smaller than the slope of
$-|p|$, as required for an s-wave bound state (see Section VC of
\cite{Piemonte:2019cbi}). The location of the pole in the scattering
matrix is shown in Fig.~\ref{fig:DD}c.  The bound state appears just below the $D\bar D$ threshold with the binding energy
\begin{equation}
\label{chic0DD-lat}
\chi_{c0}^{ D\bar D}:\quad
m-2~m_D=-4.0 ^{~ + 3.7}_{~ -5.0}\ \mathrm{ MeV}~. 
\end{equation} 
We denote this state by $\chi_{c0}^{D\bar D}$, indicating it has
 $J^{PC}\!=\!0^{++}$ and a strong  connection to the $D\bar D$ threshold. This
 state comes in addition to the conventional  $\chi_{c0}(1P)$, which is found
 significantly below threshold. Experiments cannot explore
 $D\bar D$ scattering below threshold, however, a closeby bound state below threshold
 could be identified experimentally through a sharp
 increase of the rate  just  above threshold.  Fig.~\ref{fig:DD}b shows a
 dimensionless quantity $\rho |t|^2$  related to the number of events 
 $N_{D\bar D} \propto p\sigma\propto \rho |t|^2$ expected in experiment. It
 features a peak  above threshold, which increases much more rapidly than the phase space. 

Such a $D\bar D$ bound state was not claimed by experiments so far. A similar
state was predicted in  phenomenological models 
\cite{Gamermann:2006nm,Hidalgo-Duque:2013pva,Baru:2016iwj}, and  some indication for it was suggested in
the experimental data \cite{Gamermann:2007mu,Abe:2007sya} and in data from the lattice
simulation of Ref. \cite{Lang:2015sba}. A more detailed discussion  follows in the summary in Section
\ref{sec:summary}. 

Details of the fit (\ref{DD-params}) and some variations thereof are provided in Appendix~\ref{app:DD}.
In these fits, the  ensemble average of the data gives rise to a bound state, while a very small  proportion of the bootstrap samples instead produce a virtual bound state. This indicates that our lattice results, at the employed quark masses,   favour the existence of a bound state. However, with the present statistical accuracy, one cannot completely rule out the existence of a virtual bound state. The robust conclusion is that  we observe  a significant $D\bar D$ interaction near threshold, leading to  one state just below threshold.   Such a state leads to an increase of the $D\bar D$ rate above threshold irrespective of whether it is a bound or a virtual bound state.     
Note that it is not known whether this  state would
also feature in a simulation with physical quark masses.
 
    \begin{figure}[h!]
	\begin{center}   
	\includegraphics[width=0.41\textwidth]{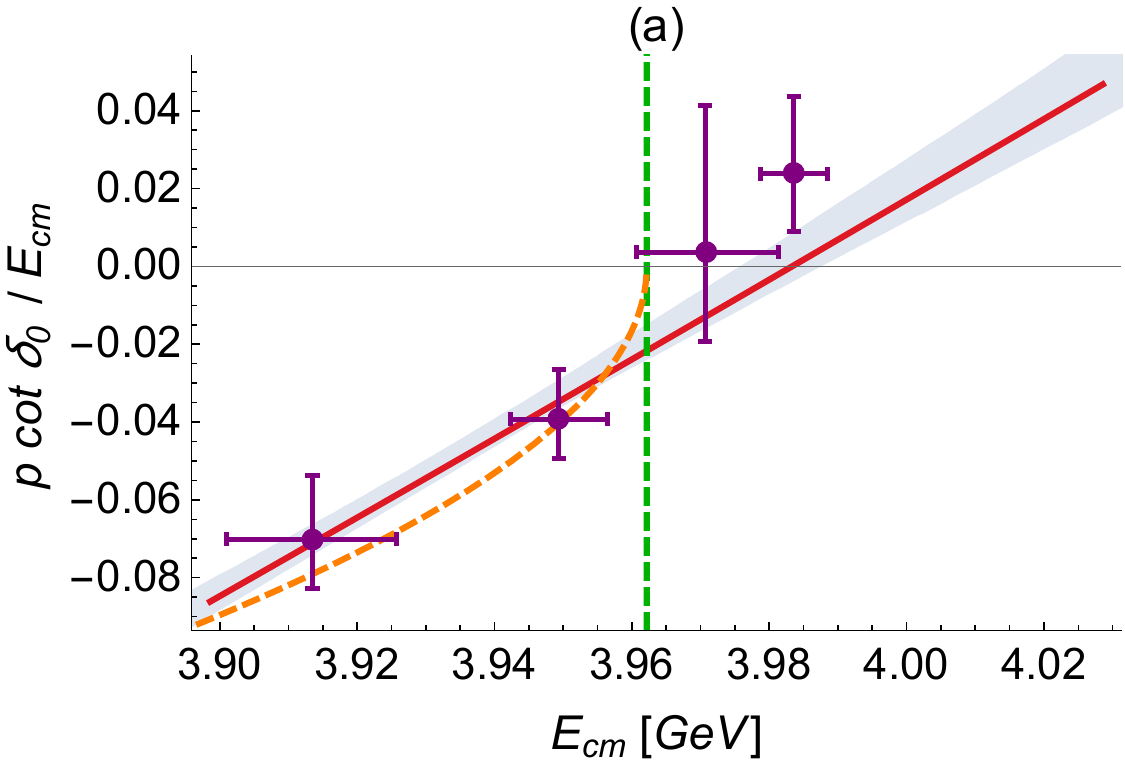} $\qquad$
	 \includegraphics[width=0.41\textwidth]{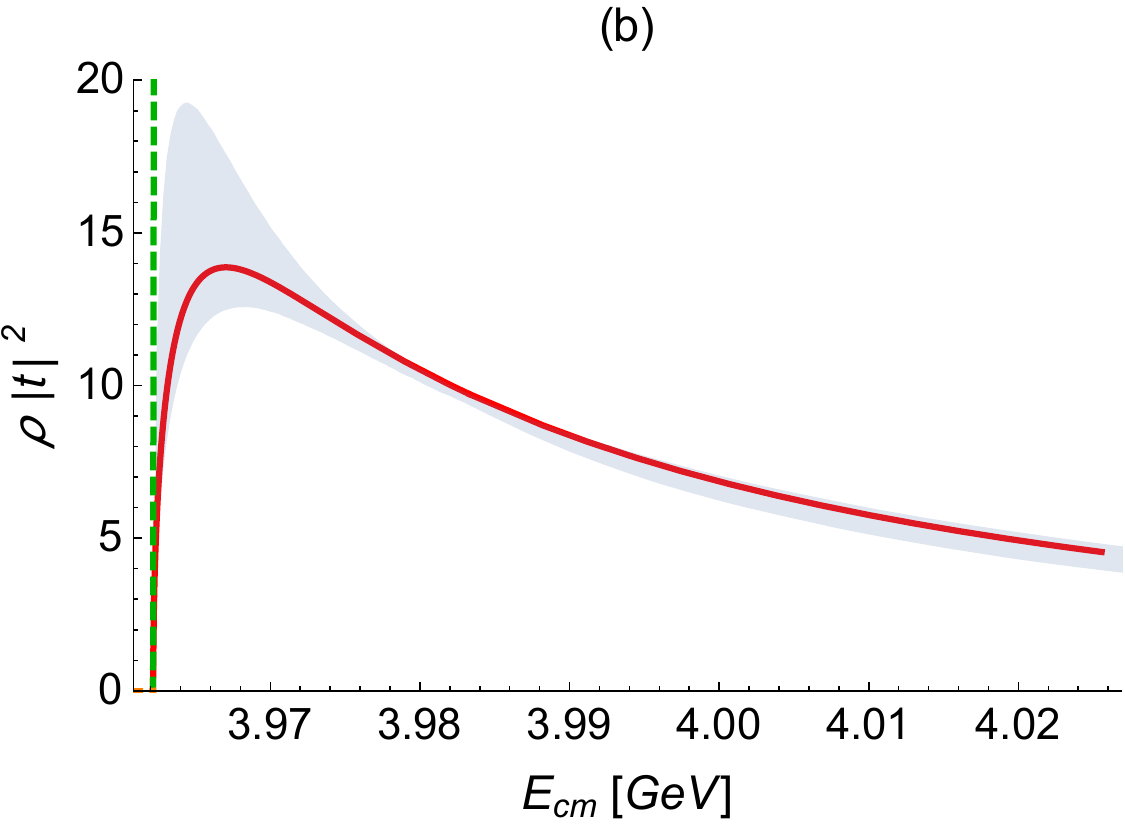}   $\quad $
	  \includegraphics[width=0.41\textwidth]{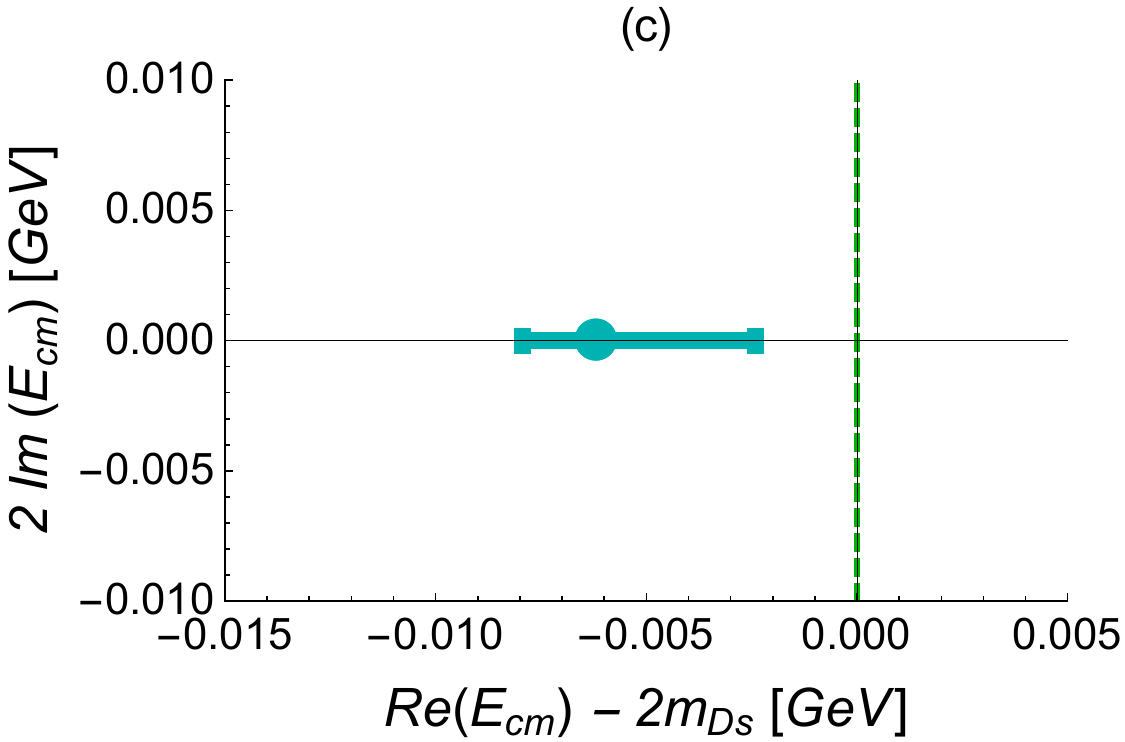}
		\caption{ $D_s\bar D_s$ scattering in partial wave
                  $l=0$ in the one-channel approximation. The green dashed
                  line denotes the $D_s\bar D_s$ threshold in the simulation,
                  where $m_{D_s}\simeq 1981$ MeV.  (a)  The violet
                  crosses show the quantity  $p\cot \delta/E_{cm}$,
                  related to the scattering phase shift $\delta$, as a function
                  of center-of-mass energy $E_{cm}$. The red line indicates the parametrization
                    of Eqs~(\ref{eq:dsdsparam}) and (\ref{DsDs-params}). The orange line
                  represents $ip/E_{cm}$. A bound state is located at the
                  energy where the red and orange curves intersect. (b) The
                  quantity $ \rho |t|^2$ that is proportional to the number of $D_s\bar D_s$ events in experiment $N_{D_s\bar D_s} \propto p\sigma\propto \rho |t|^2$
                   ($\rho=2p/E_{cm}$). (c) Position of the   pole of the scattering matrix  on sheet I: the real component corresponds to the binding energy, presented in Eq.~(\ref{DsDs-lat}). }
	\label{fig:DsDs}
	\end{center}
\end{figure}

  \subsection{$D_s\bar D_s$ scattering with $l=0$ near   threshold in the one-channel approximation}\label{sec:DsDs}
  
  The $D_s\bar D_s$  channel carries the same quantum numbers as $D\bar D$ 
  necessitating the consideration of coupled-channel scattering. In this
  subsection we aim to get a rough estimate of $D_s\bar D_s$ scattering in the
  one-channel approximation, which will also provide initial guesses for the parameters when coupled channel scattering is considered in Section \ref{sec:DD-DsDs}. The $D_s\bar D_s$ scattering near threshold is parametrized by 
   \begin{equation}
  \frac{p\cot \delta_{D_s\bar D_s}^{l=0}}{E_{cm}}=\frac{(\tilde K^{-1})_{22}^{l=0}}{E_{cm}}=a_{22}+ b_{22}  E_{cm}^2~.\label{eq:dsdsparam}
  \end{equation}
We employ the quantization condition (\ref{qc}) together with
  four lattice energies close to this threshold that are dominated by $D_s\bar
  D_s$ interpolators (listed in Appendix~\ref{app:DsDs}) and  obtain
     \begin{equation}
 \label{DsDs-params} 
    \begin{array}{cc}
 a_{22}=-2.0473 \pm 0.1513\ \  \ \ \\
  b_{22}=(0.6737 \pm 0.0514)\  a^2\\
\end{array}~,\quad \mathrm{cor}=
   \left[
 \begin{array}{cc}
 1. & -0.999 \\
   & 1. \\
\end{array}
\right],
\qquad \tfrac{\chi^2}{d.o.f.}=2.8~.
\end{equation} 
     The resulting fit is shown in Fig.~\ref{fig:DsDs}a.      The scattering matrix has a bound state pole at the energy $E_{cm}\!=\!m$ where condition (\ref{bound-state})  is satisfied, see Fig.~\ref{fig:DsDs}c. Again, the slope of $p\cot \delta$ is  smaller than the slope of $-|p|$ at the position of the pole, as required for an s-wave bound state (see Section VC of \cite{Piemonte:2019cbi}).
 
 We find a shallow $D_s\bar D_s$ bound state at
\begin{equation}
\label{DsDs-lat}
\chi_{c0}^{D_s\bar D_s}: \quad
m-2m_{D_s}=-6.2 ^{~+3.8}_{~-2.0} \ \mathrm{MeV}~
\end{equation}
that we denote $\chi_{c0}^{D_s\bar D_s}$, indicating it has
$J^{PC}\!=\!0^{++}$ and a strong connection to the $D_s\bar D_s$
threshold. This  state is responsible for the significant increase in the $D_s\bar
D_s$ rate shown in Fig.~\ref{fig:DsDs}b just above threshold. 
In order to search for the $\chi_{c0}^{D_s\bar D_s}$ in experiment an exploration of the $D_s\bar D_s$ invariant mass near threshold would be invaluable.  In  one-channel
$D_s\bar D_s$ scattering, considered here, the state is decoupled from $D\bar D$, while it will become a narrow resonance and acquire a small width when the coupling to $D\bar D$ is considered in Section~\ref{sec:DD-DsDs}.  

Two  candidates  $\chi_{c0}(3930) $ \cite{chic03930} and
$X(3915) $ \cite{pdg}   (which may correspond to the same state) have already been observed in experiment just below the  threshold
$2m_{D_s}^{\mathrm{exp}}\simeq 3936~$MeV; they   have a small coupling to $D\bar
D$ and a small width.  If the $D_s\bar D_s$ bound state (\ref{DsDs-lat})
corresponds to $\chi_{c0}(3930) $  and/or $X(3915) $, it naturally
explains both features as will be discussed in Section~\ref{sec:summary}.

   \begin{figure}[h!]
	\begin{center}
	\includegraphics[width=0.41\textwidth]{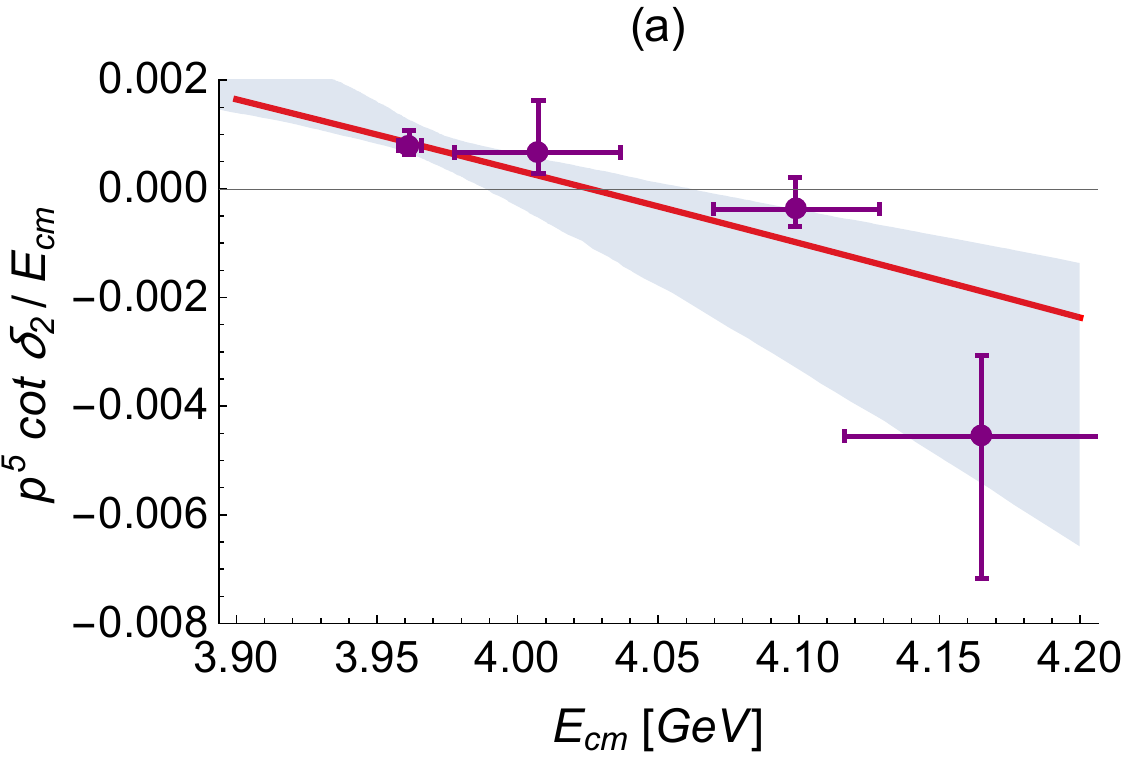} $\qquad$
	 \includegraphics[width=0.41\textwidth]{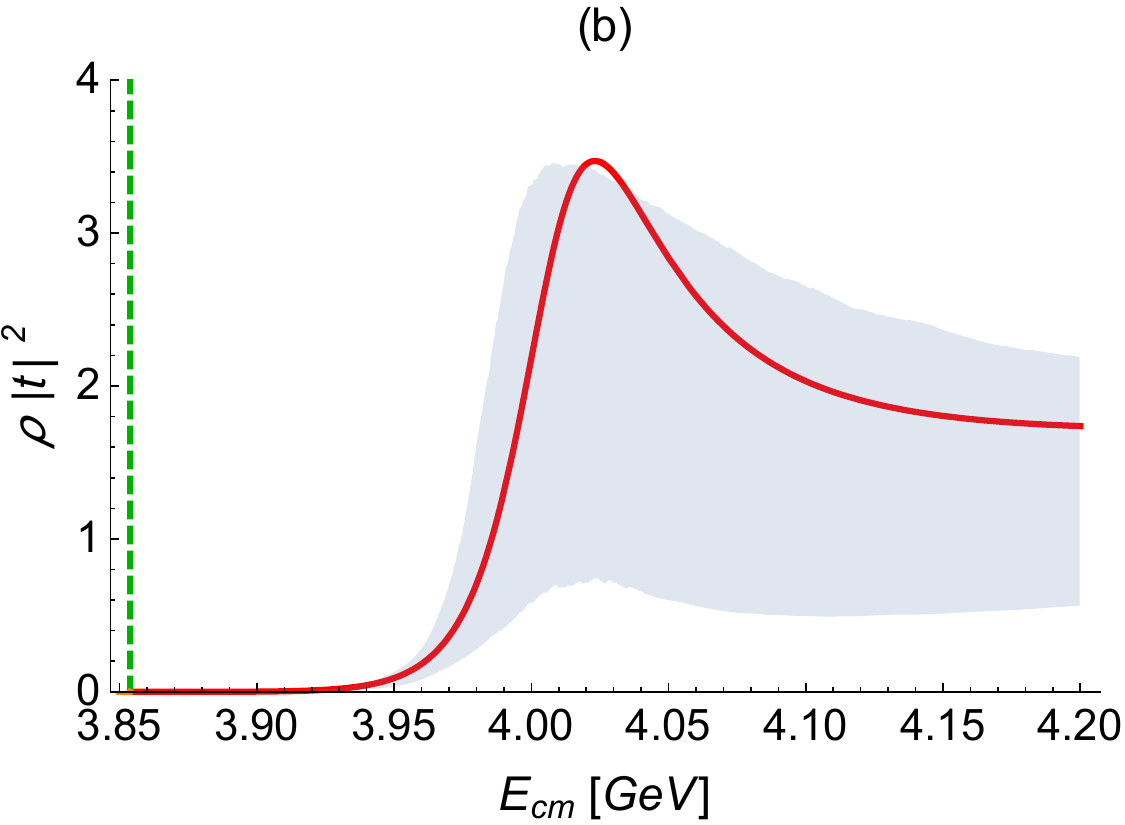}   $\quad $
	  \includegraphics[width=0.41\textwidth]{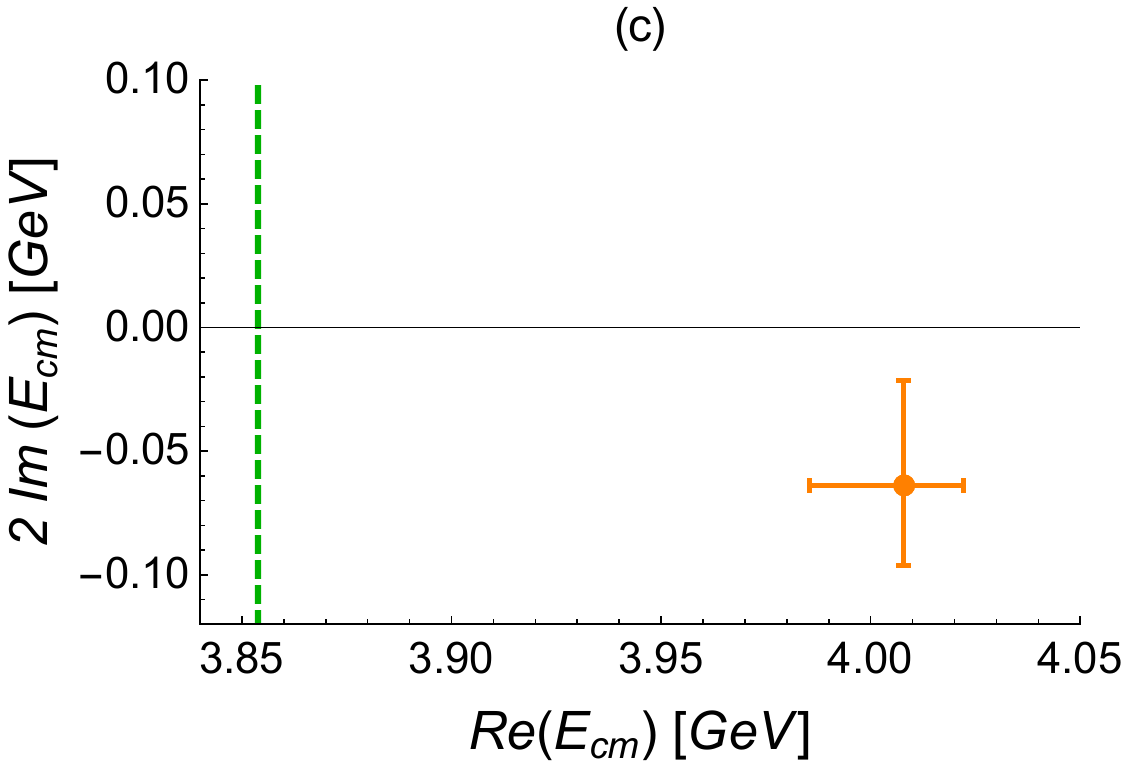} 
	  		\caption{$D\bar D$ scattering in partial wave $l=2$
                  in the energy region around the conventional $\chi_{c2}(3930)$ resonance. (a)
                  The purple crosses show the quantity  $p\cot
                  \delta/E_{cm}$  and the red line represents a simple
                  Breit-Wigner resonance fit. (b) The quantity $\rho |t|^2$
                   that is
                  proportional to the number of $D\bar D$ events in
                  experiment $N_{D_s\bar D_s} \propto p\sigma\propto \rho |t|^2$ shows a  resonance peak. 
                  (c)   The position of the $\chi_{c2}(3930)$ resonance  pole
                  $E_{cm}^p=m-\tfrac{i}{2} \Gamma $  of the scattering matrix
                  on sheet  II from Eq.~(\ref{dwave-pole}).  }
	\label{fig:dwave}
	\end{center}
\end{figure}

   \subsection{$D\bar D$   scattering  with $l=2$   and $J^{PC}=2^{++}$ resonance }\label{sec:dwave}
   
   This channel features charmonia with $J^{PC}=2^{++}$. It is  not the main
   focus of our study, however, an estimate of its scattering amplitude
   is required to extract the $l=0$ scattering amplitude using Eq.~(\ref{qc}). We
   consider the energy region encompassing the $2^{++}$ resonance and neglect the coupling to $D_s\bar D_s$ scattering with $l=2$, which we assume to be negligible in this region. The scattering amplitude is parametrized by the Breit-Wigner form (\ref{bw})
       \begin{equation}
       \label{dwave}
  \frac{p^5\cot \delta_{D\bar D}^{l=2}}{E_{cm}}=\frac{(\tilde K^{-1})_{11}^{l=2}}{E_{cm}}=a+ b  ~E_{cm}^2=\frac{m_{J2}^2-E_{cm}^2} {g_{J2}^2}~,
  \end{equation}
  and the parameters are extracted via the quantization condition (\ref{qc}). We find
   \begin{equation}
 \label{dwave-params} 
    \begin{array}{cc}
 m_{J2}=(1.762 \pm 0.016)~ a^{-1}=(4.026 \pm 0.036)~ \mathrm{GeV}, \ \  \\
  g_{J2}=(10.8 \pm 3.0)\  a=(4.7\pm 1.3)~ \mathrm{GeV}^{-1}\\
\end{array}\quad \mathrm{cor}=
   \left[
 \begin{array}{cc}
 1. & 0.54 \\
   & 1. \\
\end{array}
\right],
\quad \tfrac{\chi^2}{d.o.f.}=1.0~,
\end{equation}
  using the four lattice energies closest to the resonance region (employing
  irreducible representation $B_1$ with total momentum $|\vec P|\!=\!1$ as detailed
  in Appendix \ref{app:dwave}).   The corresponding fit is shown in Fig.~\ref{fig:dwave}a.
  The  mass $m_{J2}$ corresponds to the energy where
  the phase-shift reaches $\pi/2$, which is close-to the $2^{++}$ resonance mass   obtained from the pole position below, while the coupling $g_{J2}$ is related to its width as shown in Eq.~(\ref{bw}).  
  
  The position of the pole $E_{cm}^p $ of the scattering matrix (\ref{dwave})
  on sheet II provides a better way of determining the resonance mass $m$ and width $\Gamma$. We obtain
  \begin{equation}
  \chi_{c2}(3930): \quad E^p=( 4.008 ^{~+0.014} _{~-0.022})~-  \tfrac{i}{2} ~(0.064 ^{+0.032}_{ -0.042})~\mathrm{GeV} =m-\tfrac{i}{2} \Gamma.
  \label{dwave-pole}
  \end{equation}
The pole is plotted in Fig. \ref{fig:dwave}c. This leads to the lowest $J^{PC}=2^{++}$ resonance above $D\bar D$ threshold with   
    \begin{align}
   \label{dwave-lat} 
  \chi_{c2}(3930): &\quad m-M_{av}=905 ^{~+14} _{~-22}~\mathrm{MeV}~,\ \ \Gamma= 64 ^{+32}_{ -42}~\mathrm{MeV}\ , \quad g= 4.5 ^{+0.7} _{-1.5} \ \mathrm{GeV^{-1}}~,    \end{align} 
  where $g$ parametrizes the width  $\Gamma= g^2  p^5/m^2$. This likely corresponds to the  well-established
  resonance $\chi_{c2}(3930)= \chi_{c2}(2P)$ \cite{pdg}; a detailed
  comparison with experiment is made in Section~\ref{sec:summary}. The resonance mass and the coupling
  obtained from the pole and from Eq.~(\ref{dwave-params}) are consistent, which is expected for a narrow resonance.   
  
  The next higher $2^{++}$  charmonium is estimated\footnote{This is  based on the  second excited level in irrep $E^{++}$ with $P=0$ that employs only $\bar cc$ interpolators.} to be near $E_{cm}\simeq 4.2~$GeV, which is above our region of interest.  We assume it  to be narrow and to have a  negligible effect on the analysis of the lower-lying $2^{++}$ resonance.   
 
\begin{figure}[p!]
	\begin{center}
	\includegraphics[width=0.49\textwidth]{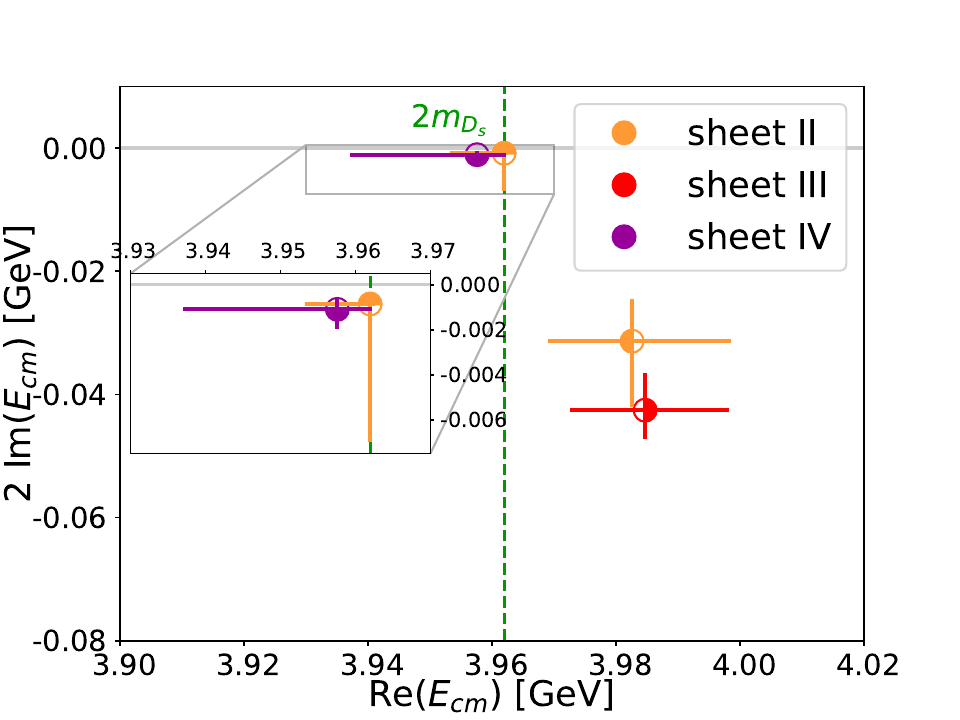}  
	\includegraphics[width=0.49\textwidth]{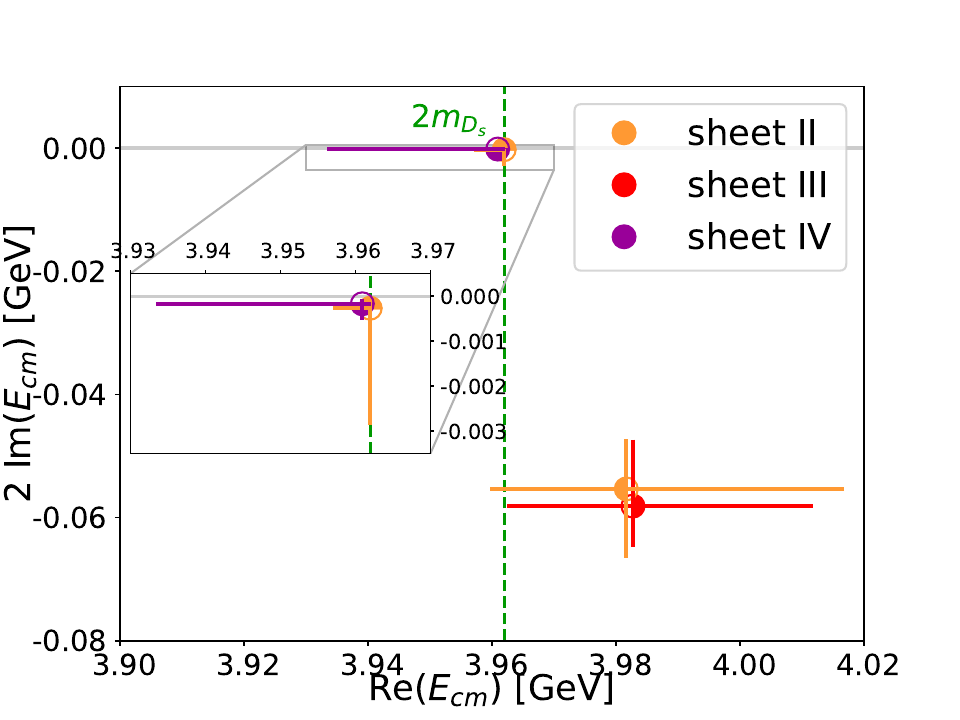}  
	\includegraphics[width=0.49\textwidth]{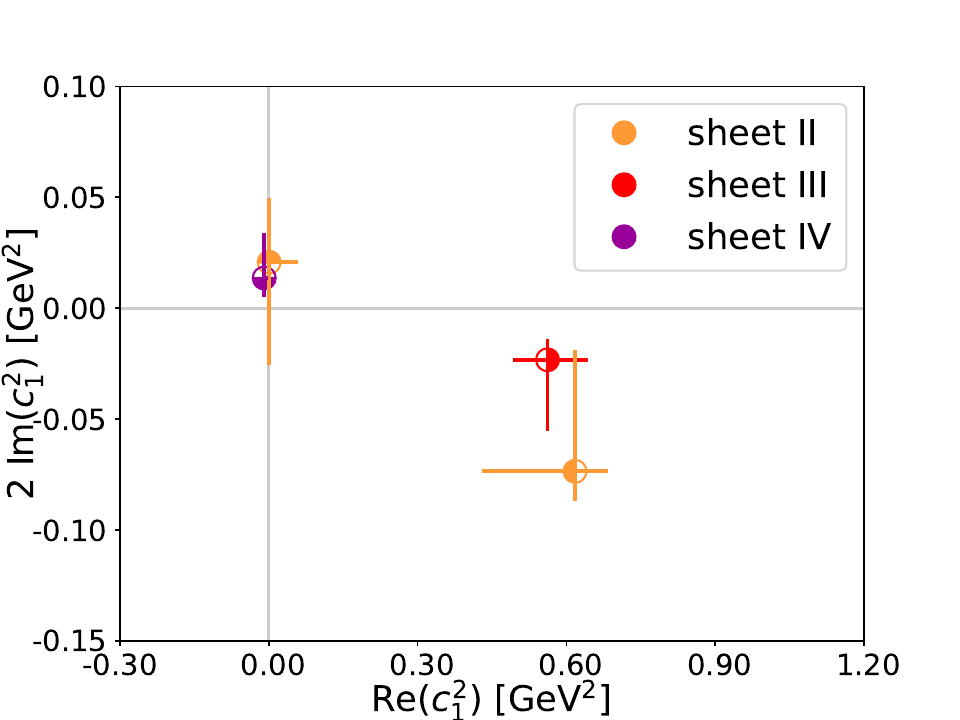}  
	\includegraphics[width=0.49\textwidth]{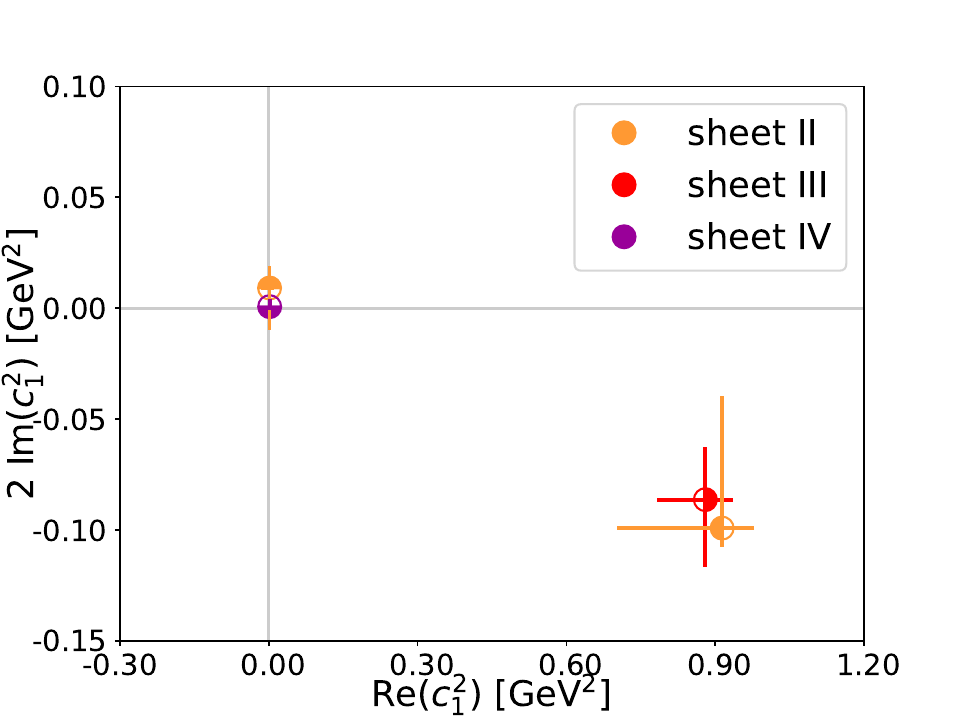} 
	\includegraphics[width=0.49\textwidth]{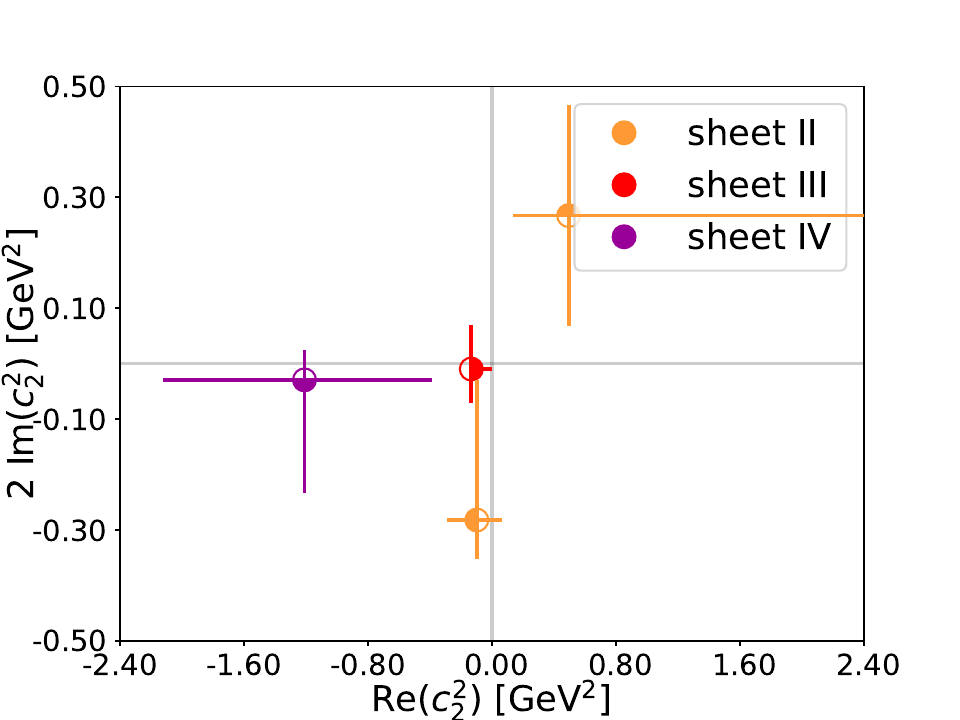}    
	\includegraphics[width=0.49\textwidth]{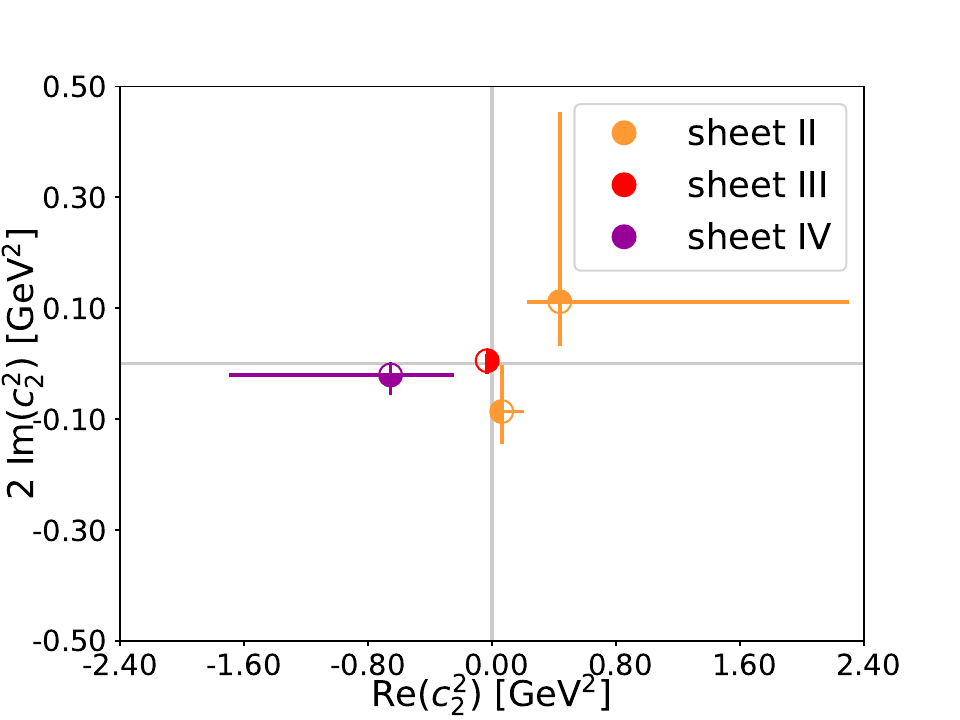}    
	\caption{   Coupled channel $D\bar D$, $D_s\bar D_s$ scattering in  partial wave $l=0$: the analysis omitting $l=2$ (left) and including $l=2$ (right).    Top:  
		Location of the poles in $t_{ij}$
                (\ref{DD-DsDs},\ref{DD-DsDs-without2-params},\ref{DD-DsDs-with2-params}) in the complex
                energy plane. The orange poles are on Riemann sheet II, the magenta
                poles are on sheet III and the violet poles are on sheet IV
                (see Eq. (\ref{sheets})). 
                There is only one pole near the $D_s\bar D_s$ threshold for each bootstrap sample and it appears either on sheet II or the nearby sheet IV. 
                The green dashed line denotes the $D_s\bar D_s$ threshold with $m_{D_s}\simeq 1981$ MeV. Middle and bottom: The  couplings $c_{i}^2$ to channels
                $i=1 ~(D\bar D)$ and  $i=2 ~(D_s\bar D_s)$ as extracted from the respective pole residues according to Eq.~(\ref{residues}). The symbol fillings distinguish the two orange
                poles. }
	\label{fig:DD-DsDs-poles}
	\end{center}
\end{figure}
\afterpage{\clearpage}

\subsection{Coupled $D\bar D$, $D_s\bar D_s$    scattering with $l=0$ for   $E_{cm} \simeq 3.93-4.13~$GeV  } \label{sec:DD-DsDs}
     
    Finally, we turn to the coupled $D\bar D-D_s\bar D_s$ scattering. We focus on the energy region $E_{cm} \simeq 3.93-4.13~$GeV
    near the   $D_s\bar D_s$ threshold and we  find an indication for several
    interesting hadrons. The  scattering matrix for partial wave $l=0$  is
    parametrized as 
  \begin{equation}
        \label{DD-DsDs}
    \frac{(\tilde K^{-1})^{l=0}}{E_{cm}}=\begin{pmatrix} a_{11}+ b_{11} E_{cm}^2 & a_{12} \\ a_{12} &  a_{22}+ b_{22} E_{cm}^2 \end{pmatrix}\equiv\begin{pmatrix} \frac{m_{J0}^2-E_{cm}^2} {g_{J0}^2}   & a_{12} \\ a_{12} &   a_{22}+ b_{22} E_{cm}^2 \end{pmatrix}~,
  \end{equation}
with the off-diagonal element held constant in $E_{cm}$.  Of the two equivalent parametrizations shown above, we will utilize the one on the rhs.
The 5 parameters in Eq.~(\ref{DD-DsDs}) are determined using all  levels of irreps $A_1^{(+)}$  within the energy region $E_{cm}=3.93-4.13~$GeV displayed  in Fig. \ref{fig:Ecm}: there are 14 levels from three frames with $P^2=0,1,2$ and  from two spatial volumes $N_L=24,32$ (see the black circles in the figure). The quantization condition (\ref{qc}) for  $A_1^{(+)}$ irreps  depends on the  scattering amplitudes for $l=0$, which we aim to determine. However, it  also  depends on the scattering amplitudes for $l=2$ when $P>0$.  Below we present analyses both including and excluding the contribution from the $l=2$ partial wave.

       \subsubsection{{Analysis omitting $l=2$} }\label{sec:DD-DsDs-without2}
       
       In the first  analysis we omit the contribution of the partial-wave $l=2$. 
       This is expected to be a fair approximation since $l\!=\!2$   effects $D\bar D$ scattering only in the narrow $2^{++}$ resonance region   that is at the upper end of the current energy range of interest. The 5 parameters of the scattering matrix (\ref{DD-DsDs}) are determined employing 
     13 out of 14 eigen-energies, where  the  level dominated by $\bar cc^{[J=2]}$
      operators is omitted\footnote{ This level corresponds to the 5th excited state in the $A_1$ irrep with $P=1$ on the smaller volume, $N_L=24$, see Fig. \ref{fig:Ecm}. } 
      \small
   \begin{align}
&\nonumber 
   \begin{array}{cc}
 m_{J0} =(1.744\pm 0.0060)~a^{-1}=(3.986\pm 0.014)~\mathrm{GeV} \\
  g_{J0}=(0.473\pm 0.023)~a^{-1}=(1.08\pm 0.053)~\mathrm{GeV}  \\
   a_{22} = - 1.474 \pm 0.019 \\
  b_{22}= (0.490 \pm 0.010)~a^2 \\
  a_{12}=-0.0362 \pm 0.0011 \\
\end{array}\ \ \mathrm{cor}=    
     \left[
\begin{array}{ccccc}
 1. & -0.034 & 0.0756 & 0.11 & 0.049 \\
   & 1. & 0.049 & 0.041 & 0.26 \\
   &   & 1. & 0.12 & 0.030 \\
   &   &   & 1. & 0.16 \\
  &  &   &   & 1. \\
\end{array}
\right], \\ 
&\qquad \qquad \qquad \frac{\chi^2}{d.o.f}=0.47~. 
 \label{DD-DsDs-without2-params} 
\end{align}
\normalsize The lattice energy levels and the levels predicted by this
parametrization are compared in Appendix \ref{app:rough}, where we
verify that the same number of levels is observed and
predicted.

The resulting scattering matrix has several poles in the energy region investigated  and their locations  ($E_{cm}^p$) in the complex energy plane are shown in the left pane of Fig. \ref{fig:DD-DsDs-poles}.  The couplings $c_i$ to both channels are extracted from the behavior of $t_{ij}$ near the poles using Eq.~(\ref{residues}) and are given in   the same figure.

                
        \subsubsection{{Analysis including $l=0$ and $l=2$} }\label{sec:DD-DsDs-with2} 
                
        In the following we present our main result for the coupled
        channel scattering in the energy region $3.93-4.13~$GeV.  
     We fix the  scattering amplitude for $D\bar D$ scattering in partial wave $l=2$
     to the values in Eq.~(\ref{dwave-params}), while $D_s\bar D_s$ scattering in partial wave $l=2$ is neglected.  
          The parameters of the coupled $D\bar D$, $D_s\bar D_s$ scattering
          matrix (\ref{DD-DsDs}) with $l=0$ are then determined using all 14
          eigen-energies in the energy region of interest (from irreps
          $A_1^{(+)}$) 
          \small
               \begin{align}
 & \nonumber     \begin{array}{cc}
 m_{J0} =(1.744\pm 0.0011)~a^{-1}=(3.986\pm 0.026)~\mathrm{GeV} \\
  g_{J0}=(0.583\pm 0.028)~a^{-1}=(1.333\pm 0.064)~\mathrm{GeV}  \\
   a_{22} = -2.409  \pm 0.018 \\
  b_{22}= (0.800 \pm 0.009)~a^2 \\
  a_{12}=-0.0176 \pm 0.0005 \\
\end{array}\  \mathrm{cor}=    
    \left[
\begin{array}{ccccc}
 1. & -0.078 & 0.13 & 0.079 & 0.023 \\
   & 1. & -0.12 & -0.14 & 0.33 \\
   &   & 1. & -0.0019 & -0.18 \\
   &   &   & 1. & -0.0096 \\
   &   &  &   & 1. \\
\end{array}
\right], \\
& \qquad \qquad \qquad \tfrac{\chi^2}{dof}=0.30~.
 \label{DD-DsDs-with2-params} 
\end{align} 
\normalsize
The coupling between channels $a_{12}$ is non-zero but small.  We also performed a study where five parameters  for  $l=0$  and  two parameters  for $l=2$ are fitted simultaneously using  18 levels  of irreps $A_1^{(+)}$ and $B_1$, and we obtain consistent results. 

\begin{figure}[h!]
	\begin{center}
	\includegraphics[width=0.32\textwidth]{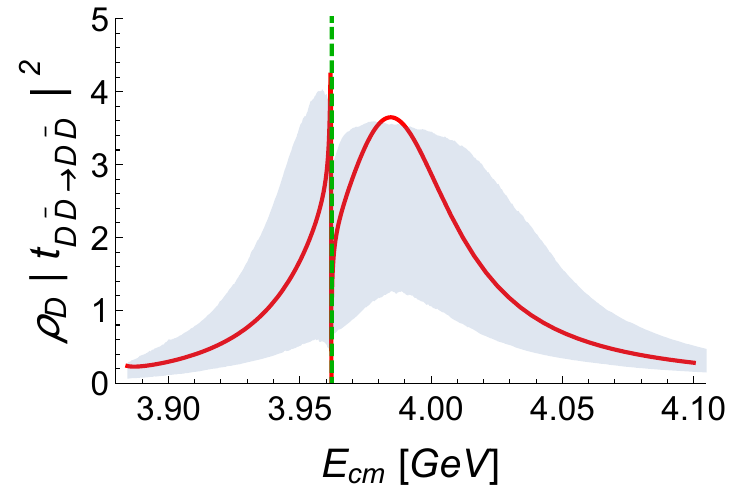}   
	\includegraphics[width=0.33\textwidth]{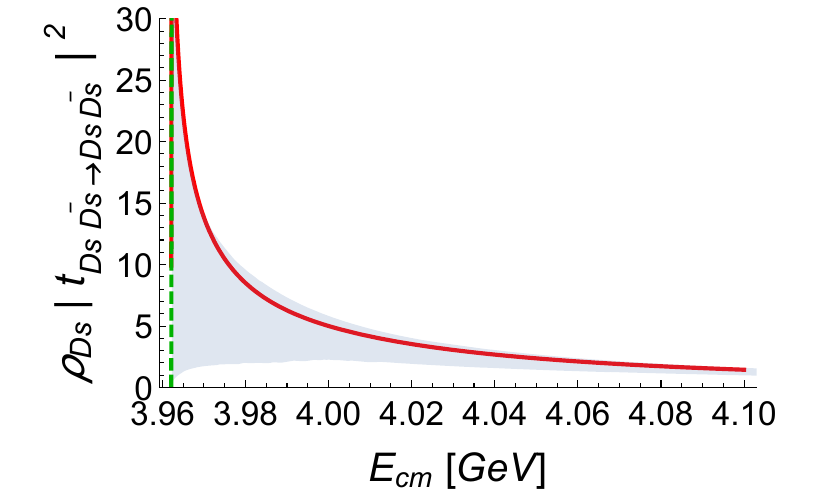}   
	\includegraphics[width=0.33\textwidth]{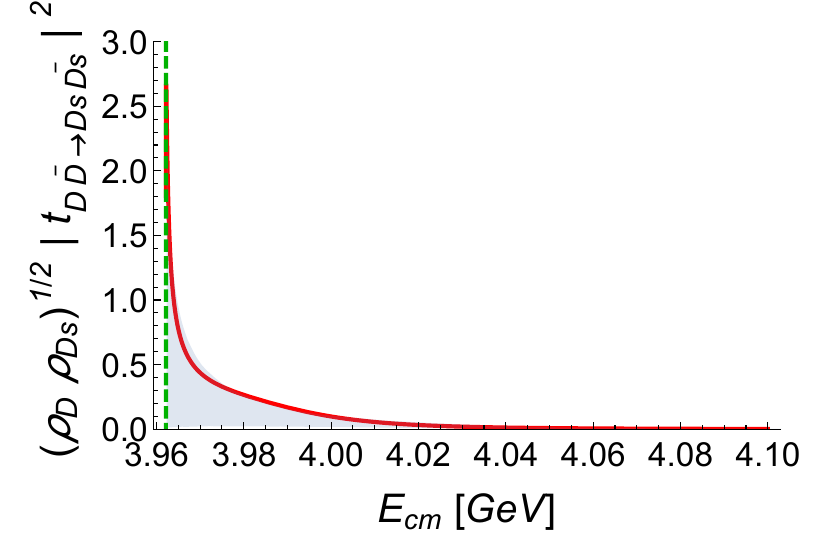}   
		\caption{   Coupled channel $D\bar D$, $D_s\bar D_s$ scattering  in  partial wave $l\!=\!0$. The three plots show the quantity  $\sqrt{\rho_i\rho_j} |t_{ij}|^2\propto p|\sigma| $ on the physical axes, that is related to the number of events in experiment  ($\rho=2p/E_{cm}$) for
		 $D\bar D\to  D\bar D$ (left), $D_s\bar D_s\to  D_s\bar D_s$ (middle) and  $D\bar D\to  D_s\bar D_s$ (right). The green lines indicate the $D_s\bar D_s$ threshold with $m_{D_s}\simeq 1981$ MeV.  The gray bands represent the uncertainty defined in Appendix~\ref{app:errors}. }
	\label{fig:DD-DsDs-with2-t}
	\end{center}
\end{figure}    
\begin{figure}[h!]
	\begin{center}
	\includegraphics[width=0.32\textwidth]{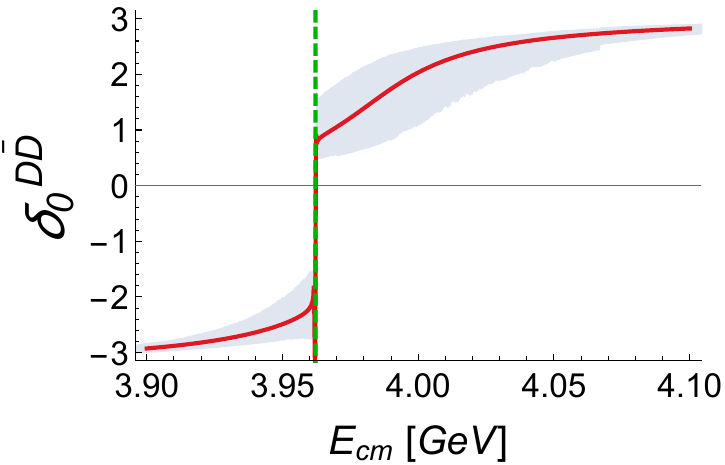}   
	\includegraphics[width=0.32\textwidth]{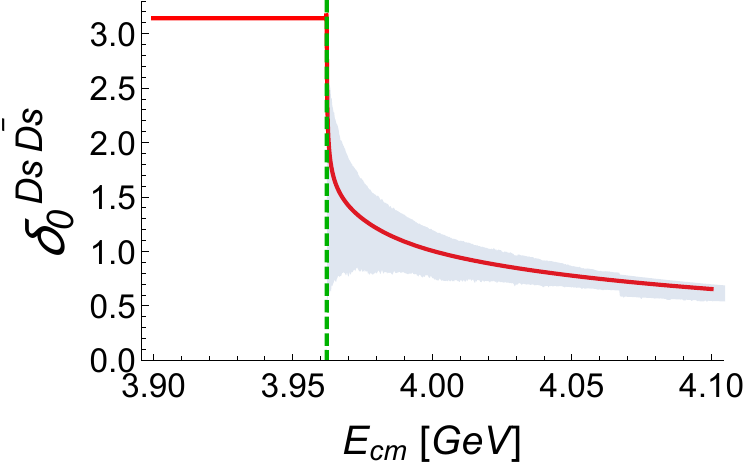}   
	\includegraphics[width=0.32\textwidth]{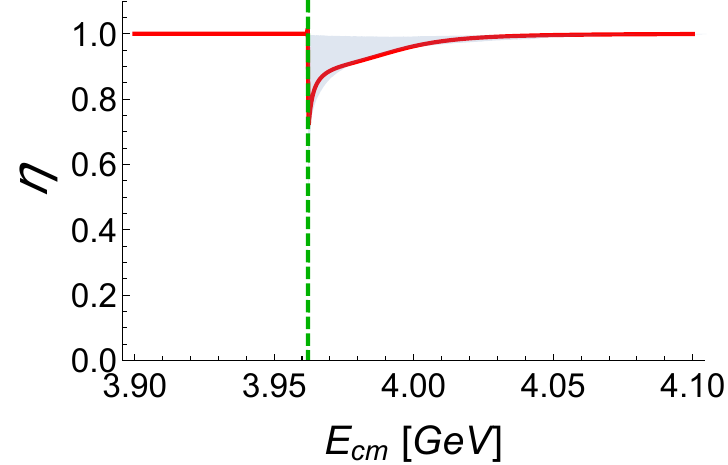}   
			\caption{
		Coupled channel $D\bar D$, $D_s\bar D_s$ scattering  in  partial wave $l\!=\!0$: two phase shifts $\delta_0^{DD}$, $\delta_0^{D_s\bar D_s}$ and inelasticity $\eta$ that parametrize the $2\times 2$ scattering matrix $S$ (\ref{S-coupled}) are shown as a function of $E_{cm}$.  The green lines  indicates the $D_s\bar D_s$ threshold with $m_{D_s}\simeq 1981$ MeV.  The gray bands represent the uncertainty.}
	\label{fig:DD-DsDs-with2-phases}
	\end{center}
\end{figure}

 The  scattering matrix has several poles in the  energy region investigated
 and their locations  ($E_{cm}^p$) in the complex energy plane are shown in
 the right pane of Fig. \ref{fig:DD-DsDs-poles}. The elements of scattering matrix $t_{ij}$ (\ref{residues})
 near the poles are parametrized by the couplings $c_i$ presented in  the same
 figure. These  give information on the couplings  of a resonance
 (associated with the  pole) to both channels $i=1,2$. The location of poles
 and the corresponding $c_i$ are similar to those extracted  in the analysis omitting $l=2$,
 shown in the left pane of Fig. \ref{fig:DD-DsDs-poles}.  The   experimental rates
 are related to the values of $\sqrt{\rho_i\rho_j} |t_{ij}|^2$ on the physical
 axes, which are
 presented for $D\bar D\to  D\bar D$, $D_s\bar D_s\to  D_s\bar D_s$ and
 $D\bar D\to  D_s\bar D_s$ in Fig. \ref{fig:DD-DsDs-with2-t}. Alternatively, the unitary
 $2\times 2$ scattering matrix $S$ (\ref{S-coupled}) on the physical axes can
 be described by the
 phase shifts $\delta^{D\bar D}$, $\delta^{D_s\bar D_s}$ and inelasticity
 $\eta$, which are shown in Fig.~\ref{fig:DD-DsDs-with2-phases}. 
 
 We find that the $D\bar D$ and $D_s\bar D_s$ channels are not strongly
 coupled, which can be seen from the  inelasticity
 $\eta\simeq 1$  in Fig.~\ref{fig:DD-DsDs-with2-phases}  (\footnote{ The inelasticity is
   equal to one below  the $D_s\bar D_s$  threshold by construction due to the $\Theta$-function in Eq.~(\ref{S-coupled}).})  and from the smallness of the off-diagonal
 element $a_{12}$ in Eqs.~(\ref{DD-DsDs-with2-params}) and (\ref{DD-DsDs}).   Our
 results suggest there are two $0^{++}$ resonances in this energy region: a
 narrow resonance dubbed $\chi_{c0}^{D_s\bar D_s}$ just below the $D_s\bar D_s$
 threshold  and a broader one denoted by  $\chi_{c0}^{\prime}$.

 {\bf The broader $\bm{0^{++}}$ resonance $\bm{\chi_{c0}^\prime}$} is related to the pole
 indicated in red on sheet III 
  \begin{equation}
  \label{chic02P-pole}  
  \chi_{c0}^\prime:\quad  E^p=( 3.983 ~ ^{+0.028}_{-0.020} )~-  \tfrac{i}{2} ~(0.058 ~ ^{+0.006}_{-0.011})~\mathrm{GeV} =m-\tfrac{i}{2} \Gamma~.
  \end{equation}
This pole affects the scattering amplitude on the physical axes above the $D_s\bar
D_s$ threshold and is responsible for a peak around
$3.98~$GeV in the $D\bar D\to D \bar D$  rate shown in the left pane of Fig.~\ref{fig:DD-DsDs-with2-t}. 
The presence of this pole is also reflected in the phase shift $\delta_0^{D\bar D}$, which  increases gradually 
starting from $2m_{D_s}$ as is evident in the left pane of Fig.~\ref{fig:DD-DsDs-with2-phases}.
The nearby pole on sheet II does not have a significant influence on the physical scattering above the second threshold. 
  The pole residues indicate that this state decays predominantly to $D\bar D$,
while the decay to $D_s\bar D_s$ is suppressed, as evidenced by $|c_1|\gg
|c_2|$, see in the last two rows of Fig.~\ref{fig:DD-DsDs-poles} for the pole presented in red.
The resonance parameters are  
 \begin{align}
   \label{chic0prime-lat} 
  \chi_{c0}^\prime: \ \ &m-M_{av}=880  ^{+28}_{-20} ~\mathrm{MeV},\ \ \Gamma = 58 ~ ^{+6}_{-11} ~\mathrm{MeV}\ , \quad g = 1.35 ~^{+0.04}_{-0.08}   \ \mathrm{GeV}~,
   \end{align}
where  $  M_{av}\!=\! \tfrac{1}{4}(3m_{J/\psi}+m_{\eta_c})$, and the coupling $g$ parametrizes the full width $ \Gamma = g^2   p_D/m^2$. 
 The possible relation of this state  to the  broad resonance  $\chi_{c0}(3860)$  discovered by Belle in 2017 \cite{Chilikin:2017evr,pdg} is discussed in Section \ref{sec:summary}.  
      
 {\bf The narrow $\bm{0^{++}}$ resonance $\bm{\chi_{c0}^{D_s\bar D_s}}$}  near the $D_s\bar
D_s$  threshold 
is related to the  pole on sheet II, indicated by the top-filled
orange symbols in the first row of Fig.~\ref{fig:DD-DsDs-poles}. Its location
relative to the threshold is given by
   \begin{equation}
  \label{chic0DsDs-pole}
  \chi_{c0}^{D_s\bar D_s}: \ \ 
 E_{cm}^p-2 m_{D_s}= ( -0.2 ^{~+0.16}_{~-4.9} )~-  \tfrac{i}{2} ~( 0.27^{~+2.5}_{~-0.15})  ~\mathrm{MeV} =m -\tfrac{i}{2} \Gamma~.
    \end{equation} 
    This resonance is related to the bound state  in the analysis of
    $D_s\bar D_s$-scattering in the one-channel approximation of Section \ref{sec:DsDs}.  
The  pole  on sheet II and the nearby pole    on sheet IV  correspond to this resonance and
      are mutually exclusive across the bootstrap samples. Further details on this can
      be found in Appendix \ref{app:rough}.
   It is clear from Figure~\ref{fig:DD-DsDs-with2-t} that the resonance pole leads to a sharp rise in the $D_s\bar D_s\to D_s\bar D_s$ and $D\bar D\to D_s\bar D_s$ rates just above $2m_{D_s}$.
       The increased $D\bar D\to D_s\bar D_s$ rate
       is also responsible for a dip in the $D\bar D\to D\bar D$ rate  at $2m_{D_s}$ and
       all three features should be  used as a signature for
       experimental searches of this  state. Note that the magnitude of the $D_s\bar D_s\to D_s\bar D_s$  
       peak above $2m_{D_s}$ is larger when the pole is closer to the threshold.

       $\chi_{c0}^{D_s\bar D_s}$ couples predominantly to $D_s\bar D_s$ and very weakly to $D\bar D$ (one can see that $|c_2|^2\gg |c_1|^2$ in Fig.~\ref{fig:DD-DsDs-poles}). The mass difference of the state with respect to the threshold and its narrow total width $\Gamma=g^2p_D/m^2$ parametrized in terms of $g$ are
    \begin{align}
   \label{chic0DsDs-lat} 
      \chi_{c0}^{D_s\bar D_s}: \ \
      &m-2m_{D_s}=-0.2 ^{~+0.2}_{~-0.3} ~\mathrm{MeV}~,\quad \Gamma=0.27^{~+2.5}_{~-0.15} ~\mathrm{MeV}~,\quad g=0.096^{~+ 0.215} _{~-0.033}   ~\mathrm{GeV}~.
      \end{align}
On the experimental side,  the newly discovered $\chi_{c0}(3930)$
\cite{chic03930} and the $X(3915)$ \cite{pdg} lie near the $D_s\bar D_s$ threshold and have very small or zero decay rate to $D\bar D$. The indication for a  $D_s\bar D_s$ state in our study  explains both properties, as detailed in Section \ref{sec:summary}.

The  parameters of the scattering matrix obtained from the analysis
         including   or excluding  $l=2$ are similar, with the exception of
         $a_{22}$ and $b_{22}$. These    parametrize $D_s\bar D_s\to
         D_s\bar D_s$   and differ between the coupled-channel analysis  and
         the one-channel approximation,  however, both analyses  lead to a state just below the $D_s\bar D_s$ threshold on the real axis (see Fig.~\ref{fig:DsDs}c) or slightly away from it.
         The   conclusion  that there is a near-threshold pole is robust, while its exact
         location and the effect on   physical scattering need to be investigated
          in a simulation with higher statistics and a better control of systematic uncertainties.

 \section{Summary of the resulting hadrons and their relation to  experiment}    \label{sec:summary}

\begin{figure}[tb]
	\begin{center}
	\includegraphics[width=0.55\textwidth]{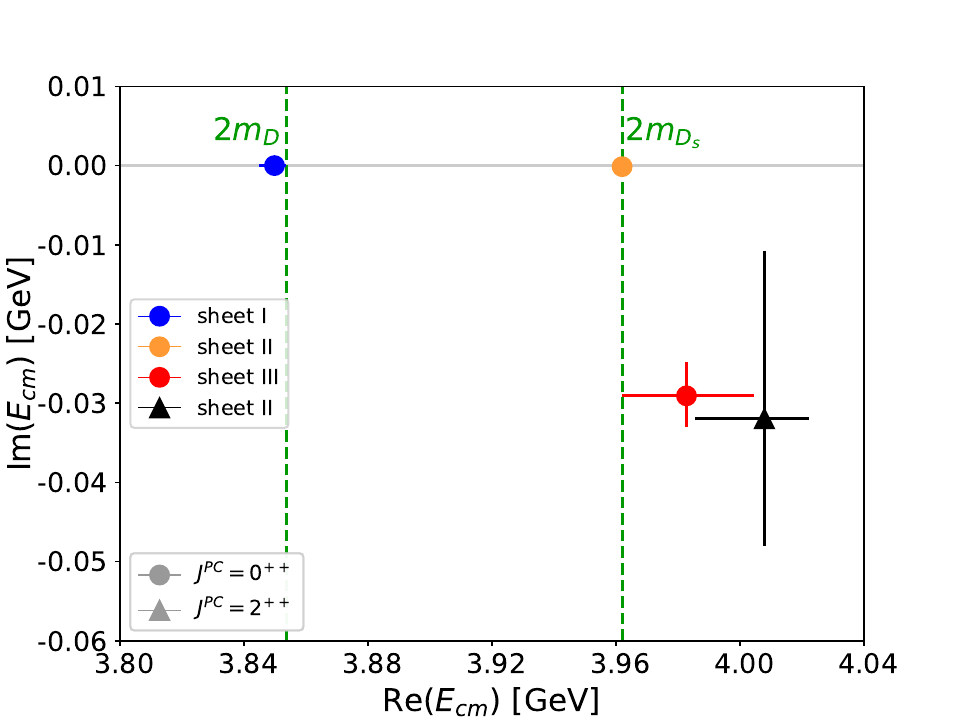}   
		\caption{ 
		 Pole singularities of the scattering amplitude/matrix in the complex energy plane, which   are  associated with the hadrons predicted in this work. The pole related to the $J^{PC}=2^{++}$ resonance appears on sheet II, as we have neglected the $l=2$ partial wave contribution from $D_s\bar D_s$ scattering.}
	\label{fig:summary-poles}
	\end{center}
\end{figure}

\begin{figure}[tbh!]
	\begin{center} 
	\includegraphics[width=0.7\textwidth,clip]{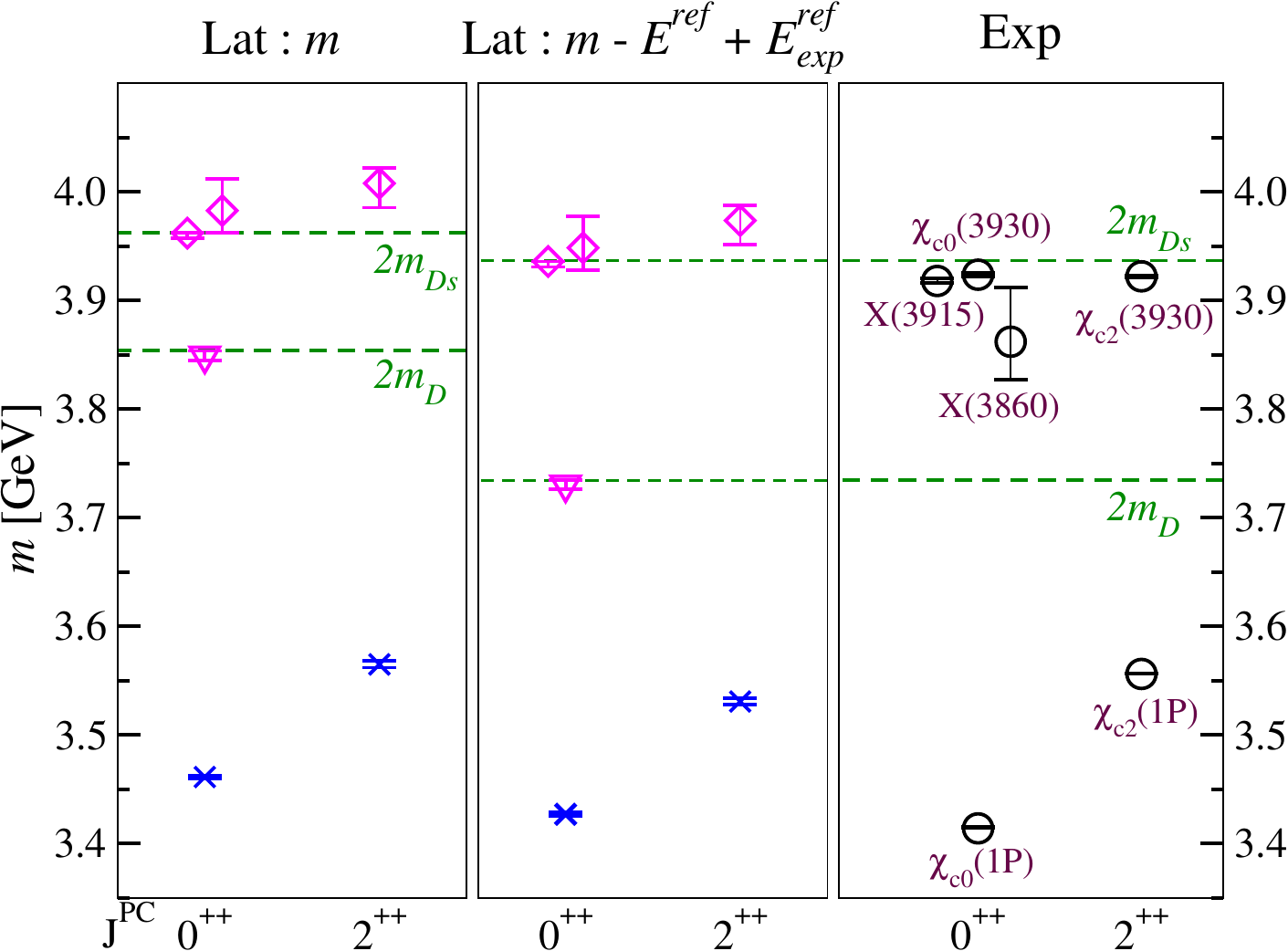}   
		\caption{  Energy
                  spectrum of hadrons predicted in this work compared with
                   experiment. The left and middle panes show the
                  lattice results, where $m$ denotes the mass  obtained
                    from the lattice study.  The magenta symbols  correspond
                    to hadrons extracted via the scattering analysis:
                    diamonds  represent resonances and triangles represent
                    bound states. The blue crosses are extracted directly from
                    the lattice energies. The reference energy in the middle
                    pane is $E^{ref}=2m_D$ ($2m_{D_s}$) for the state closest
                    to the $D\bar D$ ($D_s\bar D_s$) threshold, while
                    $E^{ref}=M_{av}=\tfrac{1}{4}(3m_{J/\psi}+m_{\eta_c})$  for
                    the remaining four states. The right  pane shows the experimental spectrum \cite{pdg}, where $\chi_{c0}(3930)$   \cite{chic03930} and $X(3915)$ \cite{pdg} may be the same state.  }
	\label{fig:summary}
	\end{center}
\end{figure}

    Below we summarize the properties of the charmonium-like hadrons
    found in this simulation. These are denoted  by the
    conventional names $\chi_{c0}(1P)$, $~\chi_{c2}(1P)$, $\chi_{c2}(3930)$
    when
    the identification with experimental states is unambiguous, while  the other states found are denoted $\chi_{c0}^\prime$, $\chi_{c0}^{D\bar D}$ and $\chi_{c0}^{D_s\bar D_s}$. The subscripts $c0$ and $c2$ indicate the assignment of $J^{PC}=0^{++}$ and  $2^{++}$, respectively.  
The  location of the poles   in the complex energy plane  related to these
  hadrons are given in Fig.~\ref{fig:summary-poles}, while the corresponding
  masses are  compared to experiment in Fig.~\ref{fig:summary}.
    Due to the unphysical quark masses employed in the simulation, the hadron
   masses obtained are not compared to experiment directly. Instead, the difference $m-E^{ref}$ is utilized, where  the reference
 energy $E^{ref}$ is either a nearby threshold or the  spin-averaged charmonium mass
 $M_{av}=\tfrac{1}{4}(3m_{J/\psi}+m_{\eta_c})$.  The positions of the $D\bar{D}$ and $D_s\bar{D}_s$ thresholds are given by the masses  $m_D\simeq 1927~$MeV and
  $m_{D_s}\simeq 1981~$MeV.


 The resonance decay widths depend on
 the phase space $p^{2l+1}$ evaluated for the  meson momenta (in the cm-frame) at the
   resonance energy, which in turn depends on the position of the
 threshold. The latter is different in the simulation and in experiment. Therefore it is customary to compare the coupling $g$ that parametrizes the full width of a  hadron 
    \begin{equation}
  \label{references}  \Gamma \equiv g^2   p_D^{2l+1}/m^2\quad \mathrm{with}\quad l=0,2\quad  \mathrm{for} \ J^{PC}=0^{++},2^{++}~,
\end{equation} 
as $g$ is expected to be less dependent on the quark masses than the width itself. \\

\subsection*{$\bm{\chi_{c0}(1P)}$ and $\bm{\chi_{c2}(1P)}$ }
    
      These states lie significantly below the $D\bar D$ threshold and their
      masses are extracted from the ground state energies
      $m\!=\!E_1(P\!=\!0)$ in irreps $A_1^{++}$ and $E^{++}$, with $J^{PC}\!=\!0^{++}$ and
    $2^{++}$, respectively. We obtain
\begin{align}
\label{chic_1P} 
  \chi_{c0}(1P):&\ m-M_{av}=358 \pm 2~\mathrm{MeV}~, \qquad  m= 3461 \pm 2~\mathrm{MeV}~, \\
\label{chic2_1P} 
\chi_{c2}(1P):&\ m-M_{av}= 462 \pm 3~\mathrm{MeV}~, \qquad  m=3565 \pm
                3~\mathrm{MeV}~,
\end{align}
which can be compared to the experimental values
\begin{align}
\label{chic_1P_exp} 
\mathrm{exp}\   \chi_{c0}(1P):& \ m-M_{av}= 346.1\pm 0.3~\mathrm{MeV}~,\\
\label{chic2_1P_exp} 
\mathrm{exp}\   \chi_{c2}(1P):& \ m-M_{av}=487.52\pm 0.14~\mathrm{MeV}~.
\end{align}

\subsection*{$\bm{0^{++}}$ state  $\bm{\chi_{c0}^{D\bar D}}$ slightly below $D\bar D$ threshold  } 
   
We find a shallow $D\bar D$ bound state 
labeled $\chi_{c0}^{D\bar D}$ with binding energy
\begin{equation} 
   m-2~m_D= -4.0 ^{~ + 3.7}_{~ -5.0}~\mathrm{MeV}~.
\end{equation}
Note that it is not known whether this bound state would
also feature in a simulation with physical quark masses.
Such a state has not been  claimed by experiments.

   The existence of a shallow $D\bar D$ bound state dubbed $X(3720)$
  was already suggested by an effective
   phenomenological model   in Ref. \cite{Gamermann:2006nm}\footnote{This state with
     $m\simeq 3.718~$GeV is listed in Table 4 of Ref. \cite{Gamermann:2006nm}.} featuring also exchanges of vector mesons. 
       Ref.~\cite{Gamermann:2007mu} indicates that
   there may already be some evidence for such a state  in the $D\bar D$ invariant mass distribution from Belle \cite{Abe:2007sya}, which shows an enhancement  just above threshold. 
The  $D\bar D$  rate  from Babar \cite{Aubert:2010ab} also shows a hint of enhancement just above threshold (see Fig. 5 of  \cite{Aubert:2010ab}).   
      In a
   molecular picture, a $0^{++}$ state is expected as a partner of $X(3872)$  via
   heavy-quark symmetry arguments \cite{Hidalgo-Duque:2013pva,Baru:2016iwj}.
   A similar virtual 
     bound state with a binding energy of 20 MeV follows from the data of 
     the only previous lattice simulation of $D\bar D$ scattering 
     \cite{Lang:2015sba}\footnote{The presence of this state was not mentioned in Ref.~\cite{Lang:2015sba}, as 
     such virtual bound states were not searched for.}.  
     
   Experimental searches for this state in inclusive final states are challenging since the region above the $D\bar{D}$ threshold can be populated by
$D\bar{D}$ from $X(3872)\to D\bar{D}^*\to D\bar{D}\pi$ (see, for example, Ref.~\cite{Aaij:2019evc}). 
Various strategies for the experimental search of such a state   in exclusive decays were proposed: $B^{0+}\to D^0\bar D^0 K^{0+}$   \cite{Dai:2015bcc}, $\psi (3770) \rightarrow  D^0 {\bar{D}}^0\gamma$ \cite{Dai:2020yfu}, $\gamma\gamma\to D\bar D$ \cite{Wang:2020elp}, $\psi(3770,4040)\to  \eta\eta^\prime  \gamma$ and $e^+e^-\to  \eta\eta^\prime J/\psi $  \cite{Xiao:2012iq}.

\subsection*{$\bm{2^{++}}$ resonance and  its relation to   $\bm{\chi_{c2}(3930)}$}
   
   We find a resonance with $J^{PC}\!=\!2^{++}$ in  $l=2$ $D\bar
   D$ scattering with the following properties 
\begin{align} 
\label{dwave-lat_compare}
\chi_{c2}(3930):\ \ &m-M_{av}=904 ^{~+14} _{~-22}~\mathrm{MeV}~,\quad g= 4.5 ^{+0.7}_{-1.5} \ \mathrm{GeV^{-1}}~.
\end{align}
   This  
is most likely related to the    conventional $\chi_{c2}(3930)$ resonance (also called $\chi_{c2}(2P)$)    discovered by Belle \cite{Uehara:2005qd}
 \begin{align}
\label{dwave-exp}
\mathrm{exp} \ \chi_{c2}(3930):\ \ & m -M_{av}=854 \pm 1 ~\mathrm{MeV}~,\quad g =2.65 \pm 0.12 ~\mathrm{GeV^{-1}}~.   
\end{align} 
Here $g$ parametrizes the width  $\Gamma \!=\! g^2  ~p_D^5/m^2$.  The masses are
reasonably close, while  the coupling from lattice QCD is  larger that in
experiment. However, the couplings are also not inconsistent given the large statistical error from our simulation and
  the unquantified systematic uncertainties discussed in Section~\ref{sec:caveats}.  

\subsection*{Broad $\bm{0^{++}}$ resonance  and  its possible relation to  $\bm{\chi_{c0}(3860)}$}
      
     This resonance couples mostly to $D\bar D$ and has a very small
     coupling to  $D_s\bar D_s$. Its resonance
     parameters are 
\begin{align}
   \label{chic0prime-lat_compare} 
  \chi_{c0}^\prime: \ \ &m-M_{av}=880  ^{+28}_{-20} ~\mathrm{MeV}\ ,\quad g = 1.35~^{+0.04}_{-0.08}   \ \mathrm{GeV}~.
\end{align}
These can be compared to the scalar resonance $\chi_{c0}(3860)$  discovered by
Belle in 2017 \cite{Chilikin:2017evr}\footnote{For the couplings calculated
  from the experimental values we vary both the mass and width by $\pm
  1\sigma$ and take the maximal positive and negative deviations as the uncertainties},  
\begin{align}
\label{chic0prime-exp} 
\mathrm{exp}\ \chi_{c0}(3860):\    &m-M_{av}=793~^{+48}_{-35} ~\mathrm{MeV}\ , \quad g=2.5 ^{+1.2}_{-0.9} ~\mathrm{GeV}~,
\end{align}
based on the following arguments:
The mass and coupling are reasonably consistent with experiment, in particular, when considering the experimental errors and the
systematic uncertainties in the lattice results.
The mass is also close to the value obtained from the only previous lattice study of $D\bar{D}$ scattering \cite{Lang:2015sba}\footnote{The value to compare is                        
  $m-M_{av}^{exp}\simeq 0.90~$GeV and $0.93~$GeV from the fits in Eqs.~(6.3) and (6.7) of Ref.~\cite{Lang:2015sba}, respectively.},
although the width and coupling are larger in the present work.

The experimental couplings $g$ of $\chi_{c0}(3860)$ and
      $\chi_{c2}(3930)$, quoted in Eqs.~(\ref{chic0prime-exp}) and
      (\ref{dwave-exp}), respectively, are similar, and their widths
      differ mainly because of the different phase space. On the lattice side, the
      $\chi_{c0}^\prime$  coupling in Eq.~(\ref{chic0prime-lat_compare}) is
      smaller than the $\chi_{c2}(3930)$ coupling in Eq.~(\ref{dwave-lat_compare}), however, both have large uncertainties.

\subsection*{Narrow $\bm{0^{++}}$ resonance  
  $\bm{\chi_{c0}^{D_s\bar D_s}}$ and its possible  relation to $\bm{\chi_{c0}(3930)}$, $\bm{X(3915)}$}
          
       We find a narrow $0^{++}$ resonance near the $D_s\bar D_s$
       threshold. It has a large coupling to $D_s\bar D_s$ and a
       very small coupling to $D\bar D$. The latter is
       responsible for its small decay rate to $D\bar D$ and the small total
       width. This state corresponds to the bound state in
       one-channel $D_s\bar D_s$ scattering discussed in Section \ref{sec:DsDs}. 
       We compare the resulting resonance parameters 
       \begin{align}
\label{chic0DsDs-compare_lat} 
     \chi_{c0}^{D_s\bar D_s}: \ \ &m-2m_{D_s}=-0.2^{~+0.16}_{~-4.9}~\mathrm{MeV}\ ,\quad g=0.10^{~+ 0.21}_{~-0.03}~\mathrm{GeV}~
\end{align}
with two experimental states that share similar features  to the state we find,
\begin{align}
\label{chic0DsDs-compare_exp} 
  \mathrm{exp}\ \chi_{c0}(3930): \ & \ m-2m_{D_s}=-12.9  \pm 1.6 ~\mathrm{MeV}~,\enskip \Gamma=17\pm 5 ~\mathrm{MeV}~,\enskip g=0.67 \pm 0.10  ~\mathrm{GeV}~,  \nonumber\\
  \mathrm{exp}\ X(3915): \ & \ m-2m_{D_s}=-18.3 \pm 1.9  ~\mathrm{MeV}~,\enskip \Gamma=20 \pm 5   ~\mathrm{MeV}~,\enskip g=0.72 \pm 0.10  ~\mathrm{GeV}~.
    \end{align}
The $\chi_{c0}(3930)$ with $J^{PC}=0^{++}$ was  very recently discovered in
$D\bar D$ decay by LHCb \cite{chic03930}. The $X(3915)$ was  observed by Belle \cite{Uehara:2009tx} and BaBar \cite{Aubert:2007vj,delAmoSanchez:2010jr,Lees:2012xs}  in $J/\psi \omega$ decay
and has $J^{PC}\!=\!0^{++}$ or  $2^{++}$, while its decay to $D\bar D$ was not
observed \cite{pdg}. They might  represent the same state if
$X(3915)$  is a scalar.    Both experimental states lie just below
the $D_s\bar D_s$ threshold.  One would naturally  expect
$0^{++}$ states with this mass to be broad, given the large phase space
available to $D\bar D$ decay. Their narrow widths can only be
explained if  their decay  to $D\bar D$ is suppressed by some
mechanism. 

If the resonance found on the lattice is
indeed related to $X(3915)$/$\chi_{c0}(3930)$, our results
indicate that this state owes its existence to a large
interaction in the $D_s\bar D_s$ channel near threshold, which naturally explains
why its  width is small and its decay to $D\bar D$ 
is suppressed.   Note
that a detailed quantitative comparison  of lattice and
experimental results in Eqs.~(\ref{chic0DsDs-compare_lat}) and
(\ref{chic0DsDs-compare_exp})  is not possible due to the unphysical masses of
the quarks in the lattice study and due to the omission of decays to
  $J/\psi \omega$ and $\eta_c\eta$, which may affect the
determination of the width.  The qualitative comparison, however, suggests
the existence of a $D_s\bar D_s$  resonance with small coupling to $D\bar D$. 
This could be further investigated experimentally by considering
the $D_s\bar D_s$ invariant mass spectrum near threshold, where a peak
(see Fig.~\ref{fig:DD-DsDs-with2-t}) would be visible for a state just below
threshold.

The $X(3915)$ was proposed to be a ground $\bar cc\bar ss$ state within the
 diquark-antidiquark approach  by Polosa and Lebed
 \cite{Lebed:2016yvr}.  The identification   $\bar cc\bar ss$ was   considered also in phenomenological studies  \cite{Giron:2020qpb,Chen:2017dpy}. 

    \section{Conclusions}\label{sec:conclusions}

We presented a lattice study of coupled-channel $D\bar{D}$-$D_s\bar{D}_s$ scattering in 
the $J^{PC}=0^{++}$ and $2^{++}$ quantum channels with isospin $0$. Using the generalized L\"uscher method and a piecewise parametrization of the energy region from slightly below 2$m_D$
to 4.13 GeV, the coupled-channel scattering matrix $S$
along the real energy axis was determined. The resulting $S$ was then analytically continued to search for pole singularities in the complex energy plane that can affect the scattering amplitudes/parameters along the physical axes. Our study utilized the spectrum in three different inertial frames 
determined on two CLS ensembles with $u/d$ and $s$ quarks, spatial extents $\sim$2.07 fm and 
$\sim$2.76 fm and a single lattice spacing $\sim$0.086 fm.

In addition to $\chi_{c0}(1P)$, the results suggest three charmonium-like
states with $J^{PC}=0^{++}$ below 4.13~GeV. One is a yet undiscovered 
$D\bar{D}$ bound state just below threshold. The second is a narrow resonance
just below the $D_s\bar{D}_s$ threshold predominantly coupled to
$D_s\bar{D}_s$. This state is possibly related to the narrow resonance
$X(3915)/\chi_{c0}(3930)$, which is also below the $D_s\bar{D}_s$ threshold in the
experiments. The third feature is a $D\bar{D}$ resonance possibly related to
the $\chi_{c0}(3860)$ observed by Belle, which is believed to be $\chi_{c0}(2P)$. 
An overview of the resulting pole structure of the coupled-channel $D\bar{D}$-$D_s\bar{D}_s$ scattering matrix 
in the complex energy plane is given in Fig.~\ref{fig:summary-poles}, and the
possible implications of this singularity structure for experiments are 
illustrated in Figs.~\ref{fig:DD-DsDs-with2-t} and
\ref{fig:DD-DsDs-with2-phases}. The   masses  are compared to experiment in Fig. \ref{fig:summary}  and   summarized in   Section \ref{sec:summary}. 
 
Turning to states with  $J^{PC}=2^{++}$, the mass of the ground state $\chi_{c2}(1P)$ 
was determined directly from the lattice energy and is compared with the experimental value in Eq.~(\ref{chic2_1P}). 
We have assumed the $2^{++}$ resonance to be coupled only with the $D\bar D$ scattering channel in the $l=2$ 
partial wave and have parametrized this with a Breit-Wigner form. The resonance parameters are extracted and compared with the experimental values of the conventional $\chi_{c2}(3930)$ in Eqs.~(\ref{dwave-lat_compare}) and (\ref{dwave-exp}). These are then fixed for the finite-volume coupled-channel analysis discussed in 
Section \ref{sec:DD-DsDs-with2}. We find the estimates for positions and residues for the poles with $J^{PC}=0^{++}$ 
to be robust with the exclusion/inclusion of the $l=2$ partial wave contribution to the analysis. The resulting 
pole positions and the residues from either study are shown in Fig. \ref{fig:DD-DsDs-poles}.
 
In this study we worked with several simplifying assumptions (detailed
in Section~\ref{sec:caveats}) necessary for a first investigation of
this coupled-channel system. The lattice QCD ensembles we used have
heavier-than-physical light and charm quarks and a
lighter-than-physical strange quark. This results in a
smaller-than-physical splitting between the $D\bar{D}$ and the
$D_s\bar{D}_s$ thresholds. In future studies, it will be important to
systematically improve upon the current results by successively
relaxing our assumptions, for example, by explicitly including
$\eta_c\eta$ interpolating fields and adding this channel as well as
$J/\psi\omega$ to the coupled-channel study. It is also essential
  to investigate additional parametrizations to test the model
  independence of our findings and this will require a larger set of
  ensembles with high statistics. With regard to the pole structure
observed in this work, it would be particularly interesting to
investigate how our observations, such as the shallow $D\bar{D}$ bound
state and our $X(3860)$ candidate evolve when simultaneously
approaching the limit of physical quark masses and the continuum
limit.

    \acknowledgments

We thank G. Bali, V. Baru, T. Gershon, F.-K. Guo, D. Johnson,  C.~B.~Lang, R. Molina, J. Nieves, E. Oset, S. Paul, 
A.~Sch\"afer and J. Simeth for useful discussions. We thank the authors of Ref.~\cite{Morningstar:2017spu} for making 
the TwoHadronsInBox package public and Ben H\"orz and C.~B.~Lang for contributions 
to the computing codes we used. We use the multigrid solver of
Refs.~\cite{Heybrock:2014iga,Heybrock:2015kpy,Richtmann:2016kcq,Georg:2017diz}
for the inversion of the Dirac operator. Our code implementing
distillation is written within the framework of the Chroma software
package~\cite{Edwards:2004sx}. The simulations were performed on the
Regensburg iDataCool and Athene2 clusters, and the SFB/TRR 55
QPACE~2~\cite{Arts:2015jia} and QPACE~3 machines. We thank our
colleagues in CLS for the joint effort in the generation of the gauge
field ensembles which form a basis for the computation. The Regensburg group was supported by the Deutsche
Forschungsgemeinschaft (collaborative research centre SFB/TRR-55), the
European Union’s Horizon 2020 Research and Innovation programme under
the Marie Sklodowska-Curie grant agreement no. 813942 (ITN EuroPLEx),
and the STRONG-2020 project under grant agreement no. 824093.
M.~P. acknowledges support from the EU under grant
no. MSCA-IF-EF-ST-744659 (XQCDBaryons).  S.~Prelovsek was supported by
Slovenian Research Agency ARRS (research core funding No. P1-0035 and
No. J1-8137). We are grateful to the Mainz Institute for Theoretical Physics (MITP) for its hospitality and its partial support during the course of this work. 

  \appendix 
  \section{Error treatment}\label{app:errors}

  Central values for all quantities $\bar Q$ are obtained from the average of correlation matrices over the gauge ensemble, while the errors are based on $N_b=999$ bootstrap samples.  
The $1\sigma$ standard error   formulae for a Gaussian-distributed quantity
$Q$ provides the range that captures the central $68\%$ of the bootstrap samples, which is represented by the gray bands in various figures.\footnote{The central value $\bar Q$ corresponds to the average over the  gauge configurations  and not to the average of the bootstrap samples, therefore it is possible that $\bar Q$ is not within the range captured by the $68\%$ of the bootstrap samples.} We present   resonance masses, widths/couplings, pole positions, phase shifts
and $|t|$    with the asymmetric errors $\bar Q^{+\sigma_+}_{-\sigma_-}$,
where the interval $[\bar Q-\sigma_-,\bar Q+{\sigma_+}]$ contains  the central
$68\%$ of the bootstrap samples.   The remaining quantities such as energies,
parameters of the scattering matrices are provided with symmetric errors  $\sigma=\sqrt{\mathrm{cov}_{ii}}$. Here $\mathrm{cov}$ is  the modified correlation matrix   defined as 
\begin{equation}
\label{cov}
\mathrm{cor}_{ij}=\frac{ \mathrm{cov}_{ij}}{\sqrt{ \mathrm{cov}_{ii} \mathrm{cov}_{jj} }}~, \quad \mathrm{cov}_{ij}=\frac{M}{N_b}  \sum_b \!\! ^\prime (Q_i^b-\bar Q_i)(Q_j^b-\bar Q_j)~, \quad M=\tfrac{1}{0.443^2}~.
\end{equation}  
The sum indicated by a prime runs over the bootstrap samples $b$ in which
$Q_i^b$ and $Q_j^b$ are not among the $16\%$ of the  values excluded  on
either end. $\mathrm{cov}_{ii}$ in Eq.~(\ref{cov}) is equal to
the standard covariance $\tfrac{1}{N_b} \sum_{b=1}^{N_b}  (Q_i^b-\bar Q_i)^2$ for
Gaussian-distributed quantities. $\mathrm{cov}_{ij}$ also coincides
with the standard covariance for completely correlated or
uncorrelated Gaussian-distributed quantities $i$ and $j$.
We have verified that  the standard covariance and $\mathrm{cov}$ render
almost identical errors on
the  energies and  most of the parameters  
of the scattering matrix. The advantage of
the modified correlation matrix is the exclusion of outliers for non-Gaussian
distributions with long tails. Such distributions might  occur for the
bootstrap samples of scattering-matrix parametrizations     due to the  highly non-linear nature of the box-functions $B(E_{cm})$ in Eq.~(\ref{qc}).  

\section{Eigen-energies}\label{app:discretization}  

\begin{figure}[tb]
        \begin{center}       
        \includegraphics[width=0.22\textwidth]{figs/A1pP0_calc.pdf}   
        \includegraphics[width=0.22\textwidth]{figs/A1P1_calc.pdf} 
        \includegraphics[width=0.22\textwidth]{figs/A1P2_calc.pdf} 
        \includegraphics[width=0.22\textwidth]{figs/B1P1_calc.pdf}  
         \includegraphics[width=0.22\textwidth]{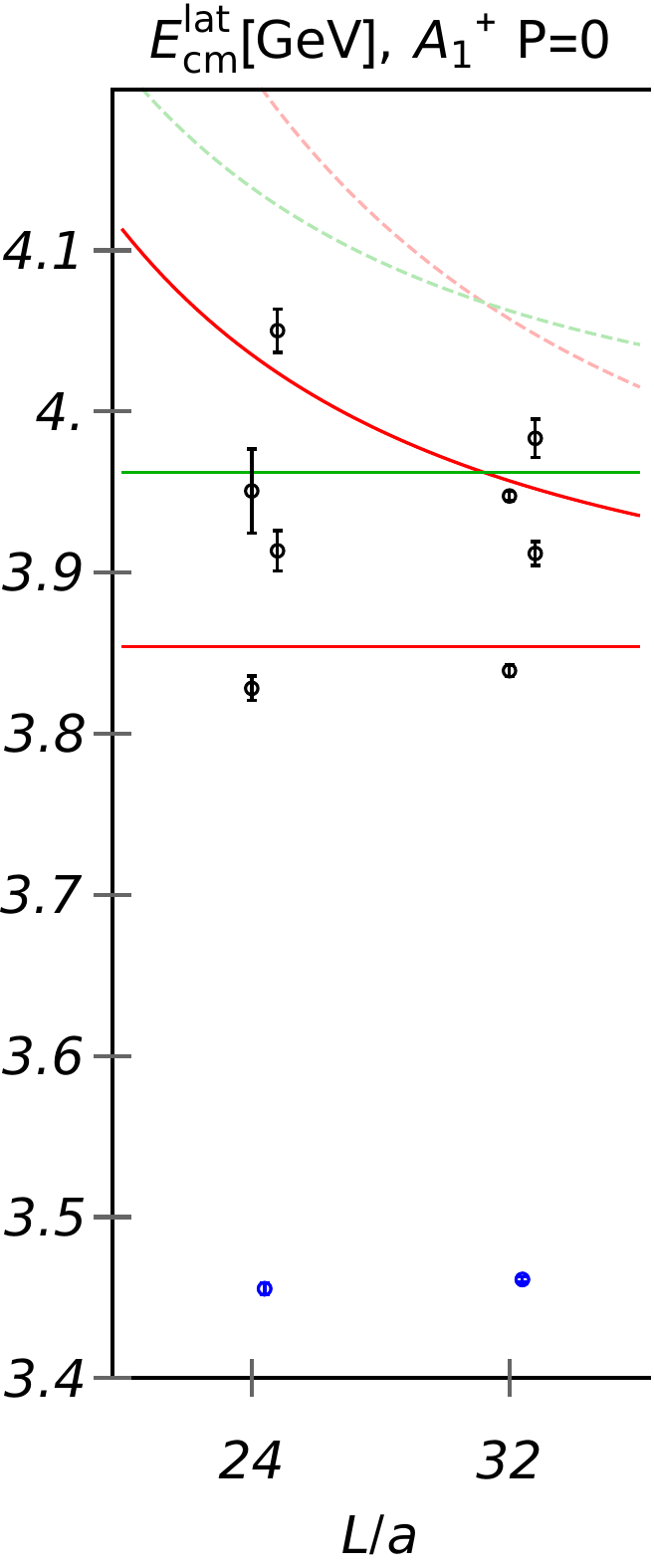}   
        \includegraphics[width=0.22\textwidth]{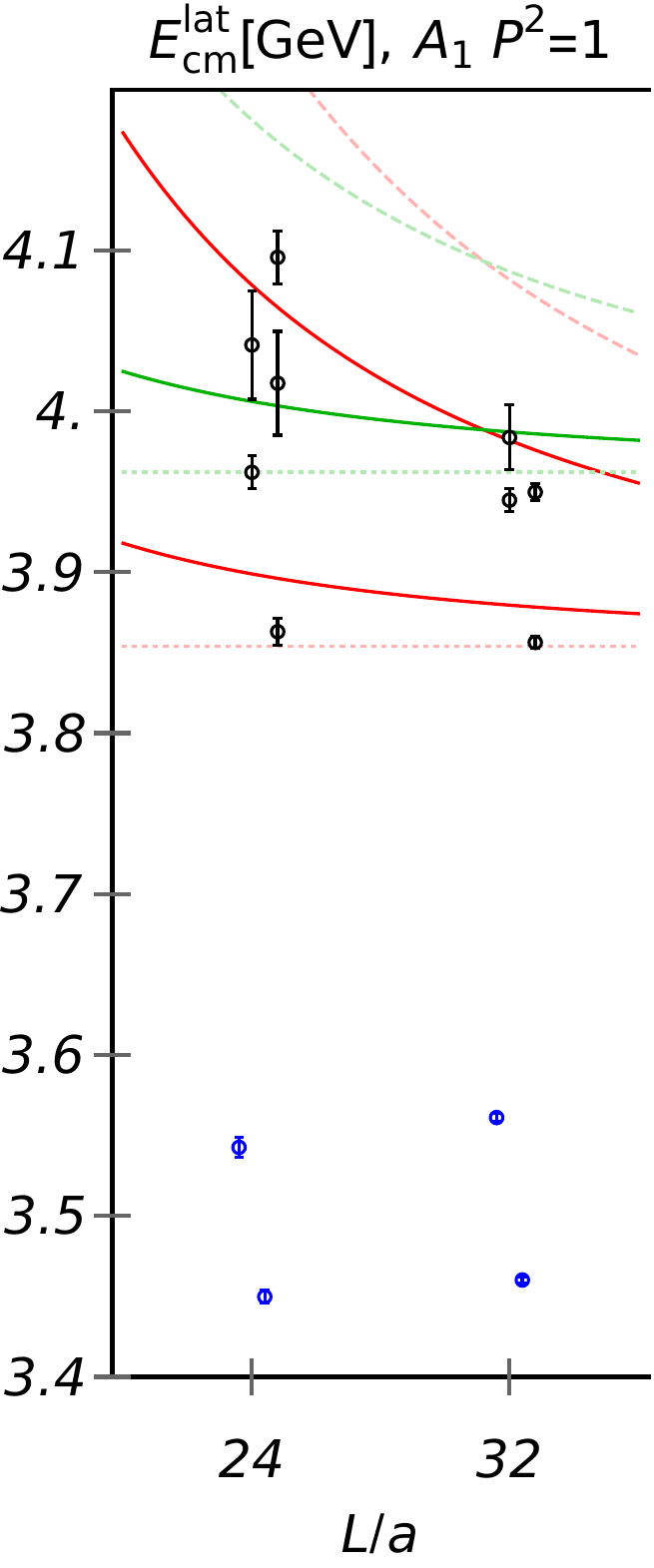} 
        \includegraphics[width=0.22\textwidth]{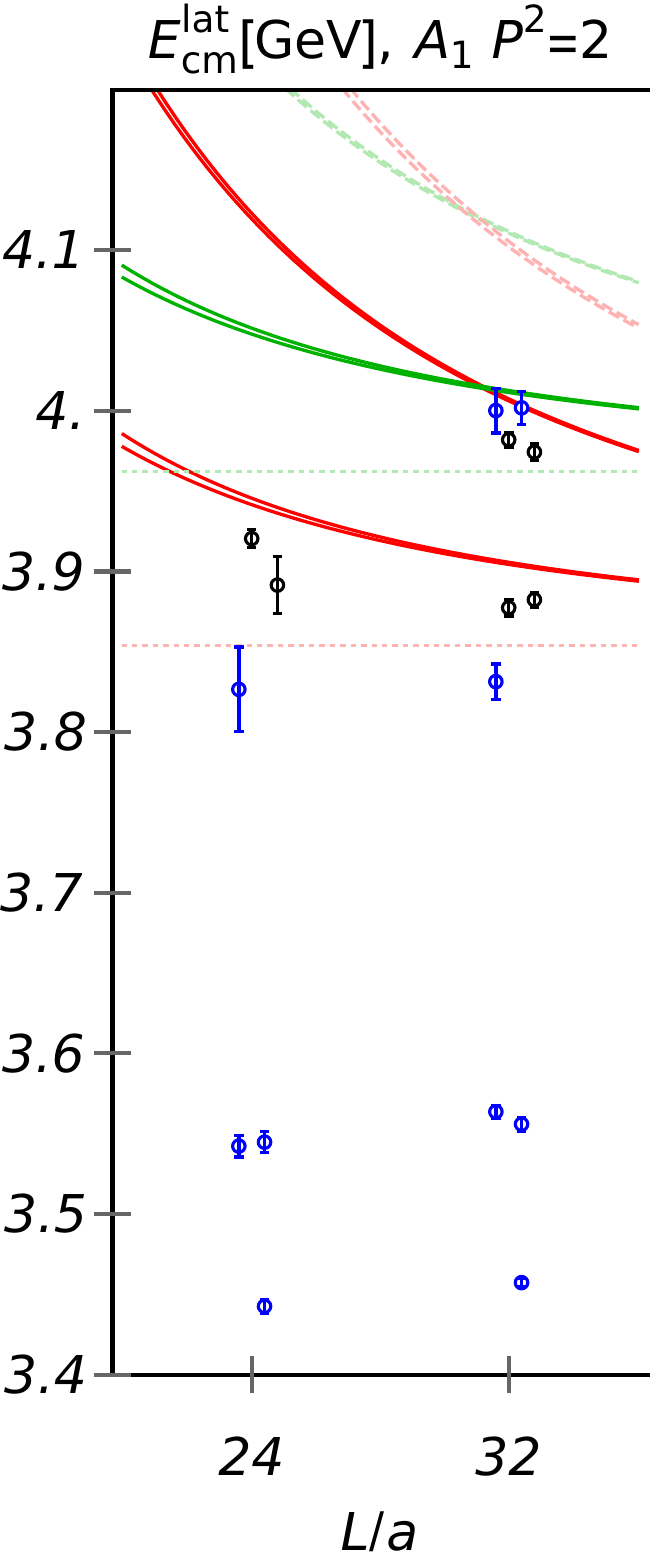} 
        \includegraphics[width=0.22\textwidth]{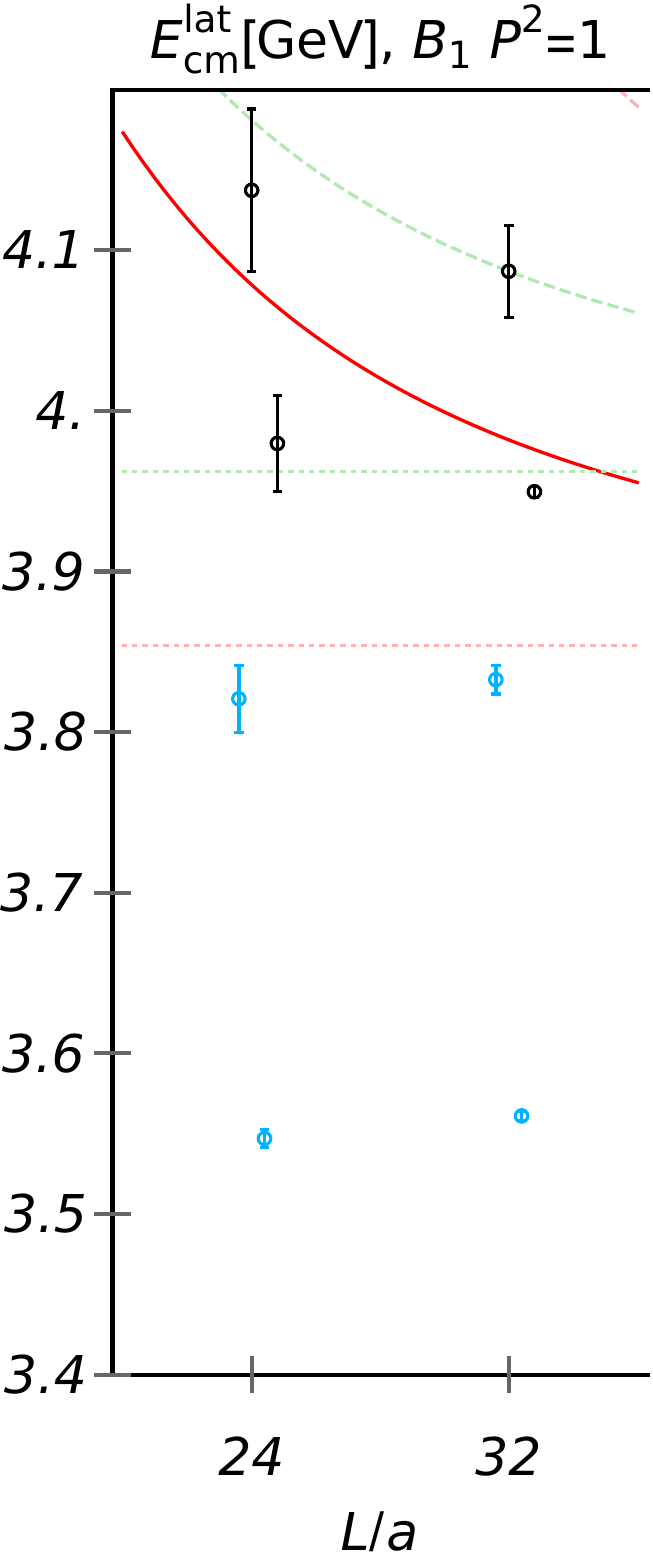}  
                \caption{   The lower figures present  the original eigen-energies $E_{cm}^{lat}$ in four irreducible representations considered. 
                The  upper figures present the eigen-energies $E_{cm}^{calc}$ obtained via  Eq. (\ref{Ecalc}) and    match  those in Fig. \ref{fig:Ecm}.   The $E^{calc}$ are   inputs to our scattering analysis, while   $E^{lat}$  are considered to be less reliable input according to the discussion   in Section \ref{sec:lattice-details}.    }
        \label{fig:discretization}
        \end{center}
\end{figure}

 The Figure \ref{fig:discretization} compares the original eigen-energies $E_{cm}^{lat}$   and the eigen-energies  $E_{cm}^{calc}$   obtained via  Eq. (\ref{Ecalc}). The  energies $E^{calc}$ are taken as  inputs to the scattering matrix according to our approach towards disretization errors, as outlined in Section \ref{sec:lattice-details}.    

\section{Fitting the parameters of the scattering matrix  }\label{app:omega}
 
 The parameters of the  matrix  $\tilde K$  are determined from the energies presented in Fig.~\ref{fig:Ecm} via the quantization condition (\ref{qc}) following the    determinant residual method proposed in  \ Ref.~\cite{Morningstar:2017spu}.  
In this method, one   determines the parameters such that the zeros of the $\Omega(E_{cm})$ function  (which  are identical to the zeros of the determinant in Eq.~(\ref{qc}))
 \begin{equation}
 \label{omega1}
 \Omega(E_{cm})=\frac{\det(A)}{\det((\mu^2+AA^\dagger)^{1/2})}~,\quad A(E_{cm})=\tilde K^{-1}(E_{cm})-B(E_{cm})~,
 \end{equation}
 are as close as possible to  $E_{cm}$ extracted from the lattice. 
 Our results are based on fits with $\mu=1$ and the parameters
   change negligibly for values of $\mu$ in the interval $[0.5,8]$. 
 Examples of $\Omega(E_{cm})$ as a function of $E_{cm}$ for the resulting
 parameters of the scattering matrix are given in
 Figs.~\ref{fig:omega-higherE} and \ref{fig:Omega_wholeE}. The values of $E_{cm}$ where $\Omega(E_{cm})$ crosses zero
 are indeed near the positions of the observed
 lattice energies. We have verified that the number of zero-crossings  equals the number of observed eigenstates in the energy-regions relevant for all our fits.  
 
\section{More details on the analysis of the scattering channels }\label{app:channels}

 Below we  specify the energy levels used for each analysis,
 referring to levels $E_{cm,n}$ plotted in Fig.~\ref{fig:Ecm} counting from
 the lowest state with $n\!=\!1$. Their values, their errors and covariance
 matrices can be obtained from the authors on  request. The masses of
the scattering particles  are $m_Da=0.8433(7)$ and $m_{D_s}a=0.8670(4)$ on
the $L/a=32$ ensemble and $m_Da=0.846(1)$ and $m_{D_s}a=0.8669(6)$ on the $L/a=24$ ensemble. For
 simplicity, we use the phase space factor from the full ensemble average
 during the pole search for all bootstrap samples. The central values are not affected by this procedure.

 \subsection{$D\bar D$   scattering with $l=0$ near threshold } \label{app:DD}

\begin{figure}[tb]
        \begin{center}
        \includegraphics[width=0.5\textwidth]{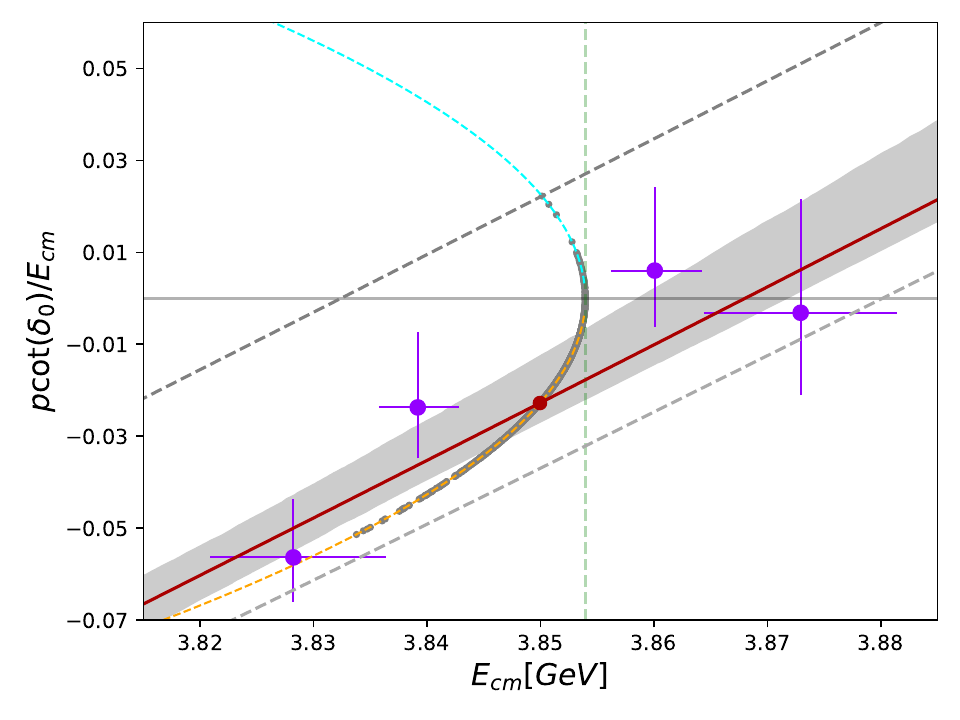}
                \caption{ Pole distribution [gray filled circles] along the (virtual) bound state constraints 
                 [(cyan) orange dashed curve] in the elastic $D\bar D$ scattering. The energy levels [violet filled 
                 circles] considered in the fit and the fit results [red solid line and the gray band] are also shown. 
                 The two gray dashed lines represent the two extreme cases, one with the virtual bound state and 
                 one with no poles on the real axes. }
        \label{fig:elasticDD_110110}
        \end{center}
\end{figure}

This analysis employs four energy levels closest to the $2m_D$ threshold    in the $A_1^{(+)}$  irreps shown in Fig.~\ref{fig:Ecm}: these are levels $n=2(3)$ from $|\vec P|=0(1)$ on both volumes.  
   The   charmonium-like state obtained lies just below
   threshold, therefore the relative error on its binding energy given in Eq.~(\ref{chic0DD-lat}) is large. We note that 6\%   of the  bootstrap samples do not   render   any poles on the real axes   ---    this   corresponds to the bootstraps  for which $p\cot\delta /E_{cm}$  just fails to cross the orange line in Fig.~\ref{fig:DD}a. 
   An additional 6\%  of the  bootstrap samples render a virtual bound
   state - this corresponds to the bootstraps  for which 
   $p\cot\delta /E_{cm}$ crosses  $ip/E_{cm}=|p|/E_{cm}$ rather than  $-
   |p|/E_{cm}$ in Fig.~\ref{fig:DD}a.  Both of these scenarios happen in
   extreme cases and end up within the 32\% of bootstrap samples that are excluded when computing the errors via (\ref{cov}).
   In Fig.~\ref{fig:elasticDD_110110}, we present the pole positions along the (virtual) bound state constraints across the bootstrap samples showing the continuous distribution of the poles along the constraint curves and hence also across the Riemann sheets. 

 The preferred fit with the scattering parameters (\ref{DD-params}) utilizes the 4 levels shown in violet in Fig.~\ref{fig:DD}a. We also performed  the fits using the 3 lowest levels, the 6 lowest levels and all 7 levels shown: the ensemble averages of the data lead to a bound state in all these fits   and the binding energy is within the error given in Eq.~(\ref{chic0DD-lat}).
  
  This analysis of $D\bar D$ scattering near threshold includes only the eigen-states with energies close to the threshold and omits the eigen-state related to $\chi_{c0}(1P)$, which is significantly below threshold.  We are unable to constrain the $D\bar D$ scattering below the  lowest violet point in  Fig.~\ref{fig:DD}a. Hence,  the pole at around $E_{cm}\simeq 3.80~$GeV, which would arise at  the  crossing of the orange and red curves (and would violate the consistency check, see Section VC of
    \cite{Piemonte:2019cbi}), is below the region in which our analysis can reasonably be applied
    and also outside of the energy range of interest.

  \subsection{$D_s\bar D_s$ scattering with $l=0$ near  threshold in the one-channel approximation}\label{app:DsDs}
  
  This analysis employs  only  those eigenstates  whose overlaps are     dominated by
  $D_s\bar D_s$ operators and that do not have significant overlap with $D\bar D$ operators. These
  are  the four levels in Fig.~\ref{fig:Ecm} near the $D_s\bar D_s$ threshold in
the $A_1^{(+)}$ irreps:   levels $n=3,4$ from   $|\vec P|^2=0,1$ on the $N_L=24$ ensemble and
  levels $n=4,7$ from  $|\vec P|^2=1,2$ on $N_L=32$.   Here 97\% of  the bootstrap  samples result
    in a bound-state pole, while 2.3\% result in a virtual
  bound-state pole and 0.7\%  do not render any poles on the real axis -- the
  latter two cases end up among the extremal 32\% of
    bootstrap samples. 
          
   \subsection{$D\bar D$   scattering  with $l=2$  }\label{app:dwave}
   
   $D\bar D$   scattering in  partial wave $l=2$ is the not the main focus of our study.  It was considered in order  to investigate and  constrain its contribution to the $A_1$ irreps  we have studied.  Since this partial wave was initially not the goal of our study, we did not  evaluate all irreps where it appears (for example $E^+$ and $T_2^+$ for $P=0$), instead we implemented only the  $B_1$ irrep   with $|\vec P|^2=1$.    The extraction of the phase shift in Eqs.~(\ref{dwave-params},\ref{dwave-pole}) employs four lattice  levels in the  $B_1$ irrep with $|\vec P|^2=1$:   these are levels $n=3,4$ on both lattice volumes   (levels $n=1,2$   correspond to the ground states with  $J^{PC}=2^{++}$ and $2^{-+}$, respectively).  
     A fit using only three lattice levels  (omitting the higher level on the smaller volume) renders the resonance pole position $E^p=( 4.013 ^{~+0.013} _{~-0.016})~-  \tfrac{i}{2} ~(0.098 ^{+0.044}_{ -0.057})~\mathrm{GeV}$. This is compatible with our main result (\ref{dwave-pole}) and has a larger central value for the width.     
      
    \subsection{Coupled $D\bar D$, $D_s\bar D_s$    scattering with $l=0$ for $E_{cm}\simeq 3.9-4.13~$GeV    } \label{app:rough}

\begin{figure}[h!]
	\begin{center} 
	\includegraphics[width=0.8\textwidth]{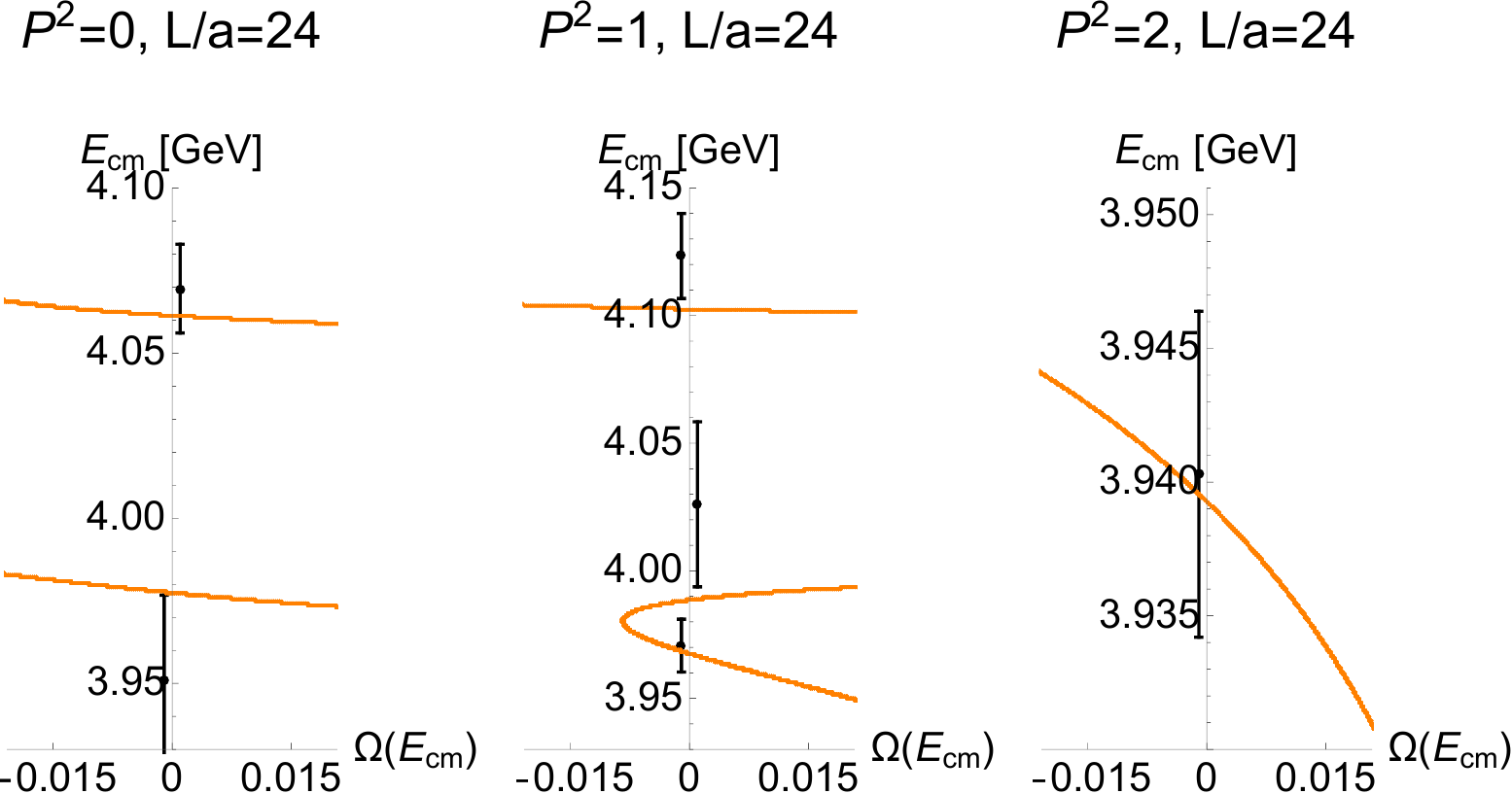}  
	
	\vspace{0.4cm}
	
	\includegraphics[width=0.8\textwidth]{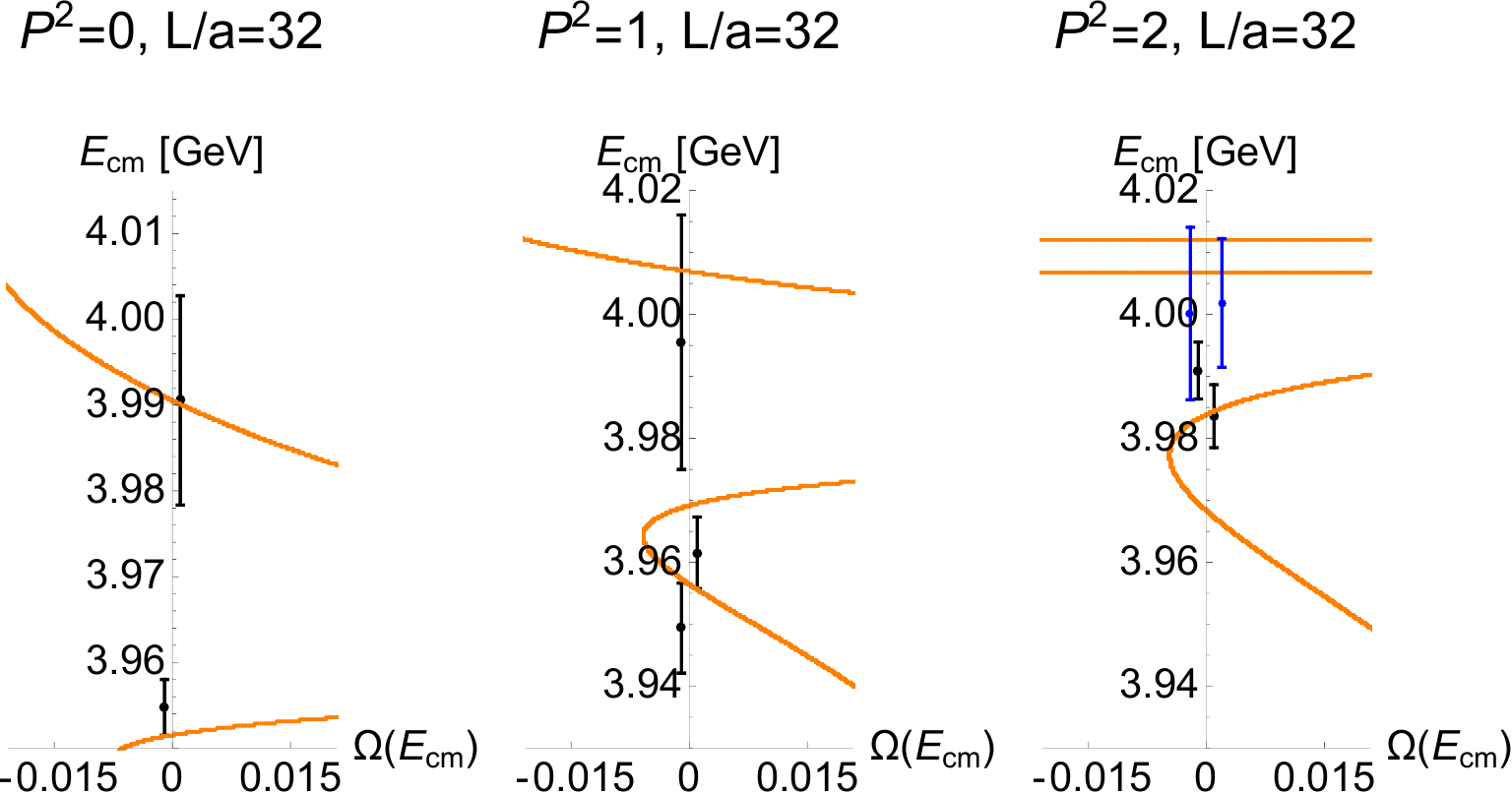}    
		\caption{ The function $\Omega(E_{cm})$ (defined in
                    Eq.~(\ref{omega1})) for the  resulting   scattering
                  matrix of the coupled channels $D\bar D-D_s\bar D_s$
                  (\ref{DD-DsDs-without2-params})   is given by the orange
                  line. The observed eigen-energies are given by the  circles: the black levels are employed to fit the parameters (\ref{DD-DsDs-without2-params}), while the blue circles are not.    }
	\label{fig:omega-higherE}
	\end{center}
\end{figure}

   Fig.~\ref{fig:omega-higherE} shows   an example of   $\Omega(E_{cm})$ (\ref{omega1})  for the   parameters of the coupled-channel scattering matrix given in Eq.~(\ref{DD-DsDs-without2-params}). The values of $E_{cm}$ at which  $\Omega$   crosses zero are indeed near the observed eigen-energies~(indicated by the black circles).  The number of crossings agrees with the number of observed levels in the relevant energy ranges.

\begin{figure}[tb]
        \begin{center}
        \includegraphics[width=0.5\textwidth]{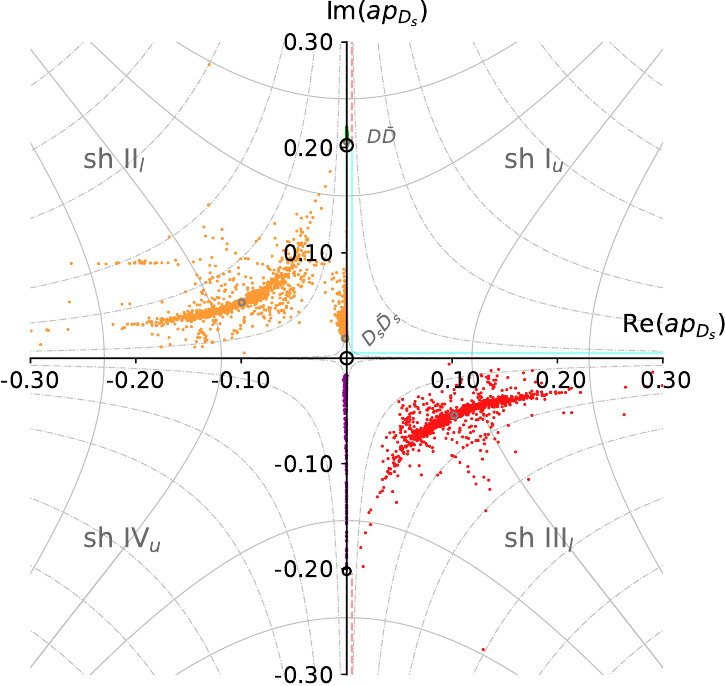}
                \caption{Pole distribution of different poles, presented in Fig. \ref{fig:DD-DsDs-poles}, across various
bootstrap samples in the complex $p_{D_s}$ plane where $p_{D_s}$ is the momentum of the $D_s$ meson in $D_s\bar D_s$
scattering in the CMF. The color coding of the poles are the same as in Fig. \ref{fig:DD-DsDs-poles}. The pole positions for the 
ensemble average of the data are shown with gray unfilled circles. The gray solid (dot-dashed) curves represent the lines of constant real (imaginary) $aE_{cm}$. The cyan line represents the physical axis presented in Fig.~\ref{fig:sketch}. The red dashed lines indicate the region below the elastic threshold, where the sheets are not connected. }
        \label{fig:pDs_plane_poles_withD}
        \end{center}
\end{figure}

In Figure~\ref{fig:pDs_plane_poles_withD}, we present the pole 
distribution across the bootstrap samples for various poles we extract in the complex $p_{D_s}$ plane, where $p_{D_s}$ is 
the momentum of the $D_s$ meson in $D_s\bar D_s$ scattering in the CMF. The two islands of poles, one in sheet II and 
the other in sheet IV, lying close to $Im(ap_{D_s})$ are mutually exclusive and hence represent the same dynamics.
The island in sheet II constitutes 70\% of the samples, while a pole appears at a similar location on sheet IV in the remaining samples. 
Hence the
results for the pole location related to  $\chi_{c0}^{D_s\bar D_s}$ given in
Eq.~(\ref{chic0DsDs-pole}) are based on the samples for which this pole
appears on sheet II, whereas the errors on the rates and phase shifts presented in Figs.~\ref{fig:DD-DsDs-with2-t} 
and \ref{fig:DD-DsDs-with2-phases} are computed from the entire bootstrap samples.
 
\section{Coupled $D\bar D$, $D_s\bar D_s$ scattering in a wider energy region } \label{sec:wholeE}
         
In Section~\ref{sec:channels}, we discussed the analysis of
rather narrow energy ranges with a linear and/or constant
parametrization in $s$ for the elements $\tilde
K_{ij}^{-1}(s)/\sqrt{s}$ (see Eq.~(\ref{DD-DsDs})). A single
description of coupled $D\bar D-D_s\bar D_s$ scattering encompassing
the whole of the energy range from $2m_D$ up to 4.13~GeV requires
additional parameters. This is difficult as, with the statistics and
the number of lattice QCD ensembles available to us, the fits become
unstable. Instead, as a cross-check, we model the
  infinite-volume scattering matrix in the wider energy range using
  the parametrizations similar to those presented in Section~\ref{sec:channels}.  
   One of the aims is to verify that the resulting scattering matrix
  predicts the same number of finite-volume energy levels as observed
  in the actual simulation. 
  
         \begin{figure}[t]
	\begin{center}
	\includegraphics[width=0.5\textwidth]{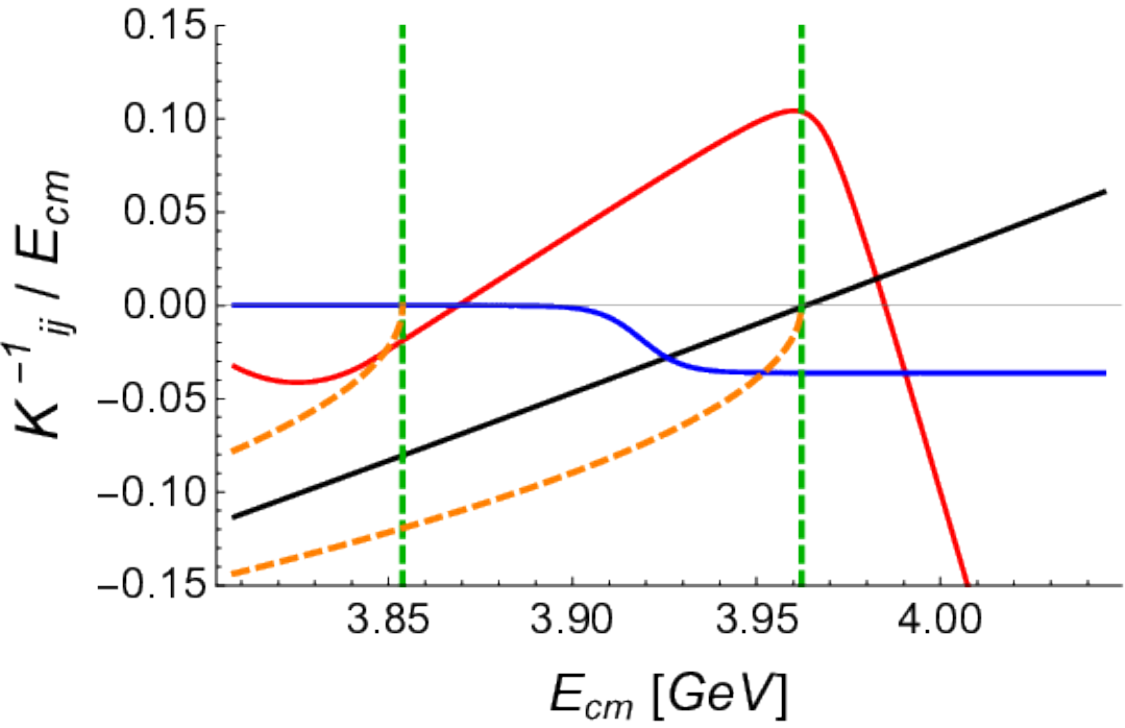}  
        \caption{The modeled $\tilde K^{-1}/\sqrt s$ -matrix elements  
for the coupled $D\bar D-D_s\bar D_s$ scattering: red for $ D\bar D\to  D\bar D$, black for $ D_s\bar D_s\to  D_s\bar D_s$ 
and blue for $ D_s\bar D_s\to D\bar D$.  The green dashed lines indicate the $ D\bar D$ and the $ D_s\bar D_s$ thresholds.  The left orange dashed line gives the bound state constraint for  $D\bar D$ channel. The right orange   line gives the bound state constrain for the $D_s \bar D_s$ channel in the limit that the two channels are decoupled.  }
	\label{fig:K-wholeE1}
	\end{center}
\end{figure}

 The $t$-matrix elements are modeled in the energy range $E_{cm}\simeq 2m_D-4.13~$GeV as shown for $\tilde K^{-1}/\sqrt s$ in Fig.~\ref{fig:K-wholeE1}.  We require that  they are continuous in energy and that they have continuous derivative. In the high energy region they asymptote to the linear dependence on $s$  (\ref{linear},\ref{DD-DsDs}) and 
the parameters are fixed to the values (\ref{DD-DsDs-without2-params}) obtained from the coupled-channel analysis.   Below we provide more details on each $t$-matrix element in turn:
\begin{itemize}
\item[$t_{11}$]   
The energy region considered is divided into three intervals, as shown by the red line in the figure.  $t_{11}$  asymptotes to the coupled and single channel results (of the main text)   in the high and middle energy intervals, respectively. In order to ensure a smooth transition between two regions, we emply hyperbola-type shape for  $\tilde K_{11}^{-1}/\sqrt{s}$ 
\begin{equation}
\biggl[\frac{\tilde K_{11}^{-1}}{\sqrt{s}}-a_{11}-b_{11}s\biggr] \biggl[\frac{\tilde K_{11}^{-1}}{\sqrt{s}}-a_{11}^\prime -b_{11}^\prime s\biggr] = c^{hyp} 
\end{equation}
 that smoothly asymptotes\footnote{The  transition is smooth for positive $c^{hyp}$, while it corresponds to an abrupt transition between two lines for $c^{hyp}=0$.} to the linear dependences $\tilde K_{11}^{-1}/\sqrt{s}=a_{11}+b_{11}s$ (\ref{DD-DsDs-without2-params}) and $\tilde K_{11}^{-1}/\sqrt{s}=a_{11}^\prime +b_{11}^\prime s$ (\ref{DD-params}). The four parameters $a,b$ are fixed to the values in the main text. The value of the smoothing parameter $c^{hyp}$ is the only free parameter of $\tilde K^{-1}_{ij}$ in this appendix. Its value 
 $c^{hyp}=0.00021(2)$ is obtained from fitting the  scattering matrix to  all energy levels and the resulting fit has $\chi^2/dof=1.8$.    
  In the  region below $D\bar D$ threshold, we choose a shape of $\tilde
  K_{11}^{-1}$ which prevents the occurrence of a
  second bound-state (this would correspond to the red and orange lines
  in the figure intersecting a second time).  The exact form of this choice is not
  important as this is beyond the region of interest. 
\item[$t_{12}$] In the upper region $\tilde K_{12}^{-1}/\sqrt{s}$   asymptotes to 
  the constant value of the coupled channel analysis (\ref{DD-DsDs-without2-params}). In the region  near $D\bar D$ threshold,   where the
  effects from $D_s\bar D_s$ channel are expected to be negligible, it asymptotes to zero. The smooth transition between two constant values is ensured by using  the sigmoid function. 
\item[$t_{22}$] This element is parametrized as 
  in Eq.~(\ref{DD-DsDs-without2-params}),  for the entire energy region, see the black line in the figure.   Note that we ignore any crossing of the $D_s\bar{D}_s$ bound state condition with the $t_{22}$ parametrization
  that occurs well below the $D\bar{D}$ threshold.
\end{itemize}

\begin{figure}[h!]
	\begin{center}
	\includegraphics[width=0.8\textwidth]{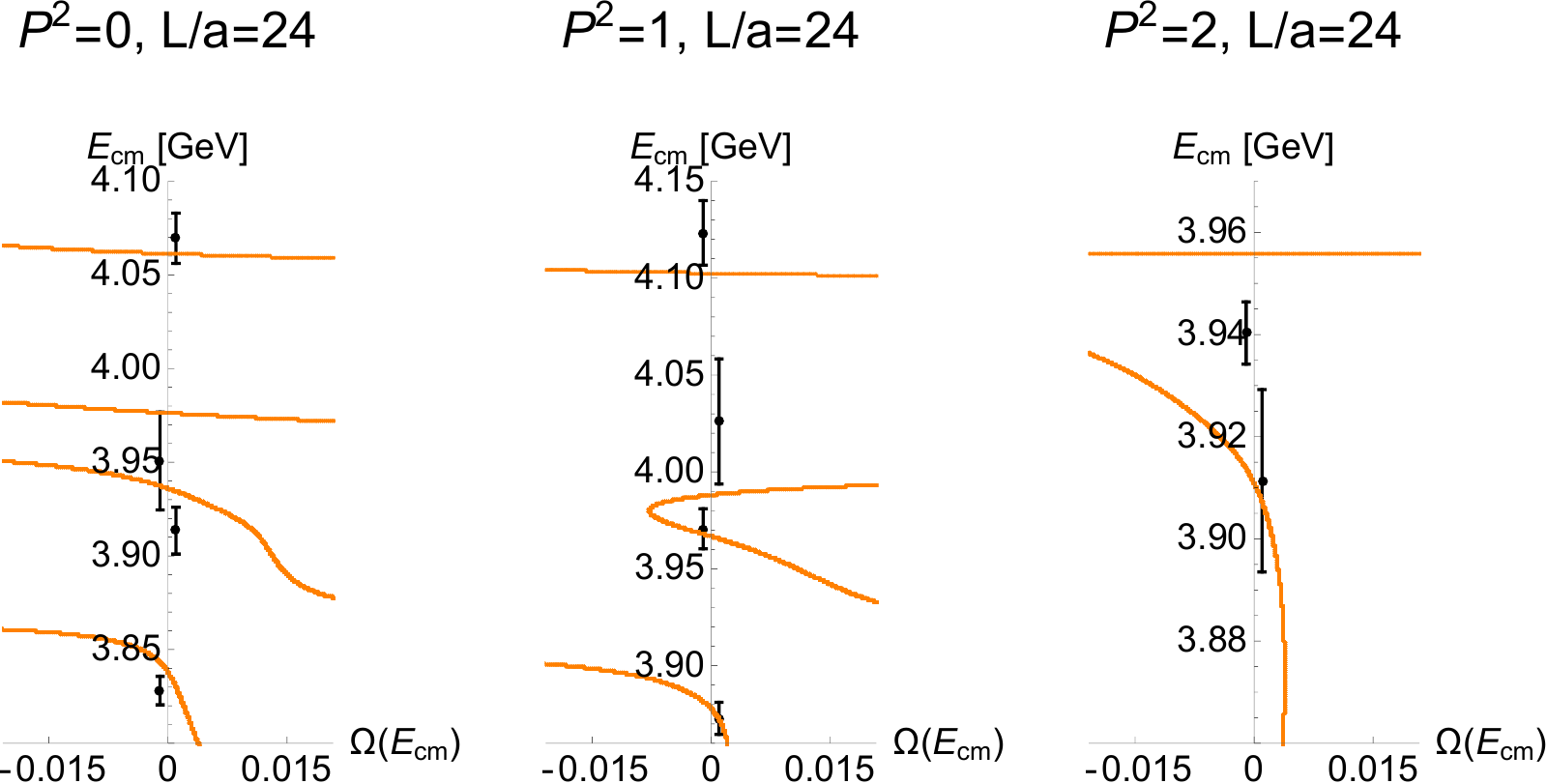}  
	
	\vspace{0.4cm}
	
	\includegraphics[width=0.8\textwidth]{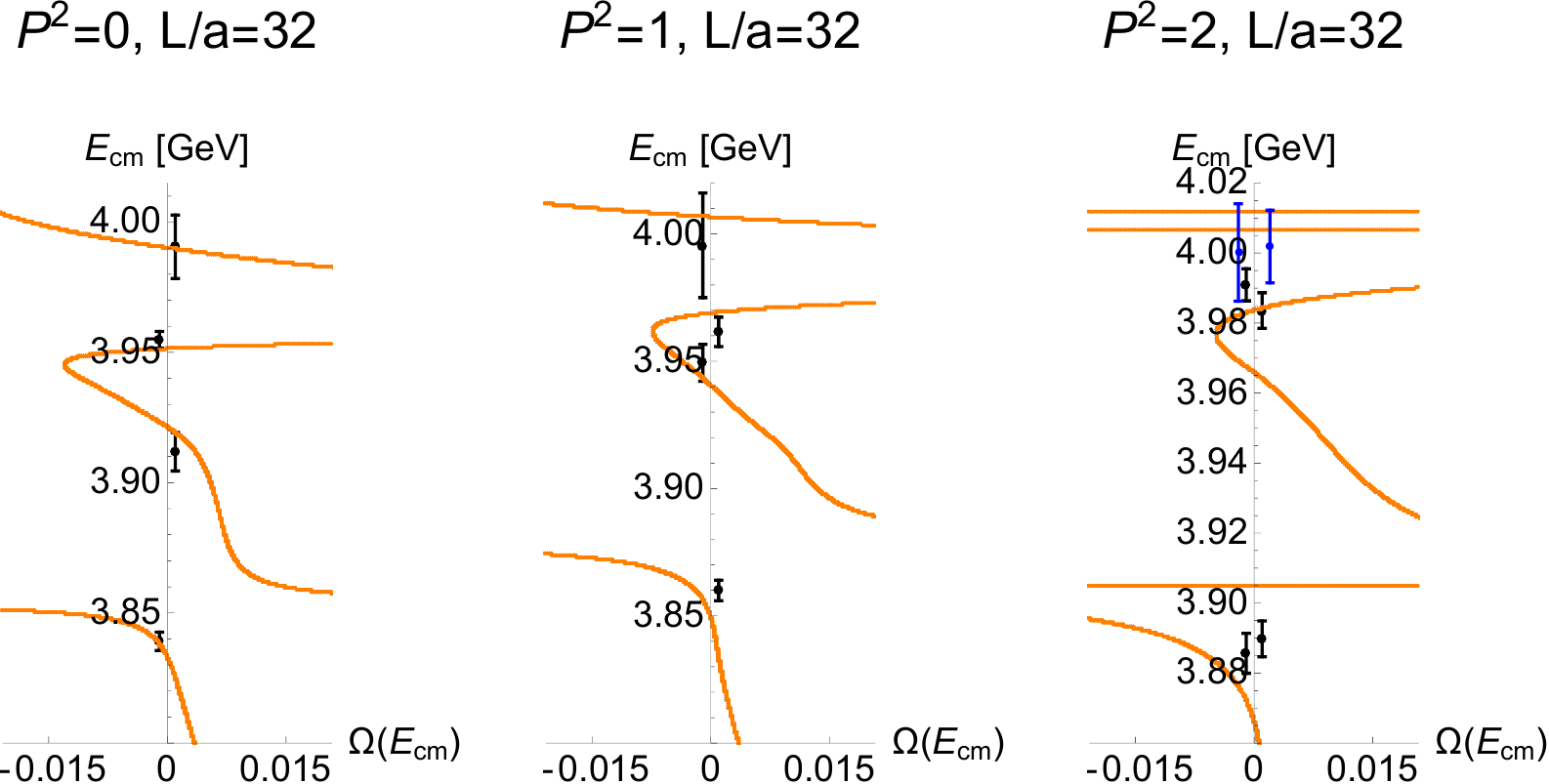}   
		\caption{The $\Omega(E_{cm})$ function (defined in
                  Eq.~(\ref{omega1})) for the scattering matrix of the coupled channels $D\bar D-D_s\bar D_s$ in the wider energy region ($E_{cm}\simeq 2m_D-4.1~$GeV) is given by the orange line. The observed eigen-energies are given by circles and the coloring is the same as in Fig. \ref{fig:Ecm}.   }
	\label{fig:Omega_wholeE}
	\end{center}
\end{figure}

We find that the number of poles with the above-designed scattering
matrix is the same as that obtained from the analysis of the
separate energy regions. The pole locations on the various complex Riemann
sheets and their residues are also almost unchanged.  Following 
L\"uscher's finite-volume analysis, we extracted the
finite-volume spectrum from this scattering matrix. 
In Fig.~\ref{fig:Omega_wholeE}, we present the $\Omega(E_{cm})$
  function defined in Eq.~(\ref{omega1}). The zeros of
  $\Omega(E_{cm})$ are the predictions for the finite-volume spectrum
  derived from the scattering matrix. The points indicate the energy
  levels observed in the actual simulation.  It is clear from the
  figure that the predicted spectrum agrees qualitatively with the
  lattice energy levels within the energy range of interest and that the
  same number of levels are obtained. 

  \providecommand{\href}[2]{#2}\begingroup\raggedright\endgroup

        
 \end{document}